\documentclass{amsart}
\usepackage{amsmath,latexsym,amssymb}

\theoremstyle{plain}
\newtheorem{definition}{Definition}[section]
\newtheorem{theorem}{Theorem}[section]
\newtheorem{proposition}{Proposition}[section]
\newtheorem*{claim}{Claim}
\newtheorem{lemma}{Lemma}[section]
\newtheorem{corollary}{Corollary}[section]
\newtheorem{remark}{Remark}[section]

\def\ds{\displaystyle}
\def\nd{\noindent}

\def\R{{\mathbb R}}

\def\C{{\mathbb C}}
\def\oH{\buildrel\circ\over H}
\def\oH1{\buildrel\circ\over H\kern-.02in{}^1}
\def\qed{{\hfill $\Box$}}
\def\l{\ell}
\def\dotf{\dot{f}}
\def\tildeR{\widetilde R}

\begin{document}
In the book "Operator theory and Applications", Fields Institute
Communications vol. 25, AMS, Providence, 2000, pp.15-76 (Ed.A.G.Ramm,
P.N.Shivakumar, A.V.Strauss).

\title{
Property C for ODE and applications to inverse problems.}

\author{A.G. Ramm}
\address{ Mathematics Department, Kansas State University, \\
  Manhattan, KS 66506-2602, USA}

\subjclass{34B25, 35R30, 81F05, 81F15,
73D25}

\email{  ramm@math.ksu.edu}

\keywords {Property C, inverse scattering, inverse problems,
incomplete data, fixed-energy phase shifts, $I$-function,
Weyl function, spectral function}

\date{}

\begin{abstract}
An overview of the author's results is given. Property C stands for
completeness of the set of products of solutions to homogeneous linear
Sturm-Liouville equations. The inverse problems discussed include
the classical ones (inverse scattering on a half-line,
on the full line, inverse spectral problem), inverse scattering problems
with incomplete data, for example, inverse scattering on the full
line when the reflection coefficient is known but no information
about bound states and norming constants  is available, but it is a
priori known that the potential vanishes for $x<0$,
or inverse scattering on a half-line when the phase shift of the
$s$-wave is known for all energies, no bound states and norming constants
are known, but the potential is a priori known to be compactly supported.
If the potential is compactly supported, then it can be uniquely
recovered from the knowledge of the Jost function
$f(k)$ only, or from $f'(0,k)$, for all $k\in \Delta$, where
$\Delta$ is an arbitrary  subset of $(0,\infty)$ of positive Lebesgue 
measure.

Inverse scattering problem for an inhomogeneous Schr\"odinger equation
is studied.

Inverse scattering problem with fixed-energy phase shifts as the data
is studied.

Some inverse problems for parabolic and hyperbolic equations are
investigated.

A detailed analysis of the invertibility of all the steps in the
inversion procedures for solving the inverse scattering and spectral
problems is presented.

An analysis of the Newton-Sabatier procedure for 
inversion of fixed-energy phase shifts is given.

Inverse problems with mixed data are investigated.

Representation formula for the $I$-function is given and
properties of this function are studied. 

Algorithms for finding the scattering data from the $I$-function,
the $I$-function from the scattering data and the potential from the
$I$-function are given.

A characterization of the Weyl solution and a formula for this solution
in terms of Green's function are obtained.

\end{abstract}

\maketitle


\vfill\pagebreak

\thispagestyle{empty}

\section*{Table of Contents}

\nd{\bf 1. Property C for ODE}

\vspace{.1in}
\nd{\bf 2. Applications of property C}

\begin{enumerate}

\item[] {2.1 Uniqueness of the solution to inverse scattering problem
with the data $I$-function.}

\item[] {2.2 Uniqueness of the solution to inverse scattering problem
on the half-line.}

\item[] {2.3 Compactly supported potentials are uniquely determined by the
phase shift of $s$-wave.}

\item[] {2.4 Recovery of $q\in L_{1,1}(\R)$ from the reflection
coefficient alone.}

\item[] {2.5 Inverse scattering with various data.}
\end{enumerate}

\vspace{.1in}
\nd{\bf
3. Inverse problems on a finite interval with mixed data.}

\vspace{.1in}
\nd{\bf
4. Property C and inverse problems for some PDE.
Recovery compactly supported potential from the knowledge
of $f(k)$ or $f'(0,k)$.}

\vspace{.1in}
\nd{\bf
5. Invertibility of the steps in the inversion procedure,
in the inverse scattering and spectral problems.}

\vspace{.1in}
\nd{\bf
6. Inverse problem for an inhomogeneous Schr\"odinger equation.}

\vspace{.1in}
\nd{\bf
7. Inverse scattering problem with fixed-energy data.}

\begin{enumerate}

\item[] {7.1 Three-dimensional inverse scattering problem.
Property C. Stability estimates.}

\item[] {7.2 Approximate inversion of fixed-energy phase shifts.}

\end{enumerate}

\vspace{.1in}
\nd{\bf
8. A uniqueness theorem for inversion of fixed-energy phase shifts.}

\vspace{.1in}
\nd{\bf
9. Discussion of the Newton-Sabatier procedure for recovery of $q(r)$
from the fixed-energy phase shifts.}

\vspace{.1in}
\nd{\bf
10. Reduction of some inverse problems to an overdetermined
Cauchy problem. An iterative method for solving this problem.}

\vspace{.1in}
\nd{\bf
11. Representation of $I$-function.}

\vspace{.1in}
\nd{\bf
12. Algorithms for finding $q(x)$ from $I(k)$.}

\vspace{.1in}
\nd{\bf
13. Remarks.}

\begin{enumerate}
\item[] {13.1. Representation of the products of solutions
to Schr\"odinger equation (1.1).}

\item[] {13.2. Characterization of Weyl's solutions.}

\item[] {13.3. Representation of the Weyl solution via the Green
function.}
\end{enumerate}
\vspace{.1in}
\nd{\bf Bibliography}

\vfill\pagebreak


\section{Property C for ODE} 

In this paper a review of the author's results is given and some
new results are included. The bibliography is not complete. Only the papers
and books used in the work on this paper are mentioned.
The contents of this paper are clear from the table of contents.

The results presented in this paper include:

\begin{enumerate}

\item Property C for ordinary differential equations
(ODE), that is, theorems about completeness of the
sets of products of solutions to homogeneous ODE.

\item Uniqueness theorems for finding the potential
  
a) from the $I$-function (which equals the Weyl function),
  
b) from the classical scattering data for the half-axis problem
    (a new very short proof which does not use the Marchenko method),
 
c) from the phase shift of $s$-wave 
in the case when the potential $q$ is compactly
  supported and no bound states or norming constants are known,
  
d) from the reflection coefficient only (when $q=0$ for $x<x_0$),
  
e) from mixed data: part of the set of eigenvalues and
  the knowledge of $q(x)$ on a part of a finite interval,
  
f) from overdetermined Cauchy data,
  
g) {\it from part of the fixed-energy phase shifts},
  
h) from various type of data which are typical in PDE problems,
  
i) from $f(k)$ or $f^\prime(0,k)$ when $q$ is compactly supported,
  

j) from the scattering data for a solution to an
inhomogeneous Schr\"odinger equation.

\item Reconstruction algorithms for finding the potential
from overdetermined Cauchy data, for finding $f(k)$ and $f'(0,k)$
from the scattering data, for finding the scattering data from the
$I$-function and the $I$-function from the scattering data.

\item Properties of the $I$-function and a representation formula for it.

\item Stability estimate for the solution of the inverse scattering problem
  with fixed-energy data. Example of two compactly supported
real-valued piecewise-constant potentials which produce practically the
same phase shifts for all values of $\ell$.

\item Discussion of the Newton-Sabatier procedure for
inversion of the fixed-energy
  phase shifts. Proof of the fact that this procedure
cannot recover generic potentials, for example,
 compactly supported potentials.  

\item Detailed analysis and proof of the invertibility of each of the
  steps in the inversion schemes of Marchenko and Gelfand-Levitan.

\item Representation of the Weyl solution via the Green function
and a characterization of this solution by its behavior for
large complex values of the spectral parameter and $x$ running
through a compact set.
 
\end{enumerate}

Completeness of the set of products of solutions to ODE has been used
for inverse problems on a finite interval in the works of Borg
\cite{B} and Levitan \cite{Lt2}, \cite{Lt3}.

Completeness of the set of products of solutions to homogeneous partial
differential equations (PDE) was introduced and used extensively
under the name property C in \cite{R9}-\cite{R14}, and \cite{R}.
Property C in these works differs essentially from
the property C defined and used in this paper:
while in \cite{R9}-\cite{R14}, and \cite{R} property C
means completeness of the set of products of solutions
to homogeneous PDE with fixed value of the spectral parameter,
in this paper we prove and use completeness of the sets
of products of solutions to homogeneous
ordinary differential equations (ODE)
with variable values of the spectral parameter.
Note that the dimension of the null-space of
a homogeneous PDE (without boundary conditions)
with a fixed value of the spectral parameter
is infinite, while the dimension of the null-space
of a homogeneous ODE (without boundary conditions)
with a fixed value of the spectral parameter
is finite. Therefore one cannot have property C
for ODE in the sense of \cite{R9}-\cite{R14}, and \cite{R},
because the set of products of solutions
to homogeneous ODE with fixed value of the spectral parameter
is finite-dimensional.

In this paper property C for ordinary differential equations
is defined,
proved and used extensively. Earlier papers are \cite{RP} and
\cite{R1}.

Let
\begin{equation}
 \l u:= u''+k^2u-q(x)u=0, \qquad x\in \R=(-\infty,\infty).
   \tag{1.1}
\end{equation}

Assume
\begin{equation} q\in  L_{1,1},
\quad L_{1,m} := \{ q:q= \overline{q},
   \int^\infty_{-\infty} (1+|x|)^m |q(x)|\,  dx < \infty \}.
   \tag{1.2} \end{equation}

It is known \cite{M}, \cite{R}
that there is a unique solution (the Jost solution)
to (1.1) with the asymptotics
\begin{equation}
  f(x,k)= e^{ikx}+ o(1),  \quad x\to +\infty.
  \tag{1.3} \end{equation}
We denote $f_+(x,k): = f(x,k)$, $f_-(x,k): = f(x,-k)$, $k\in\R$.
The function $f(0,k)=f(k)$ is called the Jost function.
The function $f(k)$ is analytic in $\C_+: =\{k: Imk> 0\}$ and has at most 
finitely many zeros in $\C_+$ which are located at the points 
$ik_j, k_j>0$, $1\leq j\leq J$. The numbers $-k_j^2$ are the negative
eigenvalues
of the selfadjoint operator defined by the differential expression
$L_q:=-\frac{d^2}{dx^2}+q(x)$ and the boundary condition $u(0)=0$ in
$L^2(\R_+)$, $\R_+=(0,\infty)$. The function $f(k)$ may have zero at $k=0$.
This zero is simple: if $f(0)=0$ then $\dot f(0)\not= 0$,
$\dot f:=\frac{\partial f}{\partial k}$.

Let $\varphi$ and $\psi$ be the solutions to (1.1) defined by the
conditions
\begin{equation}
  \varphi(0,k)=0, \quad\varphi'(0,k)=1;
 \quad  \psi(0,k)=1, \quad\psi'(0,k)=0,
  \tag{1.4}\end{equation}
where $\varphi':= \frac{\partial \varphi}{\partial x}$. It is known \cite{M},
\cite{R}, that
$\varphi(x,k)$ and $\psi(x,k)$ are even entire functions of $k$ of 
exponential type $\leq|x|$.

Let $g_\pm(x,k)$ be the unique solution to (1.1) with the asymptotics
\begin{equation}
  g_\pm(x,k)= e^{\pm ikx} +o(1),
  \quad x\to -\infty
  \tag{1.5} \end{equation}

\begin{definition} 
Let $p(x)\in  L_{1,1} (\R_+)$ and assume
\begin{equation}
  \int^\infty_0 p(x) f_1(x,k) f_2(x,k)\, dx=0,
  \qquad\forall  k>0,
  \tag{1.6} \end{equation}
where $f_j(x,k)$ is the Jost solution to (1.1) with $q(x)=q_j(x)$, $j=1,2$.
If (1.6) implies $p(x)=0$, then we say that the pair 
$\{L_1,L_2\}, L_j:=L_{q_j}: = -\frac{d^2}{dx^2} +q_j(x)$ has property $C_+$.

If $p\in  L_{1,1} (\R_-)$ and
\begin{equation}
  \int^0_{-\infty} p(x) g_1(x,k) g_2(x,k)dx=0
  \qquad\forall  k>0,
  \tag{1.7}\end{equation}
implies $p(x)=0$, then we say that the pair $\{L_1,L_2\}$
 has property $C_-$.
\end{definition}

In (1.7) $g_j:=g_{j+}(x,k)$.

Fix an arbitrary $b>0$. Assume that 
\begin{equation}
   \int^b_0 p(x)\varphi_1(x,k)\varphi_2(x,k)dx=0
   \qquad\forall k>0 \tag{1.8}
\end{equation}
implies $p(x)=0$. Then we say that the pair $\{L_1,L_2\}$ has property 
$C_{\varphi}$ and similarly $C_{\psi}$ is defined, $\psi_j$ replacing 
$\varphi_j$ in (1.8).

\begin{theorem}  
The pair $\{L_1,L_2\}$, where
$L_j:=-\frac{d^2}{dx^2} +q_j(x)$,
$q_j\in  L_{1,1} (\R_+), j=1,2$, has properties $C_+, C_{\varphi}$
and $C_{\psi}$.
If $q_j\in  L_{1,1} (\R_-)$, then $\{L_1, L_2\}$ has property $C_-$.
\end{theorem}

\begin{proof}
Proof can be found in \cite{R1}. We sketch only the idea of the proof
of property $C_+$.

Using the known formula
\begin{equation}
   f_j(x,k)=e^{ikx}+\int^\infty_x A_j(x,y)e^{iky}dy,
   \quad j=1,2,
   \tag{1.9}\end{equation}
where $A_j(x,y)$ is the transformation kernel corresponding
to the potential $q_j(x)$, $j=1,2,$ see also formula (2.17) below,
and substituting (1.9) into (1.6), one gets after a change of order of
integration a homogeneous Volterra integral equation for $p(x)$.
Thus $p(x)=0$.
\end{proof}

The reason for taking $b<\infty$ in (1.8) is: when one uses the formula
\begin{equation}
   \varphi_j(x,k)=\frac{\sin (kx)}{k} + \int^x_0K_j(x,y)
   \frac{\sin(ky)}{k}dy, \quad j= 1,2,
   \tag{1.10} \end{equation}
for the solution $\varphi_j$ to (1.1) (with $q=q_j$) satisfying first two 
conditions (1.4), then the Volterra-type integral equation for $p(x)$ contains
integrals over the infinite interval $(x,\infty)$. In this case the conclusion
$p(x)=0$ does not hold, in general. If, however, the integrals are over a
finite interval $(x,  b )$, then one can conclude $p(x)=0$.

The same argument holds when one proves property $C_\psi$, but formula (1.10)
is replaced by
\begin{equation}
   \psi_j(x,k)= \cos(kx)+ \int^x_0 \widetilde{K_j}(x,y) \cos(ky)dy,
   \quad j=1,2, \tag{1.11}
\end{equation}
with a different kernel $\widetilde{K_j}(x,y)$. 

\section {Applications of property C} 

\subsection{Uniqueness of the solution inverse
scattering problem with the data I-function.} 

The I-function  $I(k)$  is defined by the formula 
\begin{equation}
   I(k):= \frac{f'(0,k)}{f(k)}. \tag{2.1 }\end{equation}
This function is equal to the {\it Weyl function}
$m(k)$ which is defined as the
function for which
\begin{equation}
   W(x,k):= \psi(x,k) + m(k)\varphi(x,k)\in L^2(\R_+), \quad Im k>0,
   \tag{2.2}\end{equation}
where $W(x,k)$ is the Weyl solution, $W(0,k)=1$, $W^\prime(0,k)=m(k)$.
Note that $W(x,k)=\frac{f(x,k)}{f(k)}$,
as follows from formulas (1.3), (2.1) and from formula (2.3)
which says $I(k)=m(k)$.

To prove that
\begin{equation}
   I(k) = m(k), \tag{2.3}\end{equation}
one argues as follows. If $q\in L_{1,1}(\R_+)$, then there is exactly one,
up to a constant factor solution to (1.1) belonging to $L^2(\R_+)$ when 
$Im k>0$.

Since $f(x,k)$ is such a solution, one concludes that
\begin{equation}
   f(x,k)=c(k)[\psi(x,k)+m(k)\varphi(x,k)],
   \quad c(k)\neq 0. \tag{2.4}\end{equation}
Therefore,
\begin{equation}
  I(k)= \frac{\psi^\prime (0,k)+m(k)
  \varphi^\prime (0,k)}{\psi(0,k)+m(k)\psi(0,k)}=m(k),
  \notag\end{equation}
as claimed.

In sections 11 and 12 the $I$-function is studied in more detail.

The inverse problem (IP1) is:

{\it Given $I(k)$ for all $k>0$, find $q(x)$. }

\begin{theorem} 
  The IP1 has at most one solution.
\end{theorem}

\begin{proof}
Theorem 2.1 can be proved in several ways. One way \cite{R2} is to
recover the spectral function $\rho(\lambda)$ from $I(k)$,
$k=\lambda^{\frac{1}{2}}$.
This is possible since $Im\, I(k)=\frac{k}{|f(k)|^2}$, $k>0$, and 
\begin{equation}
  d\rho(\lambda)=
  \begin{cases}\frac{\sqrt{\lambda}\,d\lambda}{\pi |f(\sqrt{\lambda})|^2},
      &  \lambda>0, \\
  \sum^J_{j=1}\, c_j\delta(\lambda+k^2_j)\,d\lambda,
      & \lambda<0,\ k_j>0, \end{cases}
  \tag{2.5} \end{equation}

where $-k^2_j$ are the bound states of the Dirichlet operator
$L_q=-\frac{d^2}{dx^2} + q(x)$ in $L^2(\R_+), f(ik_j)=0$,
$\delta(\lambda)$
is the delta-function, and
\begin{equation}
  c_j= -\frac{2ik_j f^\prime (0,ik_j)}{\dotf (ik_j)},
  \qquad \dotf:=\frac{\partial f}{\partial k}.
  \tag{2.6}\end{equation}
Note that $ik_j$ and the number $J$ in (2.5) can be found as the simple 
poles of $I(k)$ in $\C_+$ and the number of these poles, and
\begin{equation}
   c_j = -2ik_j  \operatorname*{Res}_{k=k_j} I(k)=2k_jr_j,
   \tag{2.7}\end{equation}
where $ir_j:=\operatorname*{Res}_{k=k_j}I(k)$,
so $r_j=\frac{c_j}{2k_j}$.

It is well known that $d\rho(\lambda)$ determines $q(x)$ uniquely \cite{M},
\cite{R}.
An algorithm for recovery of $q(x)$ from $d\rho$ is known (Gelfand-Levitan).
In \cite{R2} a characterization of the class of $I$-functions corresponding to 
potentials in $C^m_{loc}(\R_+)$, $m\geq 0$ is given.

Here we give a very {\it simple new proof of Theorem 2.1} (cf \cite{R1}):

Assume that $q_1$ and $q_2$ generate the same $I(k)$, that is, 
$I_1 (k) = I_2 (k):=I(k)$. Subtract from equation (1.1) for $f_1(x,k)$ 
this equation for $f_2 (x,k)$ and get:
\begin{equation}
  L_1w=pf_2, \quad p(x):=q_1(x)-q_2(x),
   \quad w:=f_1(x,k)-f_2(x,k).
    \tag{2.8}\end{equation}
Multiply (2.8) by $f_1$ and integrate by parts:
\begin{equation}
  \begin{align}
  \int^\infty_0 p(x)f_2(x,k)f_1(x,k)dx
  & = (w^\prime f_1 - wf_1^\prime )   \big|^\infty_0
    = (f_1f_2^\prime -f_1^\prime f_2) \big|_{x=0} \notag \\
  & = f_1f_2(I_1{(k)} - I_2(k)) = 0 \qquad \forall  k>0,
  \tag{2.9} \end{align}
  \end{equation}
where we have used (1.3) to conclude that at infinity the boundary term 
vanishes. From (2.9) and property $C_+$ (Theorem 1.1) it follows that
$p(x)=0$. Theorem 2.1 is proved. 

\end{proof}

\subsection {Uniqueness of the solution to inverse
    scattering problem on the half axis.} 

This is a classical problem \cite{M}, \cite{R}. 
The scattering data are
\begin{equation}
   {\mathcal S} = \left\{S(k),\ k_j,\ s_j,\ 1\leq j \leq J \right\}.
   \tag{2.10} \end{equation}
Here
\begin{equation}
   S(k):= \frac {f(-k)}{f(k)} \tag{2.11} \end{equation}
is the S-matrix, $k_j>0$ are the same as in section 2.1, and the norming
constants $s_j$ are the numbers
\begin{equation}
   s_j:= -\frac {2ik_j}{\dot f(ik_j)f^\prime (0,ik_j)} >0.
   \tag{2.12} \end{equation}
Note that (2.6) implies
\begin{equation}
   c_j = s_j[f^\prime (0,ik_j)]^2 .\tag{2.13} \end{equation} 

\begin{theorem} 
Data (2.10) determine $q(x)\in  L_{1,1}(\R_+)$ uniquely.
\end{theorem}

\begin{proof}
This result is due to Marchenko \cite{M}. We give a {\it new short
proof based on property C} (\cite{R1}).
We prove that data (2.10) determine  $I(k)$
uniquely, and then Theorem 2.2 follows from Theorem 2.1. To determine  $I(k)$ 
we determine $f(k)$ and $f^\prime (0,k)$ from data (2.10).

First, let us prove that data (2.10) determine uniquely 
$f(k)$. Suppose there are
two different functions $f(k)$ and $h(k)$ with the same data (2.10). Then
\begin{equation}
   \frac{f(k)}{h(k)} = \frac{f(-k)}{h(-k)}, \qquad \forall  k\in  \R.
   \tag{2.14} \end{equation} 

The left-hand side in (2.14) is analytic in $\C_+$ since
$f(k)$ and $h(k)$ are, and
the zeros of $h(k)$ in $\C_+$ are the same as these of $f(k)$,
namely $ik_j$, and they are simple.
The right-hand side of (2.14) has similar properties in
$\C_-$. Thus $\frac{f(k)}{h(k)}$ is an entire function which tends to 1 as
$|k|\to \infty$, so, $\frac{f(k)}{h(k)} = 1$ and $f(k)=h(k)$. 
The relation
\begin{equation}
  \lim_{|k|\to\infty, k\in\C_+} f(k)=1
  \tag{2.15} \end{equation}
follows from the representation 
\begin{equation}
  f(k) = 1+ \int^\infty_0 A(0,y)e^{iky}dy,
  \quad A(0,y)\in L_1(\R_+).
  \tag{2.16} \end{equation}
Various estimates for the kernel $A(x,y)$ in the formula
\begin{equation}
  f(x,k)= e^{ikx} + \int^\infty_x A(x,y)e^{iky}dy
  \tag{2.17} \end{equation}
are given in \cite{M}. We mention the following:
\begin{equation}
  |A(x,y)| \leq c \sigma \left(\frac{x+y}{2}\right),
  \quad \sigma (x):= \int^\infty_x|q(t)|dt,
  \tag{2.18} \end{equation}
\begin{equation}
   \left|\frac{\partial A(x,y)}{\partial x} + \frac{1}{4}
    q\left( \frac {x+y}{2} \right) \right|
   \leq c \sigma(x) \sigma \left( \frac{x+y}{2} \right),
   \tag{2.19} \end{equation}
\begin{equation}
   \left| \frac{\partial A(x,y)}{\partial y} + \frac{1}{4}q
   \left( \frac {x+y}{2} \right)\right|
   \leq c \sigma(x)\sigma \left( \frac{x+y}{2} \right),
  \tag{2.20} \end{equation}

where $c>0$ here and below stands for various estimation constants.

From (2.17) and (2.18) formula (2.16) follows.

Thus, we have proved
\begin{equation}
   {\mathcal S} \Rightarrow f(k).
   \tag{2.21} \end{equation}

Let us prove
\begin{equation}
 {\mathcal S} \Rightarrow f^\prime (0,k) .
 \tag{2.22} \end{equation}

We use the Wronskian:
\begin{equation} 
   f^\prime (0,k) f(-k) - f^\prime (0,-k) f(k) = 2ik,
   \quad k\in \R.
   \tag{2.23}\end{equation}

The function $f(k)$ and therefore $f(-k)=\overline{f(k)}$,
where the overbar stands for
complex conjugate, we have already uniquely determined from data (2.10).
Assume there are two functions $f^\prime (0,k)$ and $h^\prime (0,k)$
corresponding to the same data  (2.10). Let
\begin{equation}
   w(k):=f^\prime (0,k)-h^\prime (0,k).
   \tag{2.24} \end{equation}

Subtract (2.23) with $h^\prime (0,\pm k)$ in place of
$f^\prime (0,\pm k)$ from equation (2.23) and get
\begin{equation}
   w(k) f(-k) - w(-k)f(k)=0,
   \notag\end{equation}
or
\begin{equation}
  \frac {w(k)}{f(k)} = \frac {w(-k)}{f(-k)} \qquad \forall k\in \R
  \tag{2.25} \end{equation}

\begin{claim}
 $\frac {w(k)}{f(k)}$ is analytic in $\C_+$ and vanishes at
infinity and $\frac {w(-k)}{f(-k)}$ is analytic in $\C_-$ and vanishes at 
infinity.
\end{claim}

If this claim holds, then $\frac {w(k)}{f(k)}\equiv 0, k\in  \C$,
and therefore
$w(k)\equiv 0$, so $f^\prime (0,k)=h^\prime (0,k)$.

To complete the proof, let us prove the claim.

From (2.17) one gets:
\begin{equation}
  f^\prime (0,k)=ik -A(0,0) + \int^\infty_0 A_x(0,y)e^{iky}dy.
  \tag{2.26} \end{equation}

Taking $k \rightarrow +\infty$ in (2.16), integrating by parts and using 
(2.20), one gets:
\begin{equation}
  f(k)=1 -\frac {A(0,0)}{ik} -\frac{1}{ik}\int^\infty_0 A_y(0,k)e^{iky}\,dy.
  \tag{2.27} \end{equation}
Thus
\begin{equation}
   A(0,0) = -\lim_{k\to\infty} ik[f(k) -1].
   \tag{2.28} \end{equation}

Since $f(k)$ is uniquely determined by data (2.10),
so is the constant A(0,0)
(by formula (2.28)).

Therefore (2.24) and (2.28) imply: 
\begin{equation}
  \lim_{|k|\to\infty,\ k\in\C_+} w(k)=0.
   \tag{2.29} \end{equation}

It remains to be checked that (2.10) implies
\begin{equation}
   w(ik_j)=0.
   \tag{2.30} \end{equation}
This follows from formula (2.12):
if $f(k)$ and $s_j$ are the same, so are
$f^\prime (0,ik_j)$, and $w(ik_j) = 0$ as the difference of equal numbers
\begin{equation}
  h^\prime (0,ik_j)=f^\prime (0,ik_j) = -\frac{2ik_j}{\dot f(ik_j) s_j}.
  \notag\end{equation}

Theorem 2.2 is proved.
\end{proof} 
In this section we have proved that the scattering data (2.10)
determines the $I$-function (2.1) uniquely. The converse is also true:
implicitly it follows from the fact that both sets of data
(2.1) and (2.10) determine uniquely the potential and are determined by
the potential uniquely. A direct proof is given in section 12 below.

\subsection {Compactly supported potential is uniquely determined by the phase
shift of $s$-wave.} 

Consider the inverse scattering on half-line and assume $q(x)=0$ for
$x>a>0$, where $a>0$ is an arbitrary fixed number.

The phase shift of s-wave is denoted by $\delta(k)$
and is defined by the formula
\begin{equation}
   f(k) = |f(k)|e^{-i\delta(k)},
   \tag{2.31} \end{equation}
so the S-matrix can be written as
\begin{equation}
   S(k) = \frac{f(-k)}{f(k)}=e^{2i\delta(k)}.
   \tag{2.32} \end{equation}
If $q(x)$ is real-valued, then
\begin{equation}
   \delta(-k)=-\delta(k),
   \quad k \in  \R,
   \tag{2.33} \end{equation}
and if $q\in  L_{1,1} (\R_+)$, then
\begin{equation}
   \delta(\infty) = 0 .
   \tag{2.34} \end{equation}
Note that S-matrix is unitary:
\begin{equation}
   S(-k) = \overline{S(k)}, \quad|S(k)| = 1\ \hbox{if}\  k \in \R.
   \tag{2.35} \end{equation}
Define index of $S(k)$:
\begin{equation}
   \nu:= ind \,S(k): = \frac{1}{2\pi i}
   \int^\infty_{-\infty} d \ln S(k) =
  \frac{1}{2 \pi} \Delta_{\R} arg\, S(k).
  \tag{2.36} \end{equation}

From (2.32), (2.33) and (2.34) one derives a formula for the index:
\begin{equation}
  \begin{align}
    \nu = & \frac{1}{\pi} \Delta_{\R} \delta(k)
      = \frac{1}{\pi}
          [\delta(-0)-\delta(-\infty)+\delta(+\infty)-\delta(+0)] \notag \\
     = & -\frac {2}{\pi}  \delta(+0)
      = \begin{cases} -2J & \hbox{if $f(0)\not= 0$}, \\
            -2J-1 & \hbox{if $f(0)=0$} .\end{cases}
                  \notag\end{align}
       \tag{2.37}\end{equation}

Here we have used the formula:
\begin{equation}
\frac 1{\pi}  \delta(+0) = \ \#\{ \hbox{zeros of $f(k)$ in}\ \C_+\}
+\frac 12 \delta_0,
  \tag{2.38} \end{equation}
which is the argument principle. Here $\delta_0:=0$ if $f(0)\neq 0$
and $\delta_0:=1$ if $f(0)=0$.

The zero of $f(k)$ at $k=0$ is called a resonance at zero energy.

Let us prove the following result \cite{R5}:

\begin{theorem} 
If $q \in  L_{1,1} (\R_+)$  decays
faster than any exponential:
$|q(x)| \leq ce^{-c|x|^\gamma}$, $ \gamma> 1$, then the data
$\left\{\delta(k)\ \forall  k>0 \right\}$
determines $q(x)$ uniquely.
\end{theorem}

\begin{proof}
Our proof is new and short. We prove that, if $q$ is compactly
supported or decays faster than any exponential, e.g. 
$|q(x)| \leq ce^{-c|x|^\gamma}$, $ \gamma > 1$,
then $\delta(k)$ determines uniquely
$k_j$ and $s_j$, and, by Theorem 2.2, $q(x)$ is uniquely determined.

We give the proof for compactly supported potentials.
The proof for the potentials decaying faster
than any exponentials is exactly the same.
The crucial point is: under both assumptions the Jost
function is an entire function of $k$.

If $q(x)$ is compactly supported, $q(x)=0$ for $x \geq a$,
then $f(k)$ is an
entire function of exponential type $\leq 2a$, that is 
$|f(k)| \leq ce^{2a|k|}$ (\cite[p. 278]{R}).
Therefore $S(k)$ is meromorphic in $\C_+$ (see (2.32)).
Therefore the numbers $k_j$, $1\leq j\leq J$, can be
uniquely determined as the only poles of $S(k)$ in $\C_+$.
One should check that
\begin{equation}
   f(-ik_j) \neq 0\ \hbox{ if}\  f(ik_j) = 0.
   \tag{2.39} \end{equation}

This follows from (2.23): if one takes $k=ik_j$ and uses $f(ik_j)=0$,
then (2.23) yields
\begin{equation}
  f^\prime (0,ik_j)f(-ik_j) = -2k_j < 0.
  \tag{2.40} \end{equation}
Thus $f(-ik_j) \neq 0$. Therefore $\delta(k)$
determines uniquely the numbers $k_j$ and $J$.

To determine $s_j$, note that
\begin{equation}
  \operatorname*{Res}_{k=ik_j}
  \ S(k)= \frac {f(-ik_j)}{\dot f(ik_j)} = \frac{1}{i} s_j,
  \tag{2.41} \end{equation}
as follows from (2.12) and (2.40).
Thus the data (2.10) are uniquely determined from
$S(k)$ if $q$ is compactly supported, and Theorem 2.2 implies 
Theorem 2.3.
\end{proof} 

\begin{corollary} 
If $q(r)\in L_{1,1}(\R_+)$ is compactly supported, then the knowledge
of $f(k)$ on an arbitrary small open subset of $\R_+$
(or even on an infinite sequence $k_n>0$, $k_n\not=k_m$ if $m\not= n$,
$k_n\to k$ as $n\to\infty$, $k>0$)
determines $q(r)$ uniquely.

In section 4 we prove a similar result with the data $f^\prime(0,k)$
in place of $f(k)$.

\end{corollary}

\subsection{Recovery of $q \in  L_{1,1} (\R)$ from
the reflection coefficient alone.} 

Consider the scattering problem on the full line: $u$ solves (1.1) and
\begin{equation}
  u \sim t(k)e^{ikx}, \quad x \rightarrow +\infty,
  \tag{2.42} \end{equation}
\begin{equation}
  u \sim e^{ikx} + r(k) e^{-ikx}, \quad x \rightarrow -\infty.
  \tag{2.43} \end{equation}

The coefficients $t(k)$ and $r(k)$ are called the transmission and
reflection coefficients (see \cite{M} and \cite{R}).
In general $r(k)$ alone cannot
determine $q(x)$ uniquely.

We assume
\begin{equation}
  q(x)=0 \ \hbox{for}\  x < 0,
  \tag{2.44} \end{equation}
and give a short proof, based on property C, of the following:

\begin{theorem} 
If $q \in  L_{1,1} (\R)$ and (2.44) holds, then
$r(k),\ \forall  k > 0$, determines $q(x)$ uniquely.
\end{theorem}

\begin{proof}  We claim that $r(k)$ determines uniquely $I(k)$
if (2.44) holds.
Thus, Theorem 2.4 follows from Theorem 2.1.
To check the claim, note that $u(x,k)=t(k)f(x,k)$, so
\begin{equation}
  I(k)= \frac {u^\prime (0,k)}{u(0,k)},
   \tag{2.45} \end{equation}
and use (2.44), (2.43) to get $u=e^{ikx} +r(k)e^{-ikx}$ for $x<0$, so
\begin{equation}
  \frac {u^\prime (0,k)}{u(0,k)} = \frac {ik(1- r(k))}{1 +r(k)}.
  \tag{2.46} \end{equation}
From (2.45) and (2.46) the claim follows.
Theorem 2.4 is proved.
\end{proof}

\subsection{Inverse scattering with various data} 

Consider scattering on the full line (1.1), (2.42)-(2.43), assume $q(x)=0$
if $x \not\in [0,1]$, and take as the scattering data the function
\begin{equation}
  u(0,k): = u_0 (k), \qquad \forall k > 0.
  \tag{2.47} \end{equation}

\begin{theorem} 
Data (2.47) determine $q(x)$ uniquely.
\end{theorem}

\begin{proof}
If $q=0$ for $x <0$, then $u(x,k) = e^{ikx} + r(k)e^{-ikx}$
for $x < 0$, $u(0,k)= 1+r(k)$, so data (2.47) determine $r(k)$
and, by Theorem 2.4, $q(x)$ is uniquely determined. 
Theorem 2.5 is proved. Of course, this theorem is a particular case of
Theorem 2.4.
\end{proof}  

\begin{remark} 
 Other data can be considered, for example, $u^\prime (0,k):=v(k)$.
Then $u^\prime (0,k) = ik[1-r(k)]$, and again $v(k)$ determines $r(k)$ and,
by Theorem 2.4, $q(x)$ is uniquely determined.
\end{remark} 

However, if the data are given at the right end of the support of the
potential, the inverse problem is more difficult. For example, if
$u(1,k):=u_1(k)$ is given for all $k>0$, then $u_1(k)=t(k)e^{ik}$, so $t(k)$
is determined by the data uniquely.

The problem of determining $q(x)$ from $t(k)$ does not seem to have been
studied. If $q(x) \geq 0$, then the selfadjoint operator
$L_q= -\frac{d^2}{dx^2} +q(x)$ in $L^2(\R)$ does not have bound states
(negative eigenvalues).

In this case the relation $|r^2 (k)| + |t^2(k)|=1$ allows one to find
\begin{equation}
   |r(k)| = \sqrt{1-|t(k)|^2},\quad k\in\R.
   \notag\end{equation}
   Define
\begin{equation}
  a(k):= exp \left\{ -\frac{1}{\pi i} \int^\infty_{-\infty}
  \frac {\ln|t(s)|}{s-k}ds \right\}.
  \tag{2.48} \end{equation}

The function (2.48) has no zeros in $\C_+$ if $L_q$ has no bound states.
If $q$ is compactly supported then $r(k)$ and $t(k)$
are meromorphic in $\C$.

Let us note that
\begin{equation}
  f(x,k):=f_+(x,k) = b(k)g_- (x,k) + a(k) g_+(x,k)
  \tag{2.49} \end{equation}
\begin{equation}
  g_- (x,k) = c(k)f_+(x,k) + d(k)f_- (x,k).
  \tag{2.50} \end{equation}
It is known \cite{M}, \cite{R}, that
\begin{equation}
  a(-k) = \overline{a(k)},  \quad b(-k)= \overline{ b(k)},
  \quad k\in\R,
  \tag{2.51} \end{equation}
\begin{equation}
  c(k) = - b (-k), \quad d(k) = a(k),
  \tag{2.52} \end{equation}
\begin{equation}
  a(k) = -\frac{1}{2ik} [f(x,k),g_-(x,k)],  \quad b(k) = \frac{1}{2ik}
  [f_+(x,k), g_+(x,k)],
  \tag{2.53} \end{equation}
where $[f,g]:=fg^\prime -f^\prime g$ is the Wronskian,
\begin{equation}
  |a(k)|^2 = 1+ | b (k)|^2,
  \tag{2.54} \end{equation}
\begin{equation}
  r (k) = \frac{b(k)}{a(k)}, \quad t(k) = \frac{1}{a(k)}.
  \tag{2.55} \end{equation}

The function $a(k)$ is analytic in $\C_+$. One can prove \cite[p.288]{M}
\begin{equation}
  a(k)=1 - \frac{\int^\infty_{-\infty} q(x)\, dx}{2ik}
  + o \left( \frac{1}{k} \right), \quad k \to\infty,
  \tag{2.56} \end{equation}
and
\begin{equation}
   b (k) = o \left( \frac{1}{k} \right), \quad k \rightarrow \infty.
   \tag{2.57} \end{equation}

The function $r(k)$ does not allow, in general, 
an analytic continuation
from the real axis into the complex plane.
However, if $q(x)=0$ for $x<0$, then $b(k)$ admits 
an analytic continuation
from the real axis into $\C_+$ and $r(k)$ is meromorhic in $\C_+$.
 
If $q(x)$ is compactly supported the functions
$f_\pm (x,k)$ and $g_\pm (x,k)$
are entire functions of $k$
of exponential type, so that $r(k)$ and $t(k)$ are meromorphic in $\C$.
From (2.54) one can find
$|b(k)|$ since $a(k)$ is found from formula (2.48) (assuming no bound
states). 

The conclusion is: recovery of a compactly supported potential
from the transmission coefficient is an open problem.

\section{Inverse problems on a finite interval with mixed data} 

Consider equation (1.1) on the interval [0,1]. Take some selfadjoint
boundary conditions, for example:
\begin{equation}
  -u^{\prime\prime}  + q(x)u- \lambda u=0,\quad 0\leq x\leq 1;
  \quad u(0) = u(1)=0, \quad \lambda = k^2.
  \tag{3.1}\end{equation}
Assume $q\in   L^1[0,1], q=\overline q$. Fix $ b  \in  (0,1)$ arbitrary.
Suppose $q(x)$ is known on the interval $[ b ,1]$ and the subset
$\{\lambda_{m(n)}\}$ of the eigenvalues of the problem (3.1) is known,
$n=1,2,\dots$ where $m(n)$ is a sequence with the property
\begin{equation}
  \frac{m(n)}{n} = \frac{1}{\sigma}(1+\varepsilon_n),\quad
  \varepsilon_n \rightarrow 0, \quad\sigma >0.
  \tag{3.2}\end{equation}

\begin{theorem} 
The data
$\{\lambda_{m(n)}, n= 1,2,...; q(x),  b  \leq x\leq 1 \}$
determine uniquely $q(x), 0 \leq x \leq  b $, if $\sigma >2  b $.
If $\sigma = 2 b $ and
$\sum_{n=1}^{\infty} |\varepsilon| < \infty$, then the
above data determine $q(x), 0 \leq x \leq  b $, uniquely.
\end{theorem}

\begin{proof} First, assume $\sigma > 2 b $. If there are $q_1$ and $q_2$
which produce the same data, then as above, one gets
\begin{equation}
  G(\lambda):=g(k):= \int^ b_0 p(x) \varphi_1 (x,k) \varphi_2 (x,k)\, dx
  = (\varphi_1 w^\prime - \varphi_1^\prime  w)\Big|^ b_0
  = (\varphi_1 w^\prime -\varphi_1^\prime w )\Big|_{x=b }
  \tag{3.3}\end{equation}
where $w:= \varphi_1-\varphi_2$, $p:=q_1 - q_2$, $k= \sqrt \lambda$.
Thus
\begin{equation}
  g(k) = 0\ \hbox{at}\ k
  = \pm \sqrt{\lambda_{m(n)}}: = \pm k_n.
  \tag{3.4}\end{equation}

The function $G(\lambda)$ is an entire function of $\lambda$
of order $\frac{1}{2}$
(see formula (1.10) with $k=\sqrt\lambda$), and is
an entire even function of
$k$ of exponential type $\leq 2  b $. One has
\begin{equation}
  |g(k)| \leq c \frac{e^{2 b  |Imk|}}{1+|k|^2}.
  \tag{3.5}\end{equation}
The indicator of $g$ is defined by the formula
\begin{equation}
  h(\theta):=h_g (\theta):=
  \overline{\lim_{r\to\infty}}
   \frac {\ln|g (r e^{i\theta})| }{r},
  \tag{3.6}\end{equation}
where $k=r e^{i \theta}$. Since $|Imk| = r|\sin \theta|$, one gets from
(3.5) and (3.6) the following estimate
\begin{equation}
  h(\theta) \leq 2b |\sin \theta|.
  \tag{3.7}\end{equation}

It is known \cite[formula (4.16)]{L} that for any entire function
 $g(k) \not\equiv 0$ of
exponential type one has:
\begin{equation}
  \lim_{\overline{r\to\infty}}
  \frac{n(r )}{r } \leq \frac{1}{2 \pi} \int^{2 \pi}_0 h_g
  (\theta)\, d \theta,
  \tag{3.8}\end{equation}
where $n(r )$ is the number of zeros of $g(k)$ in the disk
$|k| \leq r $. From (3.7) one gets
\begin{equation}
  \frac {1}{2\pi} \int ^{2\pi}_0 h_g(\theta)\, d\theta \leq
  \frac {2 b}{2 \pi} \int^{2 \pi}_0 |\sin \theta|\,d\theta
  = \frac{4  b }{\pi}
  \tag{3.9}\end{equation}
From (3.2) and the known asymptotics of the Dirichlet eigenvalues:
\begin{equation}
  \lambda_n = (\pi n)^2 + c + o(1),
  \quad n \rightarrow\infty, \quad c=const,
  \tag{3.10} \end{equation}
one gets for the number of zeros the estimate
\begin{equation}
  n(r) \geq 2 \sum_{ \frac{n\pi}{\sigma}
     \left[ 1+0 \left( \frac{1}{n^2} \right) \right] <r}
  1=2\frac{\sigma r }{\pi}[1+o(1)],
  \quad r  \rightarrow \infty.
  \tag{3.11} \end{equation}

From (3.8), (3.9) and (3.11) it follows that
\begin{equation}
   \sigma \leq 2 b.
   \tag{3.12} \end{equation}

Therefore, if $\sigma > 2  b $, then $g(k) \equiv 0$. If $g(k) \equiv 0$
then, by property $C_\varphi$ (Theorem 1.1), $p(x)=0$.
Theorem 3.1 is proved in
the case $\sigma > 2  b $.

Assume now that $\sigma = 2  b $ and
\begin{equation}
   \sum_{n=1}^\infty |\varepsilon_n| < \infty.
    \tag{3.13} \end{equation}
We {\it claim} that if an entire function $G(\lambda)$ in (3.3) of order
$\frac{1}{2}$ vanishes at the points
\begin{equation}
   \lambda_n = \frac{n^2 \pi^2}{\sigma^2} (1+\varepsilon_n),
    \tag{3.14} \end{equation}
and (3.13) holds, then $G(\lambda) \equiv 0$. If this is proved,
then Theorem 3.1 is proved as above.

Let us prove the claim. Define
\begin{equation}
  \Phi(\lambda):=\prod^\infty_{n=1} \left( 1-\frac{\lambda}{\lambda_n}\right)
    \tag{3.15} \end{equation}
and recall that
\begin{equation}
  \Phi_0(\lambda):=\frac{\sin(\sigma\sqrt{\lambda})}{\sigma\sqrt{\lambda}}
  =\prod^\infty_{n=1} \left( 1-\frac{\lambda}{\mu_n}\right),\quad
  \mu_n:=\frac{n^2\pi^2}{\sigma^2}.  
   \tag{3.16} \end{equation}

Since $ G(\lambda_n) = 0$, the function
\begin{equation}
  w(\lambda): = \frac {G(\lambda)}{\Phi(\lambda)}
  \tag{3.17} \end{equation}
is entire, of order $\leq\frac{1}{2}$.
Let us use a Phragmen-Lindel\"of lemma.

\begin{lemma} 
\cite[Theorem 1.22]{L}
If an entire function $w(\lambda)$ of order $<1$ has the
property $sup_{-\infty<y<\infty} |w(iy)| \leq c$,
then $w(\lambda) \equiv c$.
If, in addition $w(iy) \rightarrow 0$
as $y \rightarrow +\infty$, then $w(\lambda) \equiv 0.$
\end{lemma}

We use this lemma to prove that $w(\lambda) \equiv 0$.
If this is proved then $G(\lambda) \equiv 0$ and Theorem 3.1 proved.

The function $w(\lambda)$ is entire of order $\frac{1}{2} < 1$.

Let us check that
\begin{equation}
   \sup_{-\infty<y<\infty} |w(iy)| < \infty,
   \tag{3.18} \end{equation}
and that
\begin{equation}
   |w(iy)| \rightarrow 0\ \hbox{as}\ y \rightarrow +\infty.
   \tag{3.19} \end{equation}

One has, using (3.5), (3.15), (3.16) and taking into account that $\sigma = 2 b $:
\begin{equation}
 \begin{align}
 |w(iy)| = &
   \left| \frac{G(iy)}{\Phi(iy)} \frac{\Phi_0 (iy)}{\Phi_0(iy)} \right|
 \leq \frac{ e^{2b|Im\sqrt{iy}|} }{ (1+|y|)}
   \left(\frac{ e^{\sigma|Im\sqrt{iy}|} }{ 1+|y|^{\frac{1}{2}} }\right)^{-1}
  \left( \prod^\infty_{h=1}
    \frac{ 1+\frac{y^2}{\mu^2_n} }{ 1+\frac{y^2}{\lambda^2_n} }
       \right)^{ \frac{1}{2} }  \notag \\
 \leq &
    \frac{c}{ 1+|y|^{\frac{1}{2}} }
  \left( \prod_{ \{n:\mu_n\leq\lambda_n\} }
    \frac{\lambda_n^2}{\mu^2_n} \right)^{ \frac{1}{2} }
  \leq \frac{c}{ 1+|y|^{\frac{1}{2}} }
    \prod_{ \{n:\mu_n\leq\lambda_n\} }
    \left( 1+|\varepsilon_n|\right)
  \leq \frac{c_1}{ 1+|y|^{\frac{1}{2}} }\ . \notag\end{align}
   \tag{3.20} \end{equation}

Here we have used elementary inequalities:
\begin{equation}
   \frac {1+a}{1+ d} \leq \frac {a}{ d }\quad \hbox{if}\quad
   a \geq  d> 0;
   \quad \frac {1+a}{1+ d} \leq 1\quad \hbox{if}\quad 0\leq a \leq  d ,
   \tag{3.21} \end{equation}
with
  $a:=\frac {y^2}{\mu_n^2}$, $d:= \frac {y^2}{\lambda_n^2}$,
and the assumption (3.13).

We also used the relation:
\begin{equation}
 \left| \frac{\sin(\sigma \sqrt{iy})}{\sigma \sqrt{iy}} \right|
 \sim \frac {e^{\sigma |Im \sqrt{iy}|}}{2 \sigma |\sqrt{iy}|}
 \quad \hbox{as}\quad y \rightarrow +\infty.
 \notag \end{equation}

Estimate (3.20) implies (3.18) and (3.19).
An estimate similar to (3.20) has been used in
the literature (see e.g.\cite {D}).

Theorem 3.1 is proved.
\end{proof}

\begin{remark} 
Theorem 3.1 yields several results obtained in \cite{D}, and an
earlier result of Hochstadt-Lieberman which says that
the knowledge of all the
Dirichlet eigenvalues and the knowledge of $q(x)$ on
$\left[\frac{1}{2},1 \right]$ determine
uniquely $q(x)$ on $\left[ 0,\frac{1}{2}\right]$.
\end{remark}

In this case $ b  = \frac{1}{2}$, $\sigma = 1$.

One can also obtain the classical result of Borg \cite{B}
and its generalization due to Marchenko \cite{M}:

Two spectra (with the same
boundary conditions on one of the ends of the interval and two different
boundary conditions on the other end) determine $q(x)$ and the
boundary conditions uniquely.

\section {Property C and inverse problems for some PDE} 

\subsection{  } 
Consider the problem
\begin{equation}
   u_t = u_{xx}-q(x)u,\quad 0 \leq x \leq 1,\quad t > 0,
   \tag{4.1}\end{equation}
\begin{equation} 
  u(x,0) = 0,
  \tag{4.2}\end{equation}
\begin{equation} 
  u(0,t) = 0, \quad u(1,t) = a(t).
  \tag{4.3}\end{equation}

Assume the $a(t) \not\equiv 0$ is compactly supported,
$a(t) \in  L^1(0,\infty)$,
$q(x) \in  L^1 [0,1]$, problem (4.1) - (4.3) is solvable,
and one can measure the data
\begin{equation}
   u^\prime (1,t):=u_x(1,t):=  b  (t) \qquad \forall t>0.
   \tag{4.4}\end{equation}

The inverse problem (IP1) is:

{\it Given $\{a(t), b(t),\ \forall t > 0 \}$ find $q(x)$.}

\begin{theorem} 
IP1 has at most one solution.
\end{theorem}

\begin{proof}  Laplace-transform (4.1) - (4.3) to get
\begin{equation}
   v^{\prime\prime}-\lambda v - q(x)v=0 \quad 0 \leq x \leq 1,
   \quad v:= \int^\infty_0 e^{-\lambda t} u(x,t)\, dt,
   \tag{4.5}\end{equation}
\begin{equation}
   v(0,\lambda) = 0, \quad v(1,\lambda)
   = A(\lambda): = \int^\infty_0 e^{-\lambda t} a(t)\,dt.
   \tag{4.6}\end{equation}
\begin{equation}
   v^\prime  (1,\lambda) =  B (\lambda):= \int^\infty_0
   b (t) e^{-\lambda t}\,dt.
  \tag{4.7}\end{equation}

Assume that there are $q_1 (x)$ and $q_2 (x)$ which generate the same data
$\{A(\lambda), B(\lambda)$, $\forall  \lambda > 0 \}$. Let
$p(x): = q_1(x)-q_2 (x), w:= v_1-v_2$. Subtract from equation (4.5) with
$v=v_1, \, q=q_1$, similar equations with $v=v_2, \, q = q_2$, and get
\begin{equation}
  \l_1w:=w^{\prime\prime} - \lambda w - q_1 w= p\, v_2,
  \tag{4.8}\end{equation}
\begin{equation}
  w(0,\lambda) = 0,
  \quad w(1,\lambda) = w^\prime (1,\lambda) = 0.
  \tag{4.9}\end{equation}

Multiply (4.8) by $\varphi_1 (x,\lambda)$, where
$\l_1 \varphi_1 = 0$, $\varphi_1(0,\lambda) = 0$, $\varphi^\prime _1
(0,\lambda) = 1$,
integrate over $[0,1]$ and then
by parts on the left-hand side, using (4.9). The result is:
\begin{equation}
  \int^1_0 p(x) v_2 (x,\lambda) \varphi_1 (x,\lambda)dx = 0
  \qquad \forall \lambda > 0. \tag{4.10} \end{equation}

Note that $\varphi_1(x,\lambda)$ is an entire function of $\lambda$.

Since $a(t) \not\equiv 0$ and is compactly supported,
the function $A(\lambda)$ is an
entire function of $\lambda$, so it has a discrete set of zeros. Therefore
$v_2(x, \lambda) = c(\lambda) \varphi_2 (x,\lambda)$ where
$c(\lambda)\neq 0$ for almost all $\lambda \in  \R_+$.

Property $C_\varphi$ (Theorem 1.1) and (4.10) imply $p(x)=0$. Theorem 4.1 is
proved.
\end{proof}  

\begin{remark} 
 One can consider different selfadjoint homogeneous boundary
conditions at $x=0$, for example,
$u^\prime (0,t) = 0$ or $u^\prime (0,t)-h_0 u(0,t) = 0$, $h_0=const>0$.
\end{remark}

A different method of proof of a result similar to Theorem 4.1 can be found
in \cite{R1} and in \cite{De}. In \cite{De}
some extra assumptions are imposed on $q(x)$ and
$a(t)$.

\subsection{ } 
Consider the problem:
\begin{equation}
  u_{tt} = u_{xx} - q(x)u, \quad x \geq 0, \quad t \geq 0,
  \tag{4.11} \end{equation}
\begin{equation}
   u=u_t = 0 \ \hbox{at}\ t=0,
   \tag{4.12} \end{equation}
\begin{equation}
   u(0,t) = \delta (t),
   \tag{4.13} \end{equation}
where $\delta(t)$ is the delta-function.
Assume that
\begin{equation}
   q(x) = 0\quad \hbox{if}\quad x >1,
   \quad q=\overline q,\quad q \in  L^1 [0,1].
   \tag{4.14}\end{equation}
Suppose the data
\begin{equation}
   u(1,t):= a(t) \tag{4.15} \end{equation}
are given for all $t>0$.

The inverse problem (IP2) is:

{\it Given $a(t) \ \forall  t > 0$,  find q(x).}

\begin{theorem} 
The IP2 has at most one solution.
\end{theorem}

\begin{proof}
Fourier-transform (4.11)-(4.13), (4.15) to get
\begin{equation}
   v^{\prime\prime}  + k^2v-q(x)v = 0,
   \quad x \geq 0,
   \tag{4.16} \end{equation}
\begin{equation}
   v(0,k) = 1, \tag{4.17} \end{equation}
\begin{equation}
   v(1,k) = A(k):= \int^\infty_0 a(t) e^{ikt}dt, \tag{4.18} \end{equation}
where
\begin{equation}
   v(x,k) = \int^\infty_0 u(x,t) e^{ikt}dt. \tag{4.19} \end{equation}
It follows from (4.19) and (4.16) that
\begin{equation}
   v(x,k)=c(k) f(x,k),
   \tag{4.20} \end{equation}
where $f(x,k)$ is the Jost solution to (4.16).
From (4.20) and (4.17) one gets
\begin{equation}
    v(x,k) = \frac{f(x,k)}{f(k)},
    \tag{4.21} \end{equation}
where $f(k) = f(0,k)$. From (4.21) and (4.18) one obtains
\begin{equation}
   f(k) = \frac{f(1,k)}{A(k)}.
   \tag{4.22} \end{equation}
From (4.14) and (4.16) one concludes
\begin{equation}
   f(x,k) = e^{ikx}\ \hbox{for}\ x \geq 1.
   \tag{4.23} \end{equation}
From (4.23) and (4.22) one gets
\begin{equation}
   f(k) = \frac {e^{ik}}{A(k)}.
   \tag{4.24} \end{equation}

Thus $f(k)$ is known for all $k > 0$.

Since $q(x)$ is compactly supported, the data $\{f(k),\ \forall  k = 0\}$
determine $q(x)$ uniquely by Theorem 2.3. Theorem 4.2 is proved.

\end{proof} 

\begin{remark} 
One can consider other data at $x=1$, for example, the data
$u_x (1,t)$ or $u_x (1,t) + hu(1,t)$. The argument remains essentially the
same.
\end{remark}

However, the argument needs a modification if (4.13) is replaced by another
condition, for example, $u_x(0,t) = \delta(t)$.
In this case $v(x,k)=\frac{f(x,k)}{f^\prime (0,k)}$, and in place of $f(k)$
one obtains $f^\prime (0,k)\ \forall k > 0$ from the data (4.18).

The problem of finding a compactly supported $q(x)$ from the data
$\{f^\prime (0,k)\ \forall k > 0 \}$ was not studied, to our knowledge. We
state the following:

\begin{claim} 
The data $f^\prime (0,k)$ known on an arbitrary small open subset of
$(0,\infty)$ or even on an infinite
sequence of distinct positive numbers $k_n$ which
has a limit point $k>0$, determines a compactly supported
$q(r) \in L^1(\R_+)$ uniquely.
\end{claim}

Our approach to this problem is based on formula (2.23). If
$f^\prime (0,k)$ is known for all $k>0$, then $f^\prime (0,-k)
= \overline{f^\prime (0,k)}$
is known for all $k>0$, and (2.23) can be considered as the
  Riemann problem for
finding $f(k)$ and $f(-k)$ from (2.23) with the coefficients
 $f^\prime (0,k)$ and
$f^\prime (0,-k)$ known. If $q(x) \in L_1 (\R_+)$ is compactly supported
then $f^\prime (0,k)$ is an entire function of $k$. Thus the data determine
$f^\prime(0,k)$, for all $k>0$.

We want to prove that (2.23) defines 
$f(k)$ uniquely if $f^\prime (0,k)$ 
is known for all $k > 0$. Assume the contrary. Let $f(k)$ and
$h(k)$ be two solutions to (2.23), and $w: = f-h, \, w(k) \to 0$ as
$|k| \to \infty, \, k \in \C_+$. Then (2.23) implies
\begin{equation}
  \frac{w(k)}{f^\prime (0,k)} = \frac{w(-k)}{f^\prime (0,-k)}
  \qquad \forall k \in \R.
  \tag{4.25}\end{equation}

The function $f^\prime(0,k)$ has at most finitely many zeros in $\C_+$. All
these zeros are at the points $i\kappa_j$,
$1 \leq j \leq J_1$, where $-\kappa_j^2$
are the negative eigenvalues of the Neumann operator
$L_q := -\frac{d^2}{dx^2} + q(x)$ in $L^2(\R_+), \, u^\prime (0) =0$.
Also $f^\prime(0,0)$ may vanish. 

From (2.23) one concludes that
$w(i\kappa_j)=0$ if $f^\prime(0,i\kappa_j)=0$. Indeed, one has
$w(i\kappa_j) f^\prime(0,-i\kappa_j)
= w(-i\kappa_j)f^\prime(0,i\kappa_j)$.
If $f^\prime(0,i\kappa_j)=0$ then
$f^\prime(0,-i\kappa_j) \neq 0$ as follows from (2.23).
Therefore $w(i\kappa_j)=0$ as
claimed, and the function $\frac{w(k)}{f^\prime(0,k)}$
is analytic in $\C_+$
and vanishes at infinity in $\C_+$.
Similary, the right-hand side of (4.25) is
analytic in $\C_-$ and vanishes at infinity in $\C_-$. Thus, by analytic
continuation, $\frac{w(k)}{f^\prime (0,k)}$ is
an entire function which vanishes
at infinity and therefore vanishes identically. Therefore $w(k) \equiv 0$
and $f(k) = h(k)$. Thus, the data
$\{f^\prime (0,k),\ \forall k>0 \}$ determines
uniquely $\{f(k), \ \forall  k>0\}$.

Since $q(x)$ is compactly supported, Theorem 2.3 implies that $q(x)$ is 
uniquely determined by the above data. The claim is proved.
\qed

\section{Invertibility of the steps in the inversion provedures in the
inverse scattering and spectral problems} 

\subsection{Inverse spectral problem} 

 Consider a selfadjoint operator $L_q$ in
$L^2(\R_+)$ generated by the differential expression
$L_q = -\frac{d^2}{dx^2} + q(x)$,
$q(x) \in L^1_{loc} (\R_+), q(x) =
\overline {q(x)}$, and a selfadjoint boundary condition at $x=0$,
for example, $u(0)=0$.
Other selfadjoint conditions can be assumed. For instance:
$u^\prime (0)=h_0u(0), h_0 = const > 0$.

We assume that $q(x)$ is such that the equation (1.1) with $Im \lambda >0$, 
$\lambda: = k^2$, has exactly one solution which belongs to $L^2(\R_+)$
(the limit-point case at infinity).

In this case there is exactly one spectral function $\rho(\lambda)$ of the
selfadjoint operator $L_q$. Denote
\begin{equation} 
  \varphi_0 (x, \lambda): =
  \frac{\sin(\sqrt{\lambda} x)}{\sqrt{\lambda}}.
  \tag {5.1}\end{equation}
Let $h(x) \in L^2_0 (\R_+)$, where $L^2_0 (\R_+)$ denotes the set of 
$L^2(\R_+)$ functions vanishing outside a compact interval (this interval
depends on $h(x)$). Denote
\begin{equation} 
  H(\lambda):= \int^\infty_0 h(x) \varphi_0 (x, \lambda)\, dx.
  \tag{5.2}\end{equation}
{\it Assume that for every $h\in L^2_0 (\R_+)$ one has:}
\begin{equation}
  \int^\infty_{-\infty} H^2(\lambda)\, d\rho (\lambda)=0\Rightarrow h(x)=0.
  \tag{5.3} \end{equation}
Denote by ${\mathcal P}$ the set of nondecreasing functions $\rho (\lambda)$,
of bounded variation, such that if $\rho_1, \rho_2 \in {\mathcal P}$,
$\nu : = \rho_1-\rho_2$, and
\begin{equation}
  {\mathcal H} := \left\{ H(\lambda): h\in C^\infty_0 (\R_+) \right\},
  \tag{5.4}\end{equation}
where $H(\lambda)$ is given by (5.2), then
\begin{equation}
  \left\{ \int^\infty_{-\infty} H^2 (\lambda)\, d\nu (\lambda) = 0
  \qquad\forall h\in {\mathcal H} \right\}
  \Rightarrow \nu (\lambda) = 0.
  \tag{5.5}\end{equation}

\begin{theorem} 
 Spectral functions of the operators $L_q$, in the limit-
point at infinity case, belong to ${\mathcal  P}$.
\end{theorem}

\begin{proof}  Let $ b> 0$ be arbitrary, $f\in L^2(0, b )$, $f=0$ if 
$x> b $. Suppose
\begin{equation}
  \int^\infty_{-\infty} H^2(\lambda)\, d\nu (\lambda)
  = 0 \qquad \forall  h \in {\mathcal H}.
  \tag{5.6} \end{equation}
Denote by $I+V$ and $I+W$ the transformation operators corresponding to 
potentials $q_1$ and $q_2$ which generate spetral functions $\rho_1$ and
$\rho_2$, $\nu =\rho_1 - \rho_2$. Then
\begin{equation}
  \varphi_0 = (I+V) \varphi_1 = (I+W)\varphi_2, \tag{5.7}
  \end{equation}
where $V$ and $W$ are Volterra-type operators. Condition (5.6) implies:
\begin{equation}
  \Vert (I+V^\ast )f \Vert = \Vert(I+W^\ast )f \Vert
  \qquad\forall f \in L^2(0, b ),
  \tag{5.8}\end{equation}
where $V^\ast $ is the adjoint operator and the norm in (5.8) is
$L^2(0,b)$-norm. Note that
\begin{equation}
  Vf:= \int^x_0 V(x,y) f(y) \, dy,
  \tag{5.9}\end{equation}
and
\begin{equation}
  V^\ast f= \int^ b_s V(y,s) f(y) \, dy.
  \tag{5.10}
  \end{equation}
From (5.8) it follows that
\begin{equation}
  I+V^\ast  = U(I+W^\ast ), \tag {5.11}
\end{equation}
where $U$ is a unitary operator in the Hilbert space $H=L^2 (0,b)$.

If $U$ is unitary and $V,W$ are Volterra operators then (5.11) implies 
$V=W$.

This is proved in Lemma 5.1 below. If $V=W$ then $\varphi_1(x,\lambda)=
\varphi_2(x,\lambda)$, therefore $q_1=q_2$
 and $\rho_1(\lambda)=\rho_2(\lambda)$. Here we
have used the assumption about $L_q$ being in the limit-point at
infinity case: this assumption implies that the spectral function is
uniquely determined by the potential (in the limit-circle case
at infinity there are many spectral functions corresponding to
the given potential). Thus if $q_1=q_2$, then
$\rho_1(\lambda)=\rho_2(\lambda)$.
Theorem 5.1 is proved.
\end{proof}

\begin{lemma} 
 Assume that $U$ is unitary and $V,W$ are Volterra operators
in $H=L^2(0, b )$. Then (5.11) implies $V=W$.
\end{lemma}

\begin{proof}  From (5.11) one gets
$I+V = (I+W)U^\ast $ and, using $U^\ast  U=I$, one gets
\begin{equation}
  (I+V)(I+V^\ast ) = (I+W)(I+W^\ast ).
   \tag{5.12}  \end{equation}
Denote
\begin{equation}
  (I+V)^{-1} = I+V_1, \quad (I+W)^{-1} = I+W_1,
  \tag{5.13}
  \end{equation}
where $V_1, W_1$ are Volterra operators. From (5.12) one gets:
\begin{equation}
  (I+V^\ast )(I+W_1^\ast ) = (I+V_1)(I+W),
  \tag{5.14}\end{equation}
or
\begin{equation}
  V^\ast +W^\ast _1+V^\ast W^\ast _1 = V_1+W+V_1W.
  \tag{5.15} \end{equation}
Since the left-hand side in (5.15) is a Voleterra operator of the type (5.10)
while the right-hand side is a Volterra operator of the type (5.9), they can
be equal only if each equals zero:
\begin{equation}
  V^\ast +W_1^\ast +V^\ast W_1^\ast  = 0,
  \tag{5.16}\end{equation}
and
\begin{equation}
  V_1+W+V_1W = 0. \tag{5.17}\end{equation}
From (5.17) one gets
\begin{equation}
  V_1(I+W) = -W, \tag{5.18}\end{equation}
or
 $ [(I+V)^{-1} -I] (I+W) = -W$. 
Thus
  $(I+V)^{-1}(I+W) =I$
and
 $V=W$
as claimed. Lemma 5.1 is proved.
\end{proof}  

The inverse spectral problem consists of finding $q(x)$ given $\rho(\lambda)$.
The uniqueness of the solution to this problem was proved by Marchenko
\cite{M}
while the reconstruction algorithm was given by Gelfand and Levitan
\cite{Lt1} (see also \cite{R}).

Let us prove first the uniqueness theorem of Marchenko following \cite{R3}.
In this theorem there is no need to assume that $L_q$ is in the
limit-point at infinity case: if it is not, the spectral
function determines the potential uniquely also, but the potential
does not determine the spectral function uniquely.

\begin{theorem} 
 The spectral function determines $q(x)$ uniquely.
\end{theorem}

\begin{proof}
If $q_1$ and $q_2$ have the same spectral function $\rho(\lambda)$
then
\begin{equation}
  \Vert f \Vert^2 = \int^\infty_{-\infty} |F_1(\lambda)|^2 d\rho (\lambda)=
  \int^\infty_{-\infty} |F_2 (\lambda)|^2 d\rho (\lambda) = \Vert g \Vert^2
  \tag{5.19}\end{equation}
for any $f \in L^2 (0, b ),  b  < \infty$, where
\begin{equation}
  F_j (\lambda):= \int^ b_0 f(x) \varphi_j (x, \lambda)\, dx,
  \quad j= 1,2,
  \tag{5.20} \end{equation}
the function $\varphi_j (x, \lambda)$ solves equation (1.1) with $q=q_j$, and
$k^2=\lambda$, satisfies first two conditions (1.4), and
\begin{equation}
    g:=(I+K^\ast )f,
    \tag{5.21} \end{equation}
where $I+K$ is the transformation operator:
\begin{equation}
  \varphi_2 = (I+K) \varphi_1
  = \varphi_1 + \int^x_0 K(x,y) \varphi_1(y,\lambda)\, dy.
  \tag{5.22}\end{equation}

Note that
\begin{equation}
  F_2 (\lambda) = \int^ b_0 f(x) (I+K) \varphi_1 dx = \int^ b_0
  g(x) \varphi_1 (x, \lambda) \, dx.
  \tag{5.23}\end{equation}
From (5.19) it follows that
\begin{equation}
  \Vert f \Vert = \Vert (I+K^\ast )f\Vert
  \qquad \forall  f \in L^2 (0, b ):= H.
  \tag{5.24}\end{equation}

Since Range $(I+K^\ast ) = H$, equation (5.24) implies that $I+K$ is unitary
(an isometry whose range is the whole space $H$). Thus
\begin{equation}
  I+K = (I+K^\ast )^{-1} = I +T^\ast,
  \tag{5.25}\end{equation}
where $T^\ast $ is a Volterra operator of the type (5.10).

Therefore $K=T^\ast$ and this implies $K=T^\ast=0$. Therefore $\varphi_1 =
\varphi_2$ and $q_1 = q_2$.
Theorem 5.2 is proved.
\end{proof}
Let $d\rho_j(\lambda), j=1,2,$ be the spectral functions
corresponding to the operators $L_{q_j}$. Assume that
$d\rho_1(\lambda)=cd\rho_2(\lambda),$ where $c>0$ is a constant.
The above argument can be used with a minor change
to prove that this assumption implies: $c=1$ and $q_1=q_2$.
Indeed, the above assumption implies unitarity of
the operator $\sqrt{c}(I+K)$. Therefore $c(I+K)=I+T^*$.
Thus $c=1$ and $K=T^*=0$, as in the proof of Theorem 5.2.
Here we have used a simple claim: 

{\it If $bI+Q=0$, where $b=const$ and
$Q$ is a linear compact operator in $H$, then $b=0$ and $Q=0$.}

To prove this claim, take an arbitrary orthonormal
basis $\{u_n\}$ of the Hilbert space $H$. Then
$||Qu_n||\to 0$ as $n\to \infty$ since $Q$
is compact. Note that $||u_n||=1$, so
 $b=||bu_n||=||Qu_n||\to 0$ as $n\to \infty$.
Therefore $b=0$ and consequently $Q=0$. The claim is proved.

The Gelfand-Levitan (GL) reconstruction procedure is:
\begin{equation}
  \rho (\lambda) \Rightarrow L(x,y) \Rightarrow K(x,y) \Rightarrow q(x).
  \tag{5.26}\end{equation}
Here
\begin{equation}
  L(x,y):= \int^\infty_{-\infty} \varphi_0 (x,\lambda) \varphi_0 (y,\lambda)
  d\sigma (\lambda), \quad d\sigma = d \rho - d\rho_0,
  \tag{5.27}  \end{equation}
\begin{equation}
  d\rho_0 = \begin{cases}
  \frac{\sqrt{\lambda}\, d \lambda}{\pi}, & \lambda >0, \\
  0 & \lambda <0. \end{cases}
  \tag{5.28}\end{equation}

Compare (5.28) and (2.5) and conclude that $\rho_ 0$ is the spectral
function corresponding to the Dirichlet operator
$\l_q = -\frac{d^2}{dx^2}+q(x)$ in $L^2 (\R_+)$ with $q(x) = 0$.

The function $K(x,y)$ defines the transformation operator (cf. (1.10))
\begin{equation}
  \varphi (x, \lambda) = \varphi_0 (x, \lambda)
  + \int^x_0 K(x,y)\varphi_0(y, \lambda)\, dy,\quad 
  \varphi_0: = \frac{\sin(x \sqrt{\lambda})}{\sqrt {\lambda}},
  \tag{5.29}\end{equation}
where $\varphi$ solves (1.1) with $k^2 = \lambda$ and satisifies first two
conditions (1.4).

One can prove (see \cite{Lt1}, \cite{R}), that $K$ and $L$ are related
by the Gelfand-Levitan equation:
\begin{equation}
  K(x,y) + L(x,y) + \int^x_0 K(x,t) L(t,y)\, dt = 0, \quad
  0 \leq y \leq x.
  \tag{5.30}\end{equation}

Let us assume that the data $\rho(\lambda)$ generate the kernel
$L(x,y)$ (by formula (5.27)) such that equation (5.30) is a Fredholm-type
equation in $L^2 (0,x)$ for $K(x,y)$ for any fixed $x>0$.

Then, one can prove that assumption (5.3) implies the unique
solvability of the equation (5.30) for $K(x,y)$ in the space $L^2(0,x)$.

Indeed, the homogeneous equation (5.30)
\begin{equation}
h(y) + \int^x_0 L(t,y) h(t)\, dt = 0, \quad 0\leq y \leq x,
\tag{5.31}\end{equation}
implies $h=0$ if (5.3) holds. To see this, multiply (5.21) by
$h$ and integrate
over $(0,x)$ (assuming without loss of generaity that $h=\overline h$,
since the kernel $L(t,y)$ is real-valued). The result is:
\begin{equation}
  0= \Vert h \Vert^2 + \int^\infty_{-\infty}
  |H(\lambda)|^2 d\rho (\lambda) -\int^\infty_{-\infty}
  |H(\lambda)|^2 d \rho_0(\lambda),
  \notag\end{equation}
or, by Parseval's equality,
\begin{equation}
  \int^\infty_{-\infty} |H(\lambda)|^2 d\rho (\lambda) = 0.
  \tag{5.32}\end{equation}
From (5.32) and (5.3) it follows that $h(y) = 0$.
Therefore, by Fredholm's alternative, equation (5.30) is uniquely
solvable.

If $K(x,y)$ is its solution, then
\begin{equation}
q(x) = 2 \frac{d K(x,x)}{dx}. \tag{5.33}
\end{equation}

If $K(x,x)$ is a $C^{m+1}$ function then $q(x)$ is a $C^m$-function.

{\it One has to prove that potential (5.22) generate the spectral function
$\rho(\lambda)$ with which we started the inversion procedure (5.26).}

We want to prove more, namely the following:

\begin{theorem} 
 Each step in the diagram (5.26) is invertible, so
\begin{equation}
  \rho \Leftrightarrow L \Leftrightarrow K \Leftrightarrow q.
  \tag{5.34}  \end{equation} 
\end{theorem}

\begin{proof}
1) Step $\rho \Rightarrow L$ is done by formula (5.27).

Let us
prove $L \Rightarrow \rho$. Assume there are $\rho_1$ and $\rho_2$
corresponding to the same $L(x,y)$. Then
\begin{equation} 
  0= \int^\infty_{-\infty} \varphi_0 (x,\lambda)\varphi_0 (y,\lambda)\,d\nu
 \qquad \forall  x, y \in \R. \tag{5.35}\end{equation} 
Therefore
\begin{equation} 
  0= \int^\infty_{-\infty} H^2(\lambda)\, d\nu
  \qquad\forall h \in C^\infty_0 (\R_+).
  \tag{5.36}\end{equation} 
By Theorem 5.1 relation (5.36) implies $\nu(\lambda) =0$, so
$\rho_1 = \rho_2$.

2) Step $L \Rightarrow K$ is done by solving equation (5.30) for $K(x,y)$.
The unique solvability of this equation for $K(x,y)$ has been proved below
formula (5.30).

Let up prove $K \Rightarrow L$. From (5.27) one gets
\begin{equation} 
  L(x,y) = \frac{L(x+y) - L(x-y)}{2},
  \quad L(x):= \int^\infty_{-\infty}
  \frac{1-\cos (x \sqrt{\lambda})}{\lambda} d\sigma(\lambda).
  \tag{5.37}\end{equation} 
Let $y=x$ in (5.30) and write (5.30) as
\begin{equation}
  L(2x) + \int^x_0 K(x,t)[L(x+t)-L(t-x)] dt = -2K(x,x),
  \quad x \geq 0.
  \tag{5.38}\end{equation}
Note that $L(-x)=L(x)$. Thus (5.38) can be written as:
\begin{equation} 
  L(2x) + \int^{2x}_x K(x, s-x) L(s)ds -\int^x_0 K(x,x-s)L(s)
  ds=-2K(x,x), \quad x>0.
    \tag{5.39}\end{equation}

This is a Volterra integral equation for $L(s)$. Since it is uniquely
solvable, $L(s)$ is uniquely recovered from $K(x,y)$ and the step 
$K \Rightarrow L$ is done.

3) Step $K \Rightarrow q$ is done by equation (5.33).

The converse step
$q \Rightarrow K$ is done by solving the Goursat problem:
\begin{equation} 
  K_{xx} - q(x)K = K_{yy}
  \quad 0 \leq y \leq x, \tag{5.40}\end{equation}
\begin{equation} 
  K(x,x) = \frac{1}{2} \int^x_0 q(t)dt,
  \quad K(x,0) = 0.
  \tag{ 5.41}\end{equation}
One can prove that any twice differentiable solution to (5.30) solves
(5.40)-(5.41) with $q(x)$ given by (5.33). The Goursat problem (5.40)-(5.41)
is known to have a unique solution. Problem (5.40)-(5.41) is equivalent to a
Volterra equation (\cite{M}, \cite{R}).

Namely if $\xi = x+y$, $\eta = x-y$, $K(x,y):= B(\xi,\eta)$, then
(5.40)-(5.41) take the form
\begin{equation} 
   B_{\xi \eta} = \frac{1}{4}q (\frac{\xi + \eta}{2})  B (\xi, \eta), 
   \quad B (\xi, 0) = \frac{1}{2} \int^{\frac{\xi}{2}}_0 q(t)\,dt,
   \quad B (\xi, \xi)=0.
   \tag{5.42}\end{equation} 
Therefore
\begin{equation} 
  B (\xi,\eta)= \frac{1}{4} \int^\xi _\eta q\left(\frac{t}{2}\right)\,dt
  + \frac{1}{4}\int_{\eta}^{\xi}\,dt \int^\eta_0 q \left(
\frac{t+s}{2}\right) B (t,s)\, ds.
  \tag{5.43}\end{equation}

This Volterra equation is uniquely solvable for $B(\xi,\eta)$.

Theorem 5.3 is proved.
\end{proof} 

\subsection{Inverse scattering problem on the half-line.} 

This problem consists of finding $q(x)$ given the data (2.10). Theorem 2.2
guarantees the uniqueness of the solution of this inverse problem in the 
class $L_{1,1}: = L_{1,1} (\R_+)$ of the potentials.

The characterization of the scattering data (2.10) is known (\cite{M},
\cite{R}), that
is, necessary and sufficient conditions on ${\mathcal S}$ for
${\mathcal S}$
 to be the scattering data
corresponding to a $q(x) \in L_{1,1}$. We state the result without proof.
A proof can be found in \cite{R}.
A different but equivalent version of the result is given in \cite{M}.

\begin{theorem} 
 For the data (2.10) to be the scattering data corresponding to
a $q \in L_{1,1}$ it is necessary and sufficient that the following
conditions hold:
\begin{equation} 
  \begin{align}
  \hbox{i)}\qquad
     & \ ind\ S(k) = -\kappa \leq 0,
     \quad \kappa= 2J \quad \hbox{or} \quad \kappa= 2J +1,
     \tag{5.44}\\
  \hbox{ii)}\qquad
     &\  k_j > 0,\quad s_j >0,\quad 1 \leq j \leq J, \tag{5.45} \\
  \hbox{iii)}\qquad
     &\  \overline{S(k)}=S(-k)=S^{-1}(k),
     \  S(\infty)=1,\ k\in\R,\tag{5.46}\\
  \hbox{iv)}\qquad
    &\ \Vert F(x) \Vert_{L^\infty (\R_+)}
   + \Vert F(x) \Vert_{L^1 (\R_+)} +
   \Vert x F^\prime(x)\Vert_{L^1(\R_+)} < \infty.
   \tag{5.47}  \end{align}
  \end{equation}
Here $\kappa = 2J +1$ if $f(0) =0$ and $\kappa = 2J$ if $f(0) \neq 0$, and
\begin{equation} 
  F(x): = \frac{1}{2\pi} \int^\infty_{-\infty} [1-S(k)] e^{ikx} dk +
  \sum^J_{j=1} s_j e^{-k_jx}.
  \tag{5.48}\end{equation}
\end{theorem}

The following estimates are useful (see \cite{M}, p.209,
 \cite{R}, \cite {R19},
p.569 ):
$$
|F(2x)+A(x,x)|<c\int_x^{\infty}|q(x)|dx,\quad
|F(2x)|<c\int_x^{\infty}|q(x)|dx,
$$
$$
|F'(2x)-\frac {q(x)}4|<c(\int_x^{\infty}|q(x)|dx)^2,
$$
where $c>0$ is a constant.
The Marchenko inversion procedure for finding $q(x)$ from ${\mathcal S}$
is described by the following diagram
\begin{equation} 
 {\mathcal S} \Rightarrow F(x) \Rightarrow A(x,y) \Rightarrow q(x).
 \tag{5.49}\end{equation}

The step ${\mathcal S} \Rightarrow F$ is done by formula (5.48).

The step $F \Rightarrow A$ is done by solving the Marchenko equation for 
$A(x,y)$:
\begin{equation} 
  A(x,y) + F(x+y) + \int^\infty_x A(x,t) F(t+y)\, dt = 0,
  \quad y \geq x \geq 0.
  \tag{5.50}\end{equation} 
The step $A \Rightarrow q$ is done by the formula
\begin{equation} 
  q(x) = -2 \frac{dA(x,x)}{dx}. \tag{5.51}\end{equation} 

{\it It is important to check that the potential $q(x)$ obtained by the scheme 
(5.49) generates the same data ${\mathcal S}$
with which we started the inversion scheme
(5.49).}

Assuming $q \in L_{1,1}$ we prove:

\begin{theorem}  
 Each step of the diagram (5.49) is invertible:
\begin{equation} 
{\mathcal S} \Leftrightarrow F \Leftrightarrow A \Leftrightarrow q.
\tag{5.52}
  \end{equation} 
\end{theorem}

\begin{proof}
1. The step ${\mathcal S} \Rightarrow F$ is done by formula (5.48) as we
have already mentioned.

The step $F \Rightarrow {\mathcal S}$ is done by finding $k_j$,
$s_j$ and $J$ from the
asymptotics of the function (5.48) as $x \to -\infty$. 
As a result, one finds the function
\begin{equation} 
  F_d (x):= \sum^J_{j=1} s_j e^{-k_jx}.
  \tag{5.53}\end{equation}
If $F(x)$ and $F_d (x)$ are known, then the function
\begin{equation} 
  F_S (x):= \frac{1}{2\pi} \int^\infty_{-\infty}
  [1-S(k)] e^{ikx} dk \tag{5.54}\end{equation} 
is known. Now the function $S(k)$ can  be found by the formula
\begin{equation} 
  S(k) = 1-\int^\infty_{-\infty} F_S (x) e^{-ikx} dx.
  \tag{5.55}\end{equation}
So the step $F \Rightarrow {\mathcal S}$ is done.

2. The step $F \Rightarrow A$ is done by solving equation (5.50)
for $A(x,y)$.
This step is discussed in the literature in detail,
(see \cite{M}, \cite{R}).
If $q \in L_{1,1}$ (actually a weaker condition 
$\int^\infty_0  x|q(x)| dx < \infty$
is used in the half-line scattering theory),
then one proves that conditions i) - iv) of Theorem 5.4 are
satisified, that the operator
\begin{equation} 
 Tf: = \int^\infty_x F(y+t) f(t)\, dt,
 \quad y\geq x\geq 0, 
 \tag{5.56}\end{equation}
is compact in $L^1(x,\infty)$ and in $L^2(x,\infty)$
for any fixed $x\geq 0$,
and the homogeneous version of equation (5.50):
\begin{equation} 
  f + Tf = 0, \quad y \geq x \geq 0,
  \tag{5.57}\end{equation}
has only the trivial solution $f=0$ for every $x \geq 0$.
Thus, by the Fredholm alternative, equation (5.50)
is uniquely solvable in $L^2(x,\infty)$
and in $L^1(x,\infty)$. The step $F \Rightarrow A$ is done.

Consider the step $A(x,y) \Rightarrow F(x)$.
Define
\begin{equation} 
  A(y):= \begin{cases} A(0,y), & y\geq 0,\\ 0, & y<0. \end{cases}\ 
  \tag{5.58}\end{equation} 

The function $A(y)$ determines uniquely $f(k)$ by the formula:
\begin{equation} 
  f(k) = 1 + \int^\infty_0 A(y)e^{iky} dy, \tag{5.59}\end{equation} 
and consequently it determines
 the numbers $ik_j$ as the only zeros of $f(k)$ in $\C_+$,
the number $J$ of these zeros, and $S(k) = \frac{f(-k)}{f(k)}$. To find
$F(x)$,
one has to find $s_j$. Formula (2.12) allows one to calculate $s_j$ if $f(k)$ 
and $f^\prime (0,ik_j)$ are known. To find $f^\prime (0,ik_j)$,
use formula (2.26) and put
$k=ik_j$ in (2.26). Since $A(x,y)$ is known for $y \geq x \geq 0$, formula
(2.26) allows one to calculate $f^\prime (0,ik_j)$. Thus 
$S(k), k_j, s_j, 1\leq j \leq J$, are found and $F(x)$ can be calculated by 
formula (5.48). Step $A \Rightarrow F$ is done.

The above argument proves that the knowledge of two
functions $A(0,y)$ and $A_x(0,y)$ for all $y\geq 0$
determines $q(x)\in L_{1,1}(\R_+)$ uniquely.

Note that:

a) we have used the following scheme 
$A \Rightarrow {\mathcal S}\Rightarrow F$ in order to get
the implication $A \Rightarrow F$,

and

b)  since $ \{F(x), x\geq 0 \} \Rightarrow A$ and 
$A \Rightarrow \{F(x), -\infty < x < \infty \}$, 
we have proved also the following non-trivial 
implication $\{F(x), x\geq 0 \} \Rightarrow \{F(x), -\infty < x < \infty
\}$.
 
3. Step $A \Rightarrow q$ is done by formula (5.51). The converse step 
$q \Rightarrow A$ is done by solving the Goursat problem:
\begin{equation} 
  A_{xx} - q(x)A = A_{yy},\quad y \geq x \geq 0,
  \tag{5.60}\end{equation}
\begin{equation} 
  A(x,x) = \frac{1}{2} \int^\infty_x q(t)\, dt, \quad A(x,y) \to 0
  \ \hbox{as}\  x+y \to \infty.
  \tag{5.61}\end{equation} 
Problem (5.60)-(5.61) is equivalent to a Volterra integral equation for
$A(x,y)$ (see \cite[p.253]{R}).
\begin{equation} 
  A(x,y) = \frac{1}{2} \int^\infty_{\frac{x+y}{2}} q(t) \,dt
  + \int^\infty_{\frac{x+y}{2}}
  ds\int^{\frac{y-x}{2}}_0\, dt q(s-t)A(s-t,s+t).
  \tag{5.62}\end{equation}

One can prove that any twice differentiable solution to (5.50) solves
(5.60)-(5.61) with $q(x)$ given by (5.51).

A proof can be found in \cite{M}, \cite{Lt1} and \cite{R}.

Theorem 5.5 is proved. 
\end{proof} 

\begin{remark} 
 It follows from Theorem 5.5 that the potential obtained by the
scheme (5.49) generates the scattering data ${\mathcal S}$
with which the inversion procedure (5.49) started.
\end{remark}

Similarly, Theorem 5.3 shows that the potential obtained by the scheme (5.26)
generates the spectral function $\rho(\lambda)$ with which the inversion 
procedure (5.26) started.

{\it The last conclusion one can obtain only because
of the assumption that $q(x)$ is such that the limit-point case at infinity
is valid}.

If this is not the case then there are many spectral function
corresponding to a given $q(x)$, so one cannot claim that the $\rho(\lambda)$
with which we started is the (unique) spectral function which is generated by
$q(x)$, it is just one of many such spectral functions.

\begin{remark} 
 In \cite{R3} the following new equation is derived:
\begin{equation} 
 F(y) +A(y) + \int^\infty_0 A(t) F(t+y)dt = A(-y),
 \quad -\infty < y < \infty,
 \tag{5.63}\end{equation} 
which generalizes the usual equation (5.50) at $x=0$:
\begin{equation} 
  F(y) + A(y) + \int^\infty_0 A(t) F(t+y)dt = 0,
  \quad y \geq 0.
  \tag{5.64}\end{equation} 
Since $A(-y) = 0$ for $y>0$ (see (5.58)), equation (5.64) follows from (5.63)
for $y>0$. For
$y=0$ equation (5.64) follows from (5.63) by taking $y \to +0$ and
using (5.58).
\end{remark}

Let us prove that 
{\it equations (5.63) and (5.64) are equivalent}. Note that 
equation (5.64) is uniquely solvable if the data (2.10) correspond to a
$q \in L_{1,1}$. Since any solution of (5.63) in $L^1(\R_+)$ solves (5.64),
any solution to (5.63) equals to the unique solution $A(y)$ of (5.64) for
$y>0$.

Since we are looking for the solution $A(y)$ of (5.63) such that $A=0$ for
$y<0$ (see (5.58)) one needs only to check that (5.63) is satisfied by the 
unique solution of (5.64).

\begin{lemma} 
 Equation (5.63) and (5.64) are equivalent in $L^1(\R_+)$.
\end{lemma}

\begin{proof}
Clearly, every $L^1(\R_+)$ solution to (5.63) solves (5.64).
Let us prove the converse. Let $A(y) \in L^1(\R_+)$ solve (5.64). Define
\begin{equation} 
  f(k):=1+ \int^\infty_0 A(y) e^{iky}dy: = 1+ \widetilde A (k). \tag{5.65}
  \end{equation} 

We wish to prove that $A(y)$ solves equation (5.63). Take the Fourier 
transform of (5.63) in the sense of distributions. From (5.48) one gets
\begin{equation}
  \widetilde F(\xi)=\int^\infty_{-\infty} F(x) e^{i\xi x}dx
  =1-S(-\xi)+2\pi \sum^J_{j=1} s_j\, \delta(\xi+ik_j),
  \tag{5.66} \end{equation}
and from (5.63) one obtains:
\begin{equation}
  \widetilde F(\xi) + \widetilde A (\xi) + \widetilde A (-\xi)
  \widetilde F(\xi) = \widetilde A (-\xi).
   \tag{5.67}\end{equation}
Add $1$ to both sides of (5.67) and use (5.65) to get
\begin{equation} 
 f(\xi) + \widetilde F (\xi) f(-\xi) = f(-\xi).
 \tag{5.68}\end{equation}
From (5.66) and (5.68) one gets:
\begin{equation} 
  f(\xi) =
  \left[ S(-\xi) - 2\pi \sum^J_{j=1} s_j \delta(\xi + ik_j) \right]
  f(-\xi) = f(\xi)-2 \pi f(-\xi) \sum^J_{j=1} s_j \delta(\xi + ik_j) = f(\xi).
  \tag {5.69}\end{equation} 
Equation (5.69) is equivalent to (5.63) since all the transformations which
led from (5.63) to (5.69) are invertible. Thus, equations (5.63) and (5.69)
hold (or fail to hold) simultaneously. Equation (5.69) clearly holds because
\begin{equation} 
  f(-\xi) \sum^j_{j=1} s_j \delta(\xi +ik_j)
  = \sum^J_{j=1} s_j f(ik_j) = 0, \tag{5.70}\end{equation} 
since $ik_j$ are zeros of $f(k)$.

Lemma 5.2 is proved.
\end{proof}  

The results and proofs in this section are partly new and partly
are based on the results in \cite{R3} and \cite{R}.

\section{Inverse problem for an inhomogeneous Schr\"odinger equation} 

Consider the problem
\begin{equation} 
  u^{\prime\prime}+ k^2u-q(x)u = -\delta(x), \quad -\infty < x < \infty,
  \tag{6.1}\end{equation}
\begin{equation} 
  \frac{\partial u}{\partial |x|} - iku \to 0
  \ \hbox{as}\  |x| \to \infty.
  \tag{6.2}\end{equation} 
Assume
\begin{equation} 
  q=\overline q, \quad q = 0
  \ \hbox{for}\  |x| >1,\quad  q \in L^1[-1,1].
  \tag{6.3}\end{equation}
Suppose the data
\begin{equation} 
  \{u(-1,k),u(1,k)\}_{\forall  k > 0} 
  \tag{6.4}\end{equation} 
are given.

The inverse problem (IP6) is:

{\it Given data (6.4),  find $q(x)$.}

Let us also assume that

(A): {\it
The operator $L_q= -\frac{d^2}{dx^2} + q(x)$ in $L^2 (\R)$
has no negative eigenvalues.}

This is so if, for example, $q(x) \geq 0$.

The results of this section are taken from \cite{R4}

\begin{theorem} 
If (6.3) and (A) hold then the data (6.4) determine $q(x)$ uniquely.
\end{theorem}

\begin{proof}  The solution to (6.1)-(6.2) is
\begin{equation} 
  u=\begin{cases} \frac{g(k)}{[f,g]}\ f(x,k), &  \quad x>0,\\
  \frac{f(k)}{[f,g]}\ g(x,k), & x<0, \end{cases}
  \tag{6.5}\end{equation}

where $f=f_+(x,k)$, $g=g_-(x,k)$, $g(k):=g_-(0,k)$, $f(k):=f(0,k)$,
$[f,g]:=fg^\prime -f^\prime g=-2ik a(k)$, $a(k)$ is defined in (2.53),
$f$ is defined in (1.3) and $g$ is defined in (1.5).

The functions
\begin{equation} 
  u(1,k) = \frac{g(k) f(1,k)}{-2ik a(k)},\quad  u(-1,k) = 
  \frac{f(k) g(-1,k)}{-2ik a(k)}
  \tag{6.6}\end{equation}
are the data (6.4).

Since $q=0$ when $x \not\in [-1,1]$, condition
 (6.2) implies
$f(1,k)= e^{ik}$, so one knows
\begin{equation} 
  h_1(k):= \frac{g(k)}{a(k)}, \quad  h_2 (k):= \frac{f(k)}{a(k)},
  \qquad \forall k>0. 
  \tag{6.7}\end{equation} 
From (6.7), (2.49) and (2.50) one derives
\begin{equation}
  a(k) h_1(k) = - b  (-k) f(k) +a(k) f(-k)
  = - b(-k) h_2 (k)a(k) +h_2(-k) a(-k) a(k),
  \tag{6.8}\end{equation}
and
\begin{equation} 
  a(k)h_2(k) =  b  (k) a(k) h_1(k) + a(k) h_1 (-k) a(-k).
  \tag{6.9}\end{equation}
From (6.8) and (6.9) one gets
\begin{equation} 
  - b  (-k) h_2(k) + a(-k) h_2 (-k) = h_1(k),
  \tag{6.10}\end{equation}
and
\begin{equation}
   b  (k) h_1(k) + a (-k) h_1 (-k) = h_2(k).
   \tag{6.11}\end{equation}
Eliminate $ b (-k)$ from (6.10) and (6.11) to get
\begin{equation} 
  a(k) = m(k) a(-k) + n(k), \qquad \forall k \in \R, 
  \tag{6.12}\end{equation} 
where
\begin{equation} 
  m(k):= -\frac{h_1(-k) h_2(-k)}{h_1(k) h_2(k)},
  \quad n(k):= \frac{h_1(-k)}{h_2(k)}+ \frac{h_2(-k)}{h_1(k)}.
  \tag{6.13}\end{equation} 

Problem (6.12) is a {\it Riemann problem} for the pair $\{a(k), a(-k)\}$, the 
function $a(k)$ is analytic in $\C_+ :=\{k : k \in \C$, $Imk>0\}$
and $a(-k)$
is analytic in $\C_-$. The functions $a(k)$ and $a(-k)$ tend to one as $k$
tends to infinity in $\C_+$ and, respectively, in $\C_-$, see equation (2.55).

The function $a(k)$ has finitely many simple zeros at the points 
$ik_j, 1 \leq j \leq J$, $ k_j > 0$, where $-k_j^2$ are
the negative eigenvalues
of the operator $\l$ defined by the differential expression
$\l u = -u^{\prime\prime}  + q(x)u$
in $L^2(\R)$.

The zeros $ik_j$ are the only zeros of $a(k)$ in the upper half-plane $k$.

Define
\begin{equation} 
  ind\ a(k) := \frac{1}{2 \pi i} \int^\infty_{-\infty} d \ln a(k).
  \notag\end{equation} 
One has
\begin{equation} 
   ind\ a = J,
   \tag{6.14} \end{equation}
where $J$ is the number of negative eigenvalues of the operator $\ell$,
and, using (6.14) and (6.15), one gets
\begin{equation} 
  ind\ m(k) = -2[ind\ h_1(k) + ind\ h_2(k)]
  =-2[ind\ g(k) + ind\ f(k) - 2J].
  \tag{6.15} \end{equation} 

Since $\ell$ has no negative eigenvalues by the assumption (A), it follows
that $J=0$.

In this case $ind\ f(k) = ind\ g(k) = 0$ (see Lemma 1 below), so
$ind\ m(k) = 0$,
and {\it $a(k)$ is uniquely recovered from the data} as the solution of (6.12) 
which tends to one at infinity. If $a(k)$ is found, then $b(k)$ is uniquely 
determined by equation (6.11) and so the reflection coefficient
$r(k) := \frac{b(k)}{a(k)}$ is found. The reflection coefficient
determines
a compactly supported $q(x)$ uniquely by Theorem 2.4.

If $q(x)$ is compactly supported, then the reflection coefficient 
$r(k) := \frac{b(k)}{a(k)}$ is meromorphic. Therefore, its values for all 
$k>0$ determine uniquely $r(k)$ in the whole complex $k$-plane as a 
meromorphic function. The poles of this function in the upper half-plane are 
the numbers $ik_j = 1,2,\dots,J$. They determine uniquely the numbers 
$k_j$, $1 \leq j \leq J$, which are a part of the standard scattering data
$\{r(k)$, $k_j$, $s_j$, $1 \leq j \leq J \}$,
where $s_j$ are the norming constants.

Note that if $a(ik_j) = 0$ then $b(ik_j) \neq 0$, otherwise equation
(2.49)
would imply $f(x, ik_j) \equiv 0$ in contradiction to (1.3).

If $r(k)$ is meromorphic, then the norming constants can be calculated by the
formula $s_j = -i\frac{b(ik_j)}{\dot a(ik_j)}
= -i \operatorname*{Res}_{k=ik_j} r(k)$, where the dot
denotes differentiation with respect to $k$, and
$\operatorname*{Res}$ denotes the residue.
So, for compactly supported potential the values of $r(k)$ for all $k>0$ 
determine uniquely the standard scatering data, that is, the reflection
coefficient, the bound states $-k_j^2$, and the norming constants 
$s_j,1 \leq j \leq J$. These data determine the potential uniquely.

Theorem 6.1 is proved. 
\qed
\end{proof} 

\begin{lemma} 
 If $J=0$ then $ind\ f=ind\ g= 0$.
\end{lemma}

\begin{proof}
We prove $ind\ f = 0$. The proof of the equation $ind\ g=0$ is 
similar. Since $ind\ f(k)$ equals to the number of zeros of $f(k)$ in 
$\C_+$, we have to prove that $f(k)$ does not vanish in 
$\C_+$. If $f(z)= 0, z \, \in \C_+$, then $z= ik$, $k>0$, and
$-k^2$ is an eigenvalue of the operator $\l$ in $L^2(0, \infty)$ with the
boundary condition $u(0) = 0$.

From the variational principle one can find the negative eigenvalues of the 
operator $\l$ in $L^2(\R_+)$ with the Dirichlet condition at $x=0$ as 
consequitive minima of the quadratic functional. The minimal eigenvalue is:
\begin{equation}
 \operatorname*{inf}_{u\in\oH1 (\R_+)}
   \int^\infty_0 [u^{\prime 2}+ q(x)u^2]\, dx := \kappa_0,
  \quad u \in \oH1(\R_+),\quad \Vert u \Vert_{L^2(\R_+)} = 1, 
  \tag{6.16} \end{equation}
where $\oH1
 (\R_+)$ is the Sobolev space of
$H^1(\R_+)$-functions satisfying
the condition $u(0) = 0$.

On the other hand, if $J=0$, then
\begin{equation}
  0\leq \operatorname*{inf}_{u\in H^1(\R)}
  \int^\infty_{-\infty} [u^{\prime 2}+q(x)u^2]\, dx := \kappa_1,
  \quad u \in H^1(\R),\quad  \Vert u \Vert_{L^2(\R)} = 1. 
  \tag{6.17} \end{equation}

Since any element $u$ of $\oH1 (\R_+)$ can be considered as an element of
$H^1(\R)$ if one extends $u$ to the whole axis by setting $u=0$ for $x<0$,
it follows from the variational definitions (6.16) and (6.17) that 
$\kappa_1 \leq \kappa_0$. Therefore, if $J=0$, then $\kappa_1 \geq 0$
and therefore $\kappa_0 \geq 0$.
This means that the operator $\l$ on $L^2(\R_+)$ with the Dirichlet
condition at $x=0$ has no negative eigenvalues. Therefore $f(k)$ does 
not have zeros in $\C_+$, if $J=0$. Thus $J=0$ implies $ind\ f(k) =0$.

Lemma 6.1 is proved.
\end{proof}

The above argument shows that in general
\begin{equation}
  ind\ f \leq J\quad \hbox{and}\quad ind\ g \leq J,
  \tag{6.18} \end{equation}
so that (6.15) implies
\begin{equation}
   ind\ m(k) \geq 0. 
\tag{6.19} \end{equation}

Therefore the Riemann problem (2.17) is always solvable.
It is of interest to study the case when 
 assumption (A) does not hold.

\section{Inverse scattering problem with fixed energy data}

\subsection{Three-dimensional inverse scattering problem. Property C} 

The scattering problem in $\R^3$ consists of finding the scattering solution 
$u:=u(x, \alpha, k)$ from the equation
\begin{equation}
  \left[\nabla^2 + k^2 - q(x) \right] \psi = 0 \ \hbox{in}\ \R^3 
  \tag {7.1} \end{equation}
and the radiation condition at infinity:
\begin{equation}
  \psi = \psi_0 + v, \quad \psi_0 := e^{ik \,\alpha\cdot x},
  \quad \alpha \in S^2,
  \tag{7.2} \end{equation}
\begin{equation}
  \lim_{r \to \infty} \int_{|s|=r}
  \left| \frac{\partial v}{\partial |x|}- ik\,v \right|^2 ds = 0.
  \tag{7.3} \end{equation}

Here $k>0$ is fixed, $S^2$ is the unit sphere, $\alpha \in S^2$ is given.
One can write
\begin{equation}
  v=A(\alpha^\prime, \alpha, k) \frac{e^{ikr}}{r} + o
  \left( \frac{1}{r} \right) \ \hbox{as}\ r=|x|\to\infty,
  \quad \frac{x}{r} = \alpha^\prime.
    \tag{7.4} \end{equation}

The coefficient $A(\alpha^\prime, \alpha, k)$ is
called the scattering amplitude.
In principle, it can be measured. We consider its values for 
$\alpha^\prime$, $\alpha \in S^2$ and a fixed $k>0$ as the scattering
data. Below we take $k=1$ without loss of generality.

Assume that
\begin{equation}
  q \in Q_a := \left\{q: q= \overline q,
  \quad q=0\ \hbox{for}\ |x| >a, \quad q \in L^p (B_a) \right\},
  \tag{7.5} \end{equation}
where $a>0$ is an arbitrary large fixed number, $B_a = \{x: |x|\leq a \}$,
$p > \frac{3}{2}$.

It is known (even for much larger class of the potentials $q$) that problem 
(7.1)-(7.3) has the unique solution.

Therefore the map
$q \to A(\alpha^\prime$, $\alpha):=
A_q (\alpha^\prime, \alpha)$ is
well defined,
\begin{equation}
  A(\alpha^\prime,\alpha):= A(\alpha^\prime, \ \alpha, k)|_{k=1}.
  \notag\end{equation}

(IP7) {\it The inverse scattering problem with
fixed-energy data consists of
finding $q(x) \in Q_a$ from the scattering data 
$A(\alpha^\prime, \alpha) \ \forall \alpha^\prime$, $\alpha \in S^2$.}

Uniqueness of the solution to IP7 for $q \in Q_a$ (with $p=2$) was first 
announced in \cite{R10} and proved in \cite{R11}
by the method, based on property
C for pairs of differential operators. The essence of this method is briefly 
explained below. This method was introduced by the author \cite{R9}
and applied to many inverse problems \cite{R10}-\cite{R14}, \cite{R7},
\cite{R1} \cite{R}.

In \cite{R} and \cite{R8} a characterization of the fixed-energy
scattering amplitudes is given.

Let $\{L_1, L_2\}$ be two linear formal differential  expressions,
\begin{equation}
  L_j u= \sum^{M_j}_{|m|=0} a_{mj} (x) \partial^m u(x),
  \quad  x \in \R^n, \quad  n>1, \quad  j=1,2,
  \tag{7.6} \end{equation}
where
\begin{equation}
   \partial^m := \frac{\partial^{|m|}}{\partial x^{m_1}_1\dots x^{m_n}_n},
   \quad |m| = m_1 +\dots+m_n.
   \notag\end{equation}
Let
\begin{equation}
  N_j := N_j (D) := \{w: L_j w=0\ \hbox{in}\ D\subset R^n\} 
  \tag{7.7} \end{equation}
where $D$ is an abitrary fixed bounded domain and the equation in (7.7) is 
understood in the sense of distributions.

Suppose that
\begin{equation}
  \int_D f(x) w_1(x) w_2(x)\, dx = 0,
  \quad w_j \in N_j, \quad f\in L^2(D),
  \tag{7.8} \end{equation}
where $w_j \in N_j$ run through such subsets of $N_j$, $j=1,2$, that the 
products $w_1 w_2 \in L^2(D)$, and $f\in L^2(D)$
is an arbitrary fixed function.

\begin{definition} 
 The pair $\{L_1, L_2\}$ has property C if (7.8) implies
$f(x) = 0$, that is, the set 
$\{w_1, w_2\}_{\forall w_j \in N_j,\ w_1w_2 \in L^2(D)}$ is complete in 
$L^2(D)$. 
\end{definition}

In \cite{R14} a necessary and sufficient condition is found  for a pair 
$\{L_1,L_2\}$ with constant coefficients, $a_{mj}(x) = a_{mj} = const$,
to have property C (see also \cite{R}).

In \cite{R11} it is proved that the pair $\{L_1,L_2\}$ with $L_j = -\nabla^2 +
q(x)$, $ q_j \in Q_a$, has property C.

The basic idea of the proof of the uniqueness theorem for inverse scattering
problem with fixed-energy data, introduced in \cite{R10},
presented in detail in \cite{R11},
and developed in \cite{R}, \cite{R12}-\cite{R14}, is simple. Assume that there are
two potentials, $q_1$ and $q_2$ in $Q_a$ which generate the same scattering
data, that is, $A_1 = A_2$, where $A_j := A_{q_j} (\alpha^\prime, \alpha)$,
$j= 1,2$.

We prove that \cite[p.67]{R}
\begin{equation}
  -4\pi (A_1 - A_2) = \int_{B_a} [q_1(x) -q_2(x)] \psi_1 (x, \alpha) 
  \psi_2(x, -\alpha^\prime)\, dx
  \qquad \forall \alpha, \alpha^\prime \in S^2,
  \tag{7.9} \end{equation}
where $\psi_j (x, \alpha)$ is the scattering solution corresponding to 
$q_j$, $j=1,2$.

If $A_1 = A_2$, then (7.9) yields an orthogonality relation:
\begin{equation}
  \int_D p(x) \psi_1(x,\alpha) \psi_2 (x, \beta)\,dx = 0
  \qquad  \forall \alpha,\beta \in S^2,
  \quad p(x) := q_1 - q_2.
  \tag{7.10} \end{equation}

Next we prove \cite[p.45]{R} that
\begin{equation}
  \hbox{span}_{\alpha \in S^2} \{\psi (x, \alpha)\}
  \ \hbox{is dense in}\ L^2(D)\ \hbox{in}\ N_j (D) \cap H^2(D), 
  \tag{7.11} \end{equation}
where $H^m(D)$ is the Sobolev space. Thus (7.10) implies
\begin{equation}
  \int_D p(x) w_1(x)w_2(x)\, dx = 0
  \quad \forall w_j \in N_j(D) \cap H^2(D).
  \tag{7.12} \end{equation}

Finally, by property C  for a pair $\{L_1, L_2\}$,
$L_j = -\nabla^2 + q_j$, $q_j \in Q_a$, one concludes from (7.11)
that $p(x) = 0$, i.e. $q_1=q_2$. We have obtained

\begin{theorem}[Ramm \cite{R11}, \cite{R}] 
The data
$A(\alpha^\prime, \alpha) \quad \forall \alpha^\prime$,
$\alpha \in S^2$ determine $q \in Q_a$ uniquely.
\end{theorem}

This is the uniqueness theorem for the solution to (IP7).
In fact this theorem is proved in \cite{R} in a stonger form: the data
$A_1(\alpha^\prime, \alpha)$ and
$A_2 (\alpha^\prime, \alpha)$ are asumed to be equal
not for all $\alpha^\prime, \alpha \in S^2$ but only on a set 
$\widetilde{S}^2_1 \times \widetilde{S}^2_2$,
where $\widetilde{S}^2_j$ is an arbitrary small
open subset of $S^2$.

In \cite{R7} the stability estimates for the solution to (IP7) with noisy
data are obtained and an algorithm for finding such a solution is proposed.

The noisy data is an arbitrary function
$A_\varepsilon(\alpha^\prime, \alpha)$, not
necessarily a scattering amplitude, such that
\begin{equation}
  \sup_{\alpha^\prime, \alpha \in S^2}
  \left| A(\alpha^\prime,\alpha)-A_{\varepsilon} (\alpha^{\prime},
  \alpha) \right|< \varepsilon.
  \tag{7.13} \end{equation}
Given $A_\varepsilon (\alpha^\prime, \alpha)$,
an algorithm for computing a quantity
$\widehat{q}_{\varepsilon}$ is proposed in \cite{R7}, such that
\begin{equation}
 \sup_{\xi\in\R^3}
  \left| \widehat{q}_{\varepsilon} -\widetilde{q} (\xi)\right|
 < c \frac{(\ln |\ln \varepsilon|)^2}{|\ln \varepsilon |}.
 \tag{7.14} \end{equation}
where $c>0$ is a constant depending on the potential but not
 on $\varepsilon$,
\begin{equation}
  \widetilde{q} (\xi): = \int_{B_a} e^{i \xi\cdot x} q(x)\,dx .
  \tag{7.15} \end{equation}

The constant $c$ in (7.14)
can be chosen uniformly for all potentials $q \in Q_a$
which belong to a compact set in $L^2 (B_a)$.

The right-hand side of (7.14) tends to zero as $\varepsilon \to 0$, but 
very slowly.

The author thinks that the rate (7.14) cannot be improved for the class
$Q_a$, but this is not proved.

However, in \cite{ARS} an example of two spherically
symmetric piecewise-constant
potentials $q(r)$ is constructed such that $|q_1-q_2|$ is of order $1$,
maximal value of each of the potentials $q$ is of order 1,
the two potentials
are quite different but they generate the set of the fixed-energy $(k=1)$
phase shifts $\left\{\delta^{(j)}_\l  \right\}_{\l =0,1,2,\dots}$, $j=1,2$, such that
\begin{equation}
  \delta^{(1)}_\l  = \delta^{(2)}_\l,
  \quad 0 \leq \l  \leq 4,
  \quad \left|\delta^{(1)}_\l-\delta^{(2)}_\l\right|
  \leq 10^{-5}, \qquad \forall \l  \geq 5.
  \tag{7.16} \end{equation}

In this example $\varepsilon \sim 10^{-5}$,
$(\ln|\ln \varepsilon|)^2\sim 2.59$, $|\ln \varepsilon|\sim 5$,
so the right-hand side of (7.14) is of
order $1$ if one assumes $c$ to be of order $1$.

{\it Our point is: there are examples in which 
the left and the right sides of estimate (7.14) 
are of the same order of magnitude. Therefore
estimate (7.14) is sharp.}

\subsection{Approximate inversion of fixed-energy phase shifts.}

Let us recall that $q(x) = q(r)$, $r=|x|$ if and only if
\begin{equation}
  A(\alpha^\prime, \alpha, k) = A(\alpha^\prime\cdot\alpha,k) .
  \tag{7.17} \end{equation}
It was well known for a long time that if $q=q(r)$ then (7.17) holds. The 
converse was proved relatively recently in \cite{R15} (see also \cite{R}).

If $q=q(r)$ then
\begin{equation}
  A(\alpha^\prime, \alpha) = \sum^\infty_{\l=0} A_\l Y_\l (\alpha^\prime)
  \overline{Y_\l(\alpha)},
  \tag{7.18} \end{equation}
where $Y_\l$ are orthonormal in $L^2(S^2)$ spherical harmonics, 
$Y_\l = Y_{\l m}$, $-\l \leq m \leq \l$, summation with respect to $m$ is 
understood in (7.18) but not written for brevity, the numbers $A_\l$ are 
related to the phase shifts $\delta_\l$ by the formula
\begin{equation}
  A_\l = 4\pi e^{i\delta_\l} \sin (\delta_\l) \quad (k=1),
  \tag{7.19} \end{equation}
and the $S$-matrix is related to $A(\alpha^\prime, \alpha,k)$ by the formula
\begin{equation}
  S = I - \frac{k}{2\pi i} A.
  \notag\end{equation}

If $q=q(r)$, $r=|x|$, then the scattering solution $\psi(x,\alpha, k)$
can be written as
\begin{equation}
  \psi(x, \alpha , k) = \sum^\infty_{\l=0} \frac{4\pi}{k}\, i^\l \,
  \frac{\psi_\l(r,k)}{r}\, Y_\l (x^0)\, \overline{Y_\l (\alpha)},
  \quad x^0 := \frac{x}{r}.
  \tag{7.20} \end{equation}

The function $\psi_\l(r,k)$ solves (uniquely) the equation
\begin{equation}
  \psi_\l(r,k) = u_\l(kr) - \int^\infty_0 g_\l (r, \rho)q(\rho) 
  \psi_\l(\rho, k)\, d\rho,
  \tag{7.21} \end{equation}
where
\begin{equation}
  u_\l (kr) := \sqrt{\frac{\pi kr}{2}} J_{\ell+1/2} (kr), \quad
  v_\l := \sqrt {\frac{\pi kr}{2}} N_{\ell+1/2} (kr) .
  \tag{7.22} \end{equation}
Here $J_\l$ and $N_\l$ are the Bessel and Neumann functions, and
\begin{equation}
  g_\l(r,\rho):=\begin{cases}
  \frac{\varphi_{0\l}(k\rho)f_{0\l}(kr)}{F_{0\l}(k)}, & r\geq\rho, \\
  \frac{\varphi_{0\l}(kr) f_{0\l}(k\rho)}{F_{0\l}(k)}, & r\leq\rho,
  \end{cases}
  \tag{7.23} \end{equation}
\begin{equation}
  F_{0\l} (k) = \frac{ e^{\frac{i\l \pi}{2} }}{k^\l} ,
  \tag{7.24} \end{equation}
$\psi_{0\l}$ and $f_{0\l}$ solve the equation 
\begin{equation}
  \psi^{\prime\prime}_{0\l} + k^2 \psi_{0\l} -
\frac{\l(\l+1)}{r^2} \psi_{0\l} = 0,
  \tag{7.25} \end{equation}
and are defined by the conditions
\begin{equation}
  f_{0\l} \sim e^{ikr}\ \hbox{as}\ r \to +\infty, 
  \tag{7.26} \end{equation}
so
\begin{equation}
  f_{0\l} (kr) = i\,e^{ \frac{i\l\pi}{2} } (u_\l + iv_\l),
  \tag{7.27} \end{equation}
where $u_\l$ and $v_\l$ are defined in (7.22), and
\begin{equation}
  \varphi_{0\l} (kr) := \frac{u_\l (kr)}{k^{\l+1}} .
  \tag{$7.27^\prime$}\end{equation}

In \cite{RSch} an approximate method was proposed recently
for finding $q(r)$ given
$\{\delta_\l\}_{\l=0,1,2,\dots}$.
In \cite{RSm} numerical results based on this method are described.

In physics one often assumes $q(r)$ known for $r \geq a$ and then the data
$\{\delta_\l \}_{\l=0,1,2,\dots }$ allow one to calculate the data
$\psi_\l (a), \l=0,1,2,\dots $, by solving the equation
\begin{equation}
  \psi^{\prime\prime}_\l + \psi_\l - \frac{\l(\l +1)}{r^2}
  \psi_\l-q(r) \psi_\l = 0, \quad r>a
  \tag{7.28} \end{equation}
together with the condition
\begin{equation}
  \psi_\l \sim e^{i\delta_\l} \sin(r - \frac{\l\pi}{2} + 
  \delta_\l), \quad r \to +\infty, \quad k=1, 
  \tag{7.29} \end{equation}
and assuming $q(r)$ known for $r>a$.

Problem (7.28) and (7.29) is a Cauchy problem with Cauchy data at infinity.
Asymptotic formula (7.29) can be differentiated. If the data 
$\psi_\l (a)$ are calculated and $\psi_\l (r)$ for $r \geq a$
is found, then one uses the equation
\begin{equation}
  \psi_\l (r) = \psi_\l^{(0)} (r) - \int^a_0 g_\l (r,\rho)
  q(\rho) \psi_\l (\rho)\, d\rho, \quad 0 \leq r \leq a, 
  \tag{7.30} \end{equation}
where $g_\l$ is given in (7.23),
\begin{equation}
  \psi^{(0)}_\l (r) := u_\l (r) - \int^\infty_a g_\l 
  (r,\rho)q(\rho) \psi_\l (\rho)\, d \rho ,
  \tag{7.31} \end{equation}
and $u_\l (r)$ is given in (7.22).

Put $r=a$ in (7.30) and get
\begin{equation}
  \int^a_0 g_\l (a, \rho) \psi_\l (\rho) q(\rho)\, d\rho = 
  \psi_\l^{(0)} (a) - \psi_\l (a):= b_\l, \quad
  \l = 0,1,2,.... .
  \tag{7.32} \end{equation}

The numbers $b_\l$ are known. If $q(\rho)$ is small or
$\l$ is large then the following approximation is justified:
\begin{equation}
  \psi_\l (\rho) \approx \psi_\l^{(0)} (\rho). 
  \tag{7.33} \end{equation}
Therefore, an approximation to equation (7.32) is:
\begin{equation}
  \int^a_0 f_\l (\rho) q(\rho)\, d\rho = b_\l,
  \quad  \l = 0,1,2,\dots,  
  \tag{7.34} \end{equation}
where
\begin{equation}
  f_\l (\rho) := g_\l (a, \rho) \psi^{(0)}_\l (\rho). 
  \tag{7.35} \end{equation}
The system of functions $\{ f_\l (\rho) \}$ is linearly independent. 

Equations (7.34) constitute a moment problem which can be solved numerically 
for $q(\rho)$ (\cite[p.209]{R}, \cite{RSm}).

\section {A uniqueness theorem for inversion of fixed-energy phase 
shifts.}  

From Theorem 7.1 it follows that if $q=q(r) \in Q_a$ then the data
\begin{equation}
  \{\delta_\l \} \quad \forall \l = 0,1,2, \dots \quad k=1 
  \tag{8.1} \end{equation}
determine $q(r)$ uniquely \cite{R11}.

Suppose a part of the phase shifts is known (this is the case in practice).

{\it What part of the phase shifts is sufficient for the unique recovery of 
$q(r)$? }

In this section we answer this question following \cite{R6}.
Define
\begin{equation}
  {\mathcal L} := \left\{
  \l : \sum_{\substack{\l\not= 0\\ \l\in{\mathcal L} }}
  \frac{1}{\l} = \infty
  \right\}
  \tag{8.2} \end{equation}
to be any subset of nonnegative integers such that condition (8.2) is 
satisfied.

For instance, ${\mathcal L}=\{ 2\l \}_{\l=0,1,2,\dots }$ or 
${\mathcal L}=\{2\l + 1\}_{\l = 0,1,2,\dots }$ will be admissible.

Our main result is

\begin{theorem} [\cite{R6}]
If $q(r) \in Q_a$ and $L$ satisfies (8.2) then the set of fixed-energy 
phase shifts $\{\delta_\l\}_{\forall \l \in {\mathcal L}}$ determine $q(r)$
uniquely.
\end{theorem}

Let us outline basic steps of the proof.

\underbar{Step 1.}
Derivation of the orthogonality relation:

If $q_1,q_2 \in Q_a$ generate the same data then 
$p(r): = q_1-q_2$ satisfies the relation
\begin{equation}
  \int^a_0 p(r) \psi^{(1)}_\l (r) \psi^{(2)}_\l (r)\, dr = 
  0 \quad \forall \l \in {\mathcal L}. 
  \tag{8.3} \end{equation}
Here $\psi^{(j)}_\l (r)$ are defined in (7.20) and correspond to
$q=q_j$, $j=1,2$. Note that Ramm's Theorem 7.1
yields the following conclusion: if
\begin{equation}
  \int^a_0 p(r) \psi^{(1)}_\l (r) \psi^{(2)}_\l (r)\, dr = 0
  \qquad  \forall \l \in {\mathcal L},
  \tag {8.4} \end{equation}
then $p(r)=0$, and $q_1=q_2$.

\underbar{Step 2.}
Since $\psi^{(j)}_\l = c^{(j)}_\l \varphi^{(j)}_\l(r)$,
where $c^{(j)}_\l$ are some constants, relation (8.3) is equivalent to
\begin{equation}
  \int^a_0 p(r) \varphi^{(1)}_\l (r) \varphi^{(2)}_\l (r)\,dr = 0
  \qquad   \forall \l \in {\mathcal L}. 
  \tag{8.5} \end{equation}

Here $\varphi^{(j)}_\l (r)$ is the solution to (7.28) with $q=q_j$, which
satisfies the conditions:
\begin{equation}
  \varphi^{(j)}_\l = \frac{r^{\l + 1}}{(2\l + 1)!!} + 
  o(r^{\l + 1}), \qquad r \to 0 ,
  \tag{8.6} \end{equation}
and
\begin{equation}
  \varphi^{(j)}_\l = |F^{(j)}_\l| \sin (r - \frac{\l \pi}{2} +
  \delta^{(j)}_\l) +o(1), \qquad r \to +\infty, 
  \tag{8.7} \end{equation}
where $|F^{(j)}_\l|\neq 0$ is a certain constant, and 
$\delta^{(j)}_\l$ is the fixed-energy $(k=1)$ phase shift, which does 
not depend on $j$ by our assumption:
$\delta^{(1)}_\l = \delta^{(2)}_\l\quad \forall \l \in {\mathcal L}$.

We want to derive from (8.5) the relation (7.10).
Since $q(x)=q(r)$, $r=|x|$ in this section, relation (7.10) is equivalent
to relation (8.12) (see below).
The rest of this section contains this derivation of (8.12).

We prove existence of the transformation kernel $K(r, \rho)$,
independent of
$\l$, which sends functions $u_\l (r)$, defined in (7.22), into
$\varphi_\l (r)$:
\begin{equation}
  \varphi_\l (r) = u_\l (r) + \int^r_0 K(r,\rho) u_\l (\rho)
  \frac{d \rho}{\rho^2}, \qquad K(r,0)=0. 
  \tag{8.8} \end{equation}
Let
\begin{equation}
  \gamma_\l := \sqrt{\frac{2}{\pi}} \Gamma \left( \frac{1}{2} \right)
  2^{\l +\frac{1}{2}} \Gamma(\l + 1) ,
  \tag{8.9} \end{equation}
where $\Gamma (z)$ is the gamma function,
\begin{equation}
  H(\l) := \gamma^2_\l \int^a_0 p(r) u^2_\l(r)\, dr, 
  \tag{8.10} \end{equation}
and
\begin{equation}
  h(\l) := \gamma^2_\l \int^a_0 p(r) \varphi^{(1)}_\l (r) 
  \varphi^{(2)}_\l (r)\, dr .
  \tag{8.11} \end{equation}

We prove that:
\begin{equation}
  \left\{
  h(\l) = 0 \qquad \forall \l \in {\mathcal L}\right\}
  \Rightarrow  \left\{h(\l) = 0
  \quad \forall \l = 0,1,2,\dots   \right\}.
  \tag{8.12} \end{equation}
If $h(\l) = 0, \quad \forall \l = 0,1,2,\dots $, then (8.4) holds, 
and, by theorem 7.1, $p(r) = 0$.

Thus, Theorem 8.1 follows.

The main claims to prove are:

1) Existence of the representation (8.8) and the estimate
\begin{equation}
  \int^r_0 |K(r,\rho)| \frac{d \rho}{\rho}<c(r)<\infty \qquad \forall r>0.
  \tag{8.13} \end{equation}

2) Implication(8.12).

Representation (8.8) was used in the physical literature (\cite{CS},
\cite{N}) but,
to our knowledge, without any proof. Existence of transformation operators
with kernels depending on $\l$ was proved in the literature \cite{V}. For our
purposes it is important to have $K(r,\rho)$ independent of $\l$.

Implication (8.12) will be estblished if one checks that $h(\l)$ is a 
holomorphic function of $\l$ in the half-plane
$\prod_+ := \{\l:\l\in\C, \quad Re\,\l > 0 \}$
which belongs to N-Class (Nevanlinna class).

\begin{definition} 
A function $h(\l)$ holomorphic in $\prod_+$ belongs to N-class $iff$
\begin{equation}
  \sup_{0<r<1} \int^\pi_{-\pi} \ln^+
  \left |h \left( \frac{1-re^{i\varphi} }{1+re^{i\varphi} } \right) \right|
  \,d\varphi < \infty.
\tag{8.14}\end{equation}
\end{definition}

Here
\begin{equation}
  u^+ :=\begin{cases} u\ \hbox{if}\ u\geq 0,\\ 0\ \hbox{if}\ u<0.\end{cases}
  \notag\end{equation}
The basic result we need in order to prove (8.12) is the following uniqueness theorem:

\begin{proposition} 
 If $h(\l)$ belongs to N-class then (8.12) holds.
\end{proposition}

\begin{proof}
This is an immediate consequence of the following:

{\it Theorem (\cite[p.334]{Ru}): If $h(z)$ is holomorphic in
$D_1 := \{ z:|z| <1, \quad z \in \C \}$,
$h(z)$ is of N-class in $D_1$, that is:
\begin{equation}
  \sup_{0<r<1} \int^\pi_{-\pi} \ln^+
  |h(re^{i \varphi})|\, d\varphi < \infty,
\tag{8.15} \end{equation}
and
\begin{equation}
  h(z_n) = 0, \qquad h=1,2,3,\dots , 
  \tag{8.16} \end{equation}
where
\begin{equation}
  \sum^\infty_{h=1} (1-|z_n|) = \infty ,
  \tag{8.17} \end{equation}
then $h(z) \equiv 0$. }

The function $z=\frac{1-\l}{1+\l}$ maps conformally
$\prod_+$ onto $D_1$,
$\l=\frac{1-z}{1+z}$ and if $h(\l)=0 \quad\forall \l \in {\mathcal L}$,
then $f(z):= h(\frac{1-z}{1+z})$ is holomorphic in $D_1$,
$f(z_\l)=0$ for $\l\in{\mathcal L}$ and
$z_\l:= \frac{1-\l}{1+\l}$, and
\begin{equation}
  \sum_{\l \in {\mathcal L}}
  \left( 1- \left| \frac{1-\l}{1+\l} \right| \right)
  \leq 1+\sum_{\l \in {\mathcal L}} \left( 1-\frac{\l-1}{\l+1} \right)
  = 1 + 2\sum _{\l \in {\mathcal L}} \frac{1}{\l+1} = \infty.
  \tag{8.18} \end{equation}

From (8.18) and the above Theorem Proposition 8.1 follows.
\end{proof}

Thus we need to check that function (8.11) belongs to N-class,
that is, (8.14) holds. 

So step 2, will be completed if one proves (8.8), (8.13) and (8.14).

Assuming (8.8) and (8.13), one proves (8.14) as follows:

i) First, one checks that (8.14) holds with $H(\l)$ in place of $h(\l)$.

ii) Secondly, using (8.8) one writes $h(\l)$ as:

\begin{equation}
  \begin{align}
  h(\l) & = H(\l) + \gamma^2_\l \int^r_0 [K_1(r, \rho) + K_2(r, \rho)]\,
    u_\l(\rho)u_\l(r) \frac{d \rho}{\rho^2}+ \notag\\
  & + \gamma^2_\l \int^r_0  \int^r_0 \frac{ds}{s^2}
  \frac{dt}{t^2} K_1(r,t)K_2(r,s) u_\l(t) u_\l(s).
  \notag \end{align}
  \tag{8.19} \end{equation}

Let us now go through i) and ii) in detail.

In \cite[8.411.8]{GR} one finds the formula:
\begin{equation}
  \gamma_\l u_\l(r) = r^{\l+1} \int^1_{-1} (1-t^2)^\l e^{irt}\,dt, 
  \tag{8.20} \end{equation}
where $\gamma_\l$ is defined in (8.9).

From (8.20) and (8.10) one gets:
\begin{equation}
  |H(\l)| \leq \int^a_0 dr |p(r)| r^{2\l+2}
  \left|\int^1_{-1} (1-t^2)^\l e^{irt}\,dt \right|^2
  \leq c\, a^{2 \sigma},
  \ \l=\sigma + i\tau, \ \sigma \geq 0.
  \tag{8.21} \end{equation}

One can assume $a>1$ without loss of generality. Note that
\begin{equation}
  \ln^+ (ab) \leq \ln^+a + \ln^+b \quad \hbox {if}\quad a,b > 0 
  \tag{8.22} \end{equation}

Thus (8.21) implies
\begin{equation}
 \begin{align}
 \int^\pi_{-\pi} & \ln^+ \left| H
   \left( \frac{1-re^{i\varphi}}{1+re^{il}} \right)
       \right|\,d\varphi
   \leq \int^\pi_{-\pi} \ln^+
   \left| ca^{2Re \frac{1-re^{i\varphi}}{1+re^{i\varphi}}}
\right|\,d\varphi
 \tag {8.23}\\
 & \leq |\ln c| +2\ln a \int^\pi_{-\pi}
   \left| Re \frac{1-re^{i\varphi}}{1+re^{i\varphi}} \right|\,d\varphi
 \notag \\
 &\leq |\ln c|+2\ln a\int^\pi_{-\pi}\frac{1-r^2}{1+2r\cos\varphi+r^2}
      \,d\varphi
 \leq |\ln c| +4\pi \ln a < \infty. \notag
 \end{align}
 \end{equation}

Here we have used the known formula:
\begin{equation}
  \int^\pi_{-\pi} \frac{d \varphi}{1+2r\cos\varphi + r^2}
  = \frac{2 \pi}{1-r^2}, \quad 0<r<1 .
  \tag{8.24}\end{equation}

Thus, we have checked that $H(\l)\in N(\prod_+)$, that is (8.14) holds for
$H(\l)$.

Consider the first integral, call it $I_1(\l)$, in (8.19).
One has, using (8.20) and (8.13),
\begin{equation}
   \begin{align}
  |I_1(\l)| & \leq \int^r_0 d\rho \rho^{-1}
    (| K_1 (r, \rho)| + |K_2 (r, \rho)|) r^{\l+1} \rho^\l
    \leq c(a) a^{2 \sigma},\notag \\
  &\qquad \l = \sigma + i\tau, \qquad\sigma\geq 0 . \tag{8.25}
  \end{align}\notag \end{equation}

Therefore one checks that $I_1 (\l)$ satisfies (8.14) (with $I_1$ in
place of $h$) as in (8.23).

The second integral in (8.19), call it $I_2 (\l)$, is estimated 
similarly: one uses (8.20) and (8.13) and obtains the following estimate:
\begin{equation}
  |I_2 (\l)| \leq c(a) a^{2 \sigma}, \qquad \l= \sigma + i\tau,
  \qquad  \sigma \geq 0 .
  \tag{8.26} \end{equation}

Thus, we have proved that $h(\l) \in N(\prod_+)$.

To complete the proof one has to derive (8.4) and check (8.13).

\begin{proof}[\bf Derivation of (8.4):]

Subtract from (7.28) with $q=q_1$ this equation with $q=q_2$ and get:
\begin{equation}
  w^{\prime\prime} + w- \frac{\l(\l + 1)}{r^2} w-q_1 w= p \psi^{(2)}_\l,
  \tag{8.27} \end{equation}
where
\begin{equation}
  p: = q_1-q_2, \qquad w:= \psi_\l^{(1)} (r) - \psi_\l^{(2)} (r).
  \tag{8.28} \end{equation}
Multiply (8.27) by $\psi^{(1)}_\l (r)$, integrate over $[0,\infty)$,
and then by parts on the left, and get
\begin{equation}
  \left( w^\prime\psi^{(1)}_\l - w \psi^{(1)^\prime}_\l \right)
  \bigg\vert^\infty_0
  =\qquad  \int^a_0 p \psi^{(2)}_\l  \psi^{(1)}_\l\, dr,
  \qquad \l\in{\mathcal L}.
  \tag{8.29} \end{equation}

By the assumption $\delta^{(1)}_\l = \delta^{(2)}_\l$ if 
$\l\in{\mathcal L}$, so $w$ and $w^\prime$ vanish at infinity.
At $r=0$ the left-hand side of (8.29) vanishes since
\begin{equation}
  \psi^{(j)}_\l (r) = O(r^{\l +1})\quad \hbox{as}\ r \to 0 .
  \tag{8.30} \end{equation}
Thus (8.29) implies (8.4). 
\end{proof}

\begin{proof}[\bf Derivation of
the representation (8.8) and of the estimate (8.13).]

One can prove \cite{R6} that the kernel $K(r, \rho)$ of the transformation 
operator must solve the Goursat-type problem
\begin{equation}
  r^2K_{rr}(r,\rho) - \rho^2 K_{\rho \rho}(r,\rho)
    + [r^2-r^2 q(r)-\rho^2] K(r,\rho) = 0,
  \qquad  0 \leq \rho \leq r ,
  \tag{8.31} \end{equation}
\begin{equation}
  K(r,r) = \frac{r}{2} \int^r_0 sq(s)\, ds := g(r) ,
  \tag{8.32} \end{equation}
\begin{equation}
  K(r,0) = 0,
  \tag{8.33} \end{equation}
and conversely: the solution to this Goursat-type problem
is the kernel of the transformation operator (8.8).

The difficulty in a study of the problem comes from the fact that the 
coefficients in front of the second derivatives degenerate at 
$\rho = 0$, $r = 0$.

To overcome this difficulty let us introduce new variables:
\begin{equation}
  \xi = \ln r + \ln \rho, \qquad \eta = \ln r-\ln \rho .
  \tag{8.34} \end{equation}
Put
\begin{equation}
  K(r,\rho):=B(\xi, \eta) .
  \tag{8.35} \end{equation}
Then (8.31)-(8.33) becomes
\begin{equation}
  B_{\xi \eta} -\frac{1}{2} B_\eta + Q(\xi, \eta)B=0,
  \qquad \eta \geq 0,
  \qquad -\infty < \xi < \infty,
  \tag{8.36} \end{equation}
\begin{equation}
  B(\xi, 0) = g(e^{\frac{\xi}{2}}) := G(\xi), 
  \tag{8.37} \end{equation}
\begin{equation}
  B(-\infty, \eta) = 0, \qquad \eta \geq 0 ,
  \tag{8.38} \end{equation}
where $g(r)$ is defined in (8.32) and
\begin{equation}
  Q(\xi, \eta):= \frac{1}{4} \left[ e^{\xi+\eta}-e^{\xi+\eta}
  q \left( e^{\frac{\xi+\eta}{2}} \right) -e^{\xi-\eta} \right].
  \tag{8.39} \end{equation}
Note that
\begin{equation}
   \sup_{-\infty < \xi < \infty} e^{-\frac{\xi}{2}} G(\xi) < c,
  \tag{8.40} \end{equation}
\begin{equation}
  \sup_{0 \leq \eta \leq B} \int^A_{-\infty} |Q(s,\eta)|\,ds\leq c(A,B),
  \tag{8.41} \end{equation}
for any $A\in \R$ and any $B>0$, where $c(A,B) > 0$ is some constant.

Let
\begin{equation}
  L(\xi, \eta) := B(\xi, \eta) e^{-\frac{\xi}{2}} 
  \tag{8.42} \end{equation}

Write (8.36)-(8.38) as
\begin{equation}
  L_{\xi \eta} +Q(\xi, \eta) L = 0,
  \qquad \eta \geq 0, \qquad-\infty < \xi < \infty
  \tag{8.43} \end{equation}
\begin{equation}
  L(\xi,0) = e^{-\frac{\xi}{2}} G(\xi) := b(\xi);
  \qquad L(-\infty, \eta) = 0, \qquad \eta \geq 0 .
  \tag{8.44} \end{equation}
Integrate (8.43) with respect to $\eta$ and use (8.44), and then integrate
with respect to $\xi$ to get:
\begin{equation}
  L=VL + b,\qquad VL:= -\int^\xi_{-\infty}\,ds \int^\eta_0 dt\,Q(s,t)L(s,t).
  \tag{8.45} \end{equation}

Consider (8.45) in the Banch space $X$ of continous function $L(\xi, \eta)$ 
defined for $\eta \geq 0, -\infty < \xi < \infty$, with the norm
\begin{equation}
  \Vert L \Vert := \Vert L \Vert_{AB}
  := \sup_{\substack{0\leq t\leq B \\-\infty<s\leq A}}
   \left( e^{-\gamma t} |L(s,t)| \right) < \infty,
  \tag{8.46} \end{equation}
where $\gamma = \gamma (A,B) > 0$ is chosen so that the operator $V$ is a 
contraction mapping in $X$. Let us estimate $\Vert V \Vert$:
\begin{equation}
  \begin{align}
  \Vert VL \Vert
  &\leq
  \sup_{\buildrel{-\infty<\xi\leq A}\over{0\leq \eta\leq B}}
  \int^\xi_{-\infty} ds \int^\eta_0 dt
  |Q(s,t)| e^{-\gamma(\eta-t)} e^{-\gamma t} |L(s,t)|
        \tag{8.47}\\
  &\leq \Vert L\Vert
  \sup_{\buildrel{-\infty<\xi\leq A}\over{0\leq \eta\leq B}}
  \int^\xi_{-\infty} ds \int^\eta_0 dt
  \left( 2e^{s+t}+e^{s+t}
     \left|q \left(e^{\frac{s+t}{2}} \right)\right| \right)
  e^{-\gamma (\eta - t)} \leq\frac{c}{\gamma} \Vert L \Vert ,
  \notag\end{align}
  \notag\end{equation}
where $c>0$ is a constant which depends on $A,B$, and on
$\int^a_0 r|q(r)|\, dr$.

If $\gamma >c$ then $V$ is a contraction mapping in $X$ and equation (8.45)
has a unique solution in $X$ for any $-\infty <A < \infty$ and $B>0$.

Let us now prove that estimate (8.13) holds for the constructed function
$K(r, \rho)$.

One has
\begin{equation}
  \int^r_0 |K(r, \rho)| \rho^{-1} d\rho
   = r \int^\infty_0 |L(2\ln r - \eta, \eta)|
  e^{-\frac{\eta}{2}} d\eta < \infty 
  \tag{8.48} \end{equation}

The last inequality follows from the estimate:
\begin{equation}
  |L(\xi, \eta)| \leq c e^{(2+ \varepsilon_1)[\eta \mu_1 (\xi + \eta)]^
  {\frac{1}{2} + \varepsilon_2}}
  \tag{8.49} \end{equation}
where $\varepsilon_1$ and $\varepsilon_2 > 0$ are arbitrarily small numbers,
\begin{equation}
  \mu_1(\xi) := \int^\xi_{-\infty}\, ds \mu (s),
  \qquad \mu (s) := \frac{e^s}{2}
  \left( 1 + \left| q \left( e^{\frac{s}{2}}\right) \right| \right)
\tag{8.50} \end{equation}

The proof of Theorem 8.1 is complete when (8.49) is proved.

\begin{lemma} 
Estimate (8.49) holds.
\end{lemma}

\begin{proof}
From (8.45) one gets
\begin{equation}
  m(\xi, \eta) \leq c + Wm 
  \tag{8.57} \end{equation}
where
\begin{equation}
  c_0:= \sup_{-\infty<\xi<\infty}|b(\xi)|\leq\frac{1}{2}\int^a_0 s(q(s)\,ds,
  \qquad m(\xi, \eta):= |L(\xi, \eta)|,
  \tag{8.52} \end{equation}
and
\begin{equation}
  Wm := \int^\xi_{-\infty} ds \int^\eta_0 dt\, \mu(s+t) m(s,t) .
  \tag{8.53} \end{equation}
Without loss of generality we can take $c_0 =1$ in (8.51): If (8.49) is
derived from (8.51) with $c_0=1$, it will hold for any $c_0>0$ (with a 
different $c$ in (8.49)). Thus, consider (8.51) with $c_0=1$ and solve this
inequality by iterations.

One has
\begin{equation}
  W1= \int^\xi_{-\infty} ds \int^\eta_0 \mu(s+t)\,dt = \int^\eta_0
  \mu_1 (\xi + t)\,dt \leq \eta \mu_1 (\xi + \eta) .
  \tag{8.54} \end{equation}

One can prove by induction that
\begin{equation}
  W^n1 \leq \frac{\eta^n}{h!} \frac{\mu_1^n (\xi + \eta)}{n!} .
  \tag{8.55} \end{equation}
Therefore (8.57) with $c_0 = 1$ implies
\begin{equation}
  m(\xi, \eta) \leq 1 + \sum^\infty_{h=1} \frac{\eta^n}{n!} 
  \frac{\mu^n_1(\xi + \eta)}{n!} .
  \tag{8.56} \end{equation}

Consider
\begin{equation}
  F(z) := 1+\sum^\infty_{n=1} \frac{z^n}{(n!)^2} .
  \tag{8.57} \end{equation}

This is an entire function of order $\frac{1}{2}$ and type 2.

Thus
\begin{equation}
  |F(z)| \leq c e^{(2+ \varepsilon_1)|z|^{\frac{1}{2} +\varepsilon_2} }.
  \tag{8.58} \end{equation}
From (8.56) and (8.58) estimate (8.49) follows.

Lemma 8.1 is proved. \end{proof}

Theorem 8.1 is proved. \end{proof}

\section {Discussion of the Newton-Sabatier procedure for recovery of
$q(r)$ from the fixed-energy phase shifts} 

In \cite{CS} and \cite{N} the following procedure is proposed for inversion of the
fixed-energy phase shifts for $q(r)$. We take $k=1$ in what follows.

\underbar{Step 1}. Given
$\{\delta_\l \}_{\forall \l = 0,1,2,\dots} $ one
solves an infinite linear algebraic system ((12.2.7) in \cite{CS})
\begin{equation}
  \tan \delta_\l = \sum^\infty_{\l^\prime = 0}
    M_{\l\l^\prime}
  (1 +\tan \delta_\l\tan \delta_{\l^\prime}) a_{\l^\prime} 
  \tag{9.1} \end{equation}
for constants $a_\l$. Here
\begin{equation}
  M_{\l\l^\prime} =
  \begin{cases} 0\ \  \hbox{if\ \ $|\l-\l^\prime|$
    is even or zero},\notag\\
  \frac{1}{\left(\l^\prime+\frac{1}{2}\right)^2
  - \left( \l+\frac{1}{2}\right)^2} \quad\hbox{if\quad
    $|\l-\l^\prime|$ is odd}.
  \end{cases}
  \tag{9.2} \end{equation}
{\it Assuming that (9.1) is solvable and $a_\l$ are found},
one calculates (see formula (12.2.8) in \cite{CS})
\begin{equation}
  c_\l = a_\l \left( 1 + \tan^2 \delta_\l \right)
  \left\{ 1-\frac{\pi a_\l \left(1+ \tan^2 \delta_\l \right)}{4\l + 2}
  - \sum^\infty_{\l^\prime=0} M_{\l \l^\prime}
  a_{\l^\prime}(\tan \delta_{\l^\prime}-\tan \delta_\l) \right\}^{-1} 
  \tag{9.3} \end{equation}

\underbar{Step 2}. 
If $c_\l$ are found, one solves the equation for $K(r,\rho)$
(see formula (12.1.12) in \cite{CS})
\begin{equation}
  K(r, s) = f(r,s)-\int^r_0 K(r,t) f(t,s) t^{-2}\, dt,
  \tag{9.4} \end{equation}
where
\begin{equation}
  f(r,s) := \sum^\infty_{\l=0} c_\l u_\l(r) u_\l (s) 
  \tag{9.5}, \end{equation}
and $u_\l (r)$ are defined in (7.22).

Note that in this section the notations from \cite{CS}
are used and by this reason the kernel $K(r,t)$
in formulas (9.4) and (9.6) differs by sign from the kernel
$K(r,\rho)$ in formula (8.8). This explains the minus sign
in formula (9.6).

{\it Assuming that (9.4) is solvable for all $r>0$}, one calculates
\begin{equation}
  q(r) = -\frac{2}{r}\ \frac{d}{dr}\quad \frac{K(r,r)}{r} .
  \tag{9.6} \end{equation}
Note that if (9.4) is not solvable for some $r=r > 0$,
then the procedure
breaks down because the potential (9.6) is no longer locally integrable in
$(0, \infty)$. In \cite{N} it is argued that for sufficiently small $a$
equation (9.4) is uniquely solvable by iterations for all $0<r<a$,
but no discussion of the
global solvability, that is, solvability for all $r>0$, is given. It is 
assumed in \cite{CS} and \cite{N}
that the sequence $\{c_\l\}$ in (9.3) does not
grow fast. In (\cite[(12.2.2)]{CS}) the following is assumed:
\begin{equation}
  \sum^\infty_{\l=1} |c_\l| \l^{-2} < \infty. 
  \tag{9.7} \end{equation}

Under this assumption, and also under much weaker assumption
\begin{equation}
  |c_\l| \leq ce^{b\l}
  \tag{$\hbox{9.7}^\prime$} \end{equation}
for some $b>0$ arbitrary large fixed,
one can prove that the kernel (9.5)
is an entire function of $r$ and $s$. This follows from the known asymptotics
of $u_\l (r)$ as $\l \to \infty$:
\begin{equation}
  u_\l (r) = \sqrt{\frac{r}{2}}
  \left( \frac{er}{2 \l +1} \right)^{\frac{2\l+1}{2}}
  \frac{1}{\sqrt{2\l+1}} [1 + o(1)],  \qquad \l \to +\infty.
  \tag{9.8} \end{equation}

Thus, equation (9.4) is a Fredholm-type equation
with kernel which is an entire
function of $r$ and $s$.
Since $K(r,0)=0$ and $f(r,s)=f(s,r) = 0$ at $s=0$,
equation (9.4) is a Fredholm equation in the space of continuous functions
$C(0,r)$ for any $r>0$.

If (9.4) is uniquely solvable for all $r>0$, then one can prove the 
following:

\begin{claim}
{\it $K(r,s)$ is an analytic function of $r$ and $s$ in a
neighborhood $\Delta$
 of the positive semiaxis $(0,\infty)$ on the complex plane
of the variables $r$ and $s$.}
\end{claim}

This claim is proved below, at the end of this section.

Therefore the potential (9.6) has to have the following:

{\it Property P: $q(r)$ is an analytic function in $\Delta$
with a possible simple pole at $r=0$.}

Most of the potentials do not have this property. Therefore, if one takes any
potential which does not have property $P$, for example, a compactly supported
potential $q(r)$, and if it will be possible to carry through the 
Newton-Sabatier procedure, that is,
{\it (9.1) will be solvable for $a_\l$}
and generate $c_\l$ by formula (9.3) such that (9.7) or
($\hbox{9.7}^\prime$) hold, and (9.4)
{\it will be uniquely solvable for all $r>0$ then the potential (9.6),
which this procedure yields, cannot coincide with the potential with which
we started.}

An important open question is: assuming that the Newton-Sabatier procedure
can be carried through, is it true that the reconstructed potential (9.6)
generates the scattering data, that is, the set of the fixed-energy phase
shifts $\{\delta_\l \}$ with which we started.

In \cite[pp.203-205]{CS} it is claimed that this is the case.  But the
arguments in  \cite{CS}  are not convincing. In particular, 
the author was not able
to verify equation (12.3.12) in  \cite{CS}  and in the argument on p.205
it is not clear why
$A^\prime_\l$ and $\delta^\prime_\l$ satisfy the same equation (12.2.5) as
the orginal $A_\l$ and $\delta_\l$.

In fact, it is claimed in \cite[(12.5.6)]{CS} that
$\delta^\prime_\l = O(\frac{1}{\l})$,
while it is known \cite{RAI} that if $q(r) = 0$ for $r>a$ and $q(r)$
does not change
sign in some interval $(a-\varepsilon, a)$, $q \neq 0$ if 
$r \in (a- \varepsilon, a)$, where $\varepsilon >0$
is arbitrarily small fixed number, then
\begin{equation}
  \lim_{\l \to \infty}
  \left( \frac{2\l}{e} \left|\delta_\l \right|^{\frac{1}{2\l}} \right)
   = a.
  \tag{9.9} \end{equation}

It follows from (9.9) that for the above potentials the phase shifts decay 
very fast as $\l \to \infty$, much faster that $\frac{1}{\l}$. Therefore 
$\delta^\prime_\l$ decaying at the rate $\frac{1}{\l}$ cannot be equal to
$\delta_\l$, as it is claimed in \cite[p.205]{CS}, because
formula (9.9) implies:
\begin{equation}
  |\delta_\l|\sim\left( \frac{ea[1+o(1)]}{2\l}\right)^{2\l}.
  \notag\end{equation}

It is claimed in \cite{CS}, p.105, that the Newton-Sabatier procedure
leads to "one (only one) potential which decreases faster
that $r^{-\frac 32}$" and yields the original phase shifts.
However, if one starts with a compactly
supported integrable potential (or any other rapidly
decaying potential which does not have property $P$
and belongs to the class of the potentials for which 
the uniqueness of the solution to inverse scattering
problem with fixed-energy data is established), then the 
Newton-Sabatier procedure will not lead to this potential as is
proved in this section. Therefore, either the potential
which the Newton-Sabatier procedure yields does not
produce the original phase shifts, or there are 
at least two potentials which produce the same
phase shifts.

A more detailed analysis of the Newton-Sabatier procedure is given 
by the author in \cite{ARS}.

\begin{proof}[{\it Proof of the claim.}] 
 The idea of the proof is to consider $r$ in (9.4) as
a parameter and to reduce (9.4) to a Fredholm-type equation with constant
integration limits and kernel depending on the parameter $r$. Let 
$t=r\tau$, $s = r \sigma$,
\begin{equation}
  \frac{K(r, r\tau)}{\tau} := b(\tau; r), \quad 
  \frac{f(r \tau, r \sigma)}{r \tau \sigma} := a(\sigma, \tau, r),
  \qquad \frac{f(r, \sigma)}{\sigma} := g(\sigma; r) 
  \tag{9.10} \end{equation}
Then (9.4) can be written as:
\begin{equation}
  b(\sigma;r) = g(r;\sigma) -
  \int^1_0 a(\sigma, \tau ;r) b(\tau; r)\, d\tau.
  \tag{9.11}\end{equation}
Equation (9.11) is equivalent to (9.4), it is a Fredholm-type
equation with kernel $a(\sigma,\tau;r)$ which is an entire
function of $\sigma$ and of the parameter $r$. The free term $g(r,\sigma)$
is an entire function of $r$ and $\sigma$.
This equation is uniquely solvable for all $r>0$ by the assumption.
Therefore its solution $b(\sigma;r)$ is an analytic function of $r$
in a neighborhood of any point $r>0$, and it is an entire function of
$\sigma$ \cite{R18}.
Thus $K(r,r)=b(1,r)$ is an analytic function of $r$ in a a neighborhood
of the positive semiaxis $(0,\infty)$.

\end{proof}


\section{Reduction of some inverse problems to an overdetermined Cauchy
problem} 

Consider, for example, the classical problem of finding $q(x)$ from
the knowledge of
two spectra. Let $u$ solve (1.1) on the interval [0,1] and satisfy the
boundary conditions
\begin{equation}
  u(0) = u(1) = 0,
  \tag{10.1}\end{equation}
and let the corresponding eigenvalues $k_n^2 := \lambda_n$,
$ n=1,2,\dots ,$ be given. If
\begin{equation}
  u(0) = u^\prime(1) + hu(1) =0, \quad
  \tag{10.3}\end{equation}
then the corresponding eigenvalues are $\mu_n, \quad n=1,2,\dots $.

The inverse problem (IP10) is:

{\it Given the two spectra
$\{\lambda_n \} \cup\{ \mu_n \}$, $n=1,2,3,\dots$, find $q(x)$ (and
$h$ in (10.3)).}

Let us reduce this problem to an overdetermined Cauchy problem. Let
\begin{equation}
  u(x,k) = \frac{\sin (kx)}{k} + \int^x_0 K(x,y) \frac{\sin (ky)}{k}\,dy:=
  (I + K) \left( \frac{\sin (kx)}{k} \right)
  \tag{10.4}\end{equation}
solve (1.1). Then (10.1) and (10.2) imply:
\begin{equation}
  0= \frac{\sin \sqrt{\lambda_n}}{\sqrt{\lambda_n}} + \int^1_0 K(1,y) 
  \frac{\sin (\sqrt{\lambda_n}y)}{\sqrt{\lambda_n}}\, dy,
  \quad n=1,2,\dots
  \tag{10.5}\end{equation}
and
\begin{equation}
  0= \cos \sqrt{\mu_n} + K(1,1) \frac{\sin \sqrt {\mu_n}}{\sqrt{\mu_n}} +
  \int^1_0 K_x (1,y) \frac{\sin (\sqrt {\mu_n} y)}{\sqrt{\mu_n}}\,dy,
  \quad  n=1,2,\dots
  \tag{10.6}\end{equation}
It is known \cite{M} that
\begin{equation}
  \lambda_n = (n \pi)^2 + c_0 + o(1), \quad n \to \infty,
  \tag{10.7} \end{equation}
and
\begin{equation}
  \mu_n = \pi^2 \left(n + \frac{1}{2} \right)^2 + c_1 + o(1),
  \quad n \to \infty,
  \tag{10.8}\end{equation}
where $c_0$ and $c_1$ can be calculated explicitly, they are proportional
to $\int^1_0 q(x)\, dx$.

Therefore,
\begin{equation}
  \sqrt{\lambda_n} = \pi n \left[1 + O \left(\frac{1}{n^2}\right)\right],
  \quad \sqrt{\mu_n} =
  \pi \left( n +\frac{1}{2} \right)
  \left[1 + O\left(\frac{1}{n^2}\right)\right], \quad n \to \infty.
  \tag{10.9}\end{equation}

It is known \cite{L} that if (10.9) holds then each of the systems of functions:
\begin{equation}
  \{\sin (\sqrt{\lambda_n }x)\}_{n=1,2,\dots } ,
  \quad \{\sin (\sqrt{\mu_n }x)\}_{n=1,2,\dots }
  \tag{10.10}  \end{equation}
is complete in $L^2[0,1]$.

Therefore equation (10.5) determines uniquely $\{K(1,y)\}_{0 \leq y \leq 1}$
and can be used for an efficient numerical procedure for finding $K(1,y)$
given the set $\{\lambda_n\}_{n=1,2,\dots }$
Note that the system
$\{\sin(\sqrt{\lambda_n}x)\}_{n=1,2,\dots}$
forms a Riesz basis of
$L^2[0,1]$ since the operator $I+K$, defined by (10.4) is boundedly invertible
and the system $\{u(x, \sqrt{\lambda_n})\}_{n=1,2,\dots }$
forms an orthornormal
basis of $L^2(0,1)$.

Equation (10.6) determines uniquely $\{K_x(1,y)\}_{0 \leq y \leq 1}$ if
$\{\mu_n\}_{n=1,2,\dots }$ are known. Indeed, the argument is the same as above.
The constant $K(1,1)$ is uniquely determined by the data 
$\{\mu_n\}_{n=1,2,\dots }$ because, by formula (5.41),
\begin{equation}
  K(1,1) = \frac{1}{2} \int^1_0 q(x)\, dx .
  \tag{10.11}\end{equation}

We have arrived at the following
{\it overdetermined Cauchy problem:

Given the Cauchy data
\begin{equation}
  \{K(1,y), K_x(1,y) \}_{0 \leq y \leq 1}
  \tag{10.12}\end{equation}
and the equations (5.40) - (5.41), find $q(x)$. }

It is easy to derive \cite[eq. (4.36)]{R5} for the unknown vector function
\begin{equation}
  U := \begin{pmatrix}q(x)\\K(x,y)\end{pmatrix}
  \tag{10.13}\end{equation}
the following equation
\begin{equation}
  U = W(U) + h,
  \tag{10.14}\end{equation}
where
\begin{equation}
  W(U) :=\begin{pmatrix}-2\int^1_x q(s) K(s,2x-s)\,ds\\
           \frac{1}{2} \int_{D_{xy}} q(s)K(s,t)\,ds\,dt\end{pmatrix},
  \tag{10.15}\end{equation}
$D_{xy}$ is the region bounded by the straight lines on the $(s,t)$ plane:
$s=1, \quad t-y=s-x$ and $t-y=-(s-x)$, and
\begin{equation}
  h:= \begin{pmatrix}f\\g\end{pmatrix},
  \tag{10.16}\end{equation}
\begin{equation}
  f(x) := 2[K_y(1, 2x-1) + K_x(1,2x-1)],
  \tag{10.17}\end{equation}
\begin{equation}
  g(x) := \frac{K(1,y+x -1) + K(1, y-x+1)}{2}\ - \frac{1}{2}
  \int^{y-x+1}_{y+x-1}K_s(1,t)\,dt.
  \tag{10.18} \end{equation}

Note that $f$ and $g$ are computable from data (10.12), and (10.14) is a 
nonlinear equation for $q(x)$ and $K(x,y)$.

Consider the iterative process:
\begin{equation}
  U_{n+1}= W(U_n) + h, \quad U_0 = h.
  \tag{10.19}\end{equation}

Assume that
\begin{equation}
  q(x) = 0\ \hbox{for}\ x>1,
  \quad q=\overline{q}, \quad q \in L^\infty [0,1].
  \tag{10.20}\end{equation}

Let $x_0 \in (0,1)$ and define the space of functions:
\begin{equation}
  L(x_0) := L^\infty (x_0, 1) \times L^\infty (\Delta_{x_0}),
  \tag{10.21}\end{equation}
where
\begin{equation}
  \Delta_{x_0} := \{ x,y : x_0 \leq x \leq 1, \quad 0 \leq |y| \leq x\}.
  \tag{10.22}\end{equation}
Denote
\begin{equation}
  \Vert u \Vert := \operatorname*{esssup}_{x_0\leq x\leq 1} |q(x)|
  + \sup_{x,y \in \Delta_{xo}} |K(x,y)|.
  \tag{10.23}\end{equation}
Let
\begin{equation}
  \Vert h \Vert \leq R.
  \tag{10.24}\end{equation}

\begin{theorem}[\cite{R5}]     
Let (10.17), (10.18), (10.20), and (10.24) hold, and choose any
$\tildeR >R$.
Then process (10.19) converges in $L(x_0)$ at the rate of a geometrical 
progression for any $x_0 \in (1-\mu, 1)$, where
\begin{equation}
  \mu := min \left( \frac{8 (\tildeR-R)}{5\tildeR^2},
  \frac{2}{5\tildeR} \right).
  \notag\end{equation}

One has
\begin{equation}
  \lim_{n \to \infty} U_n = \begin{pmatrix}q(x)\\K(x,y)\end{pmatrix},
  x,y \in \Delta_{x_0}.
  \tag{10.25}\end{equation}
If one starts with the data
\begin{equation}
  \{K(x_0, y), \quad K_x(x_0, y) \}_{0 \leq |y| \leq x_0},
  \tag{10.26}\end{equation}
replaces in (10.19) $h$ by
$h_0 := \left( \begin{smallmatrix}f_0\\g_0 \end{smallmatrix} \right)$,
where $f_0$ and $g_0$ are calculated by formulas (1.17)
and (1.18) in which the first argument in
$K(1,y), x=1$, is replaced by $x=x_0$,
then the iterative process (10.19) with the new $h=h_0$, in the new space
\begin{equation}
  L(x_1) := L^\infty (x_1, x_0) \times L^\infty (\Delta_{x_1}),
  \notag\end{equation}
with
\begin{equation}
  \Delta_{x_1} = \{x,y : x_1 \leq x \leq x_0,
  \qquad 0 \leq |y| \leq x\},
  \notag\end{equation}
converges to
$\left(\begin{smallmatrix}q(x)\\K(x,y)\end{smallmatrix}\right)$
in $L(x_1)$.

In finite number of steps one can uniquely reconstruct $q(x)$ on [0,1] from
the data (10.12) using (10.19).
\end{theorem}

\begin{proof}
First, we prove convergence of the process (10.19) in $L(x_0)$.

The proof makes it clear that this process will converge in $L(x_1)$ and
that in final number of steps one recovers $q(x)$ uniquely on [0,1].
Let $B(R) := \{U: \Vert U \Vert \leq \tildeR$, $U \in L(x_0)\}$,
$\tildeR>R$.

Let us start with

\begin{lemma}   
The map $U \in W(U) +h$ maps $B(\tildeR)$ into itself
and is a contraction on
$B(\tildeR)$ if $x_0 \in (1-\mu, 1)$,
$\mu := min \left( \frac{8 (\tildeR-R)}{5\tildeR^2} ,
\frac{2}{5\tildeR}\right)$.
\end{lemma}

\begin{proof}[{\it Proof of the lemma.}]
Let $U=\left( \begin{smallmatrix}q_1\\K_1\end{smallmatrix} \right)$,
$V= \left( \begin{smallmatrix}q_2\\K_2\end{smallmatrix}\right)$
One has:
\begin{equation}\begin{align}
 \Vert W (U) - W(V)\Vert & \leq \left\Vert
 \begin{array}{l}
   2\int^1_x \left( |q_1-q_2|\,|K_1|+|q_2|\,|K_1-K_2| \right)\,ds
     \\
   \frac{1}{2}\int_{D_{xy}} \left( |q_1-q_2|\,|K_1|+|q_2|\,|K_1-K_2|
      \right)\,ds\,dt
   \end{array} \right\Vert
   \notag \\
 & \leq \Vert U-V\Vert
 \left[2(1-x_0) \tildeR +\frac{\tildeR}{2} (1-x_0)^2 \right]
 \leq \Vert U-V\Vert (1-x_0) \frac{5}{2} \tildeR.
   \tag{10.27}\end{align}
 \end{equation}

Here we have used the estimate $(1-x_0)^2 < 1-x_0$ and the assumption
$\Vert U \Vert \leq \tildeR$, $\Vert V \Vert \leq \tildeR$.

If
\begin{equation}
  1-x_0 < \frac{2}{5\tildeR},
  \tag{10.28}\end{equation}
then $W$ is a contraction on $B(\tildeR)$.

Let us check that the map $T(U) = W(U)+h$ maps $B(\tildeR)$
into itself if $1-x_0<\mu$.

Using the inequality
$ab \leq \frac{\tildeR^2}{4}$ if $a+b=\tildeR$, $a,b \geq 0$,
one gets:
\begin{equation}
  \begin{align}
  &\Vert W(U) +h \Vert \leq \Vert W(U) \Vert + \Vert h \Vert \notag\\
   & \quad \leq 2(1-x_0)\frac{\tildeR^2}{4}+\frac{1}{2} (1-x_0)^2
    \frac{\tildeR^2}{4} +R  \notag\\
  &\leq \frac{5}{2}
  (1-x_0) \frac{\tildeR^2}{4} +R < \tildeR
   \quad \hbox{if}\quad 1-x_0 < \frac{8(\tildeR-R)}{5\tildeR^2}.
  \notag\end{align}
  \tag{10.29}\end{equation}

Thus if
\begin{equation}
 \mu = min \left( \frac{8(\tildeR-R)}{5\tildeR^2},
 \quad\frac{2}{5\tildeR} \right)
 \tag{10.30}\end{equation}
then the map $U \to W(U) + h$ is a contraction on $B(\tildeR)$ in the
space $L(x_0)$.
Lemma 10.1 is proved.
\end{proof}

From Lemma 10.1 it follows that process (10.19) converges at the rate of
geometrical progression with common ratio (10.30). The solution to
(10.14) is therefore unique in $L(x_0)$.

Since for the data $h$ which comes from a potential $q\in L^\infty (0,1)$
the vector
$\left(\begin{smallmatrix}q(x)\\K(x,y)\end{smallmatrix} \right)$
solves (10.14) in $L(x_0)$,
it follows that this vector satisfies (10.25). Thus, process (10.19)
allows one to reconstruct $q(x)$ on the interval from data (10.12),
$x_0 = 1-\mu$, where $\mu$ is defined in
(10.30).

If $q(x)$ and $K(x,y)$ are found on the interval $(x_0, 1)$, then $K(x_0,y)$
and $K_x(x_0,y)$ can be calculated for $0 \leq |y| \leq x_0$. Now one can 
repeat the argument for the interval $(x_1, x_0), \quad x_0-x_1 < \mu$, and
in finite number of the steps recover $q(x)$ on the whole interval [0,1].

Note that one can use a fixed $\mu$ if one chooses $R$ so that (10.24) holds
for $h$ defined by (10.16) and (10.17) with any $x\in L^\infty [0,1]$.
Such $R$ does exist if $q\in L^\infty[0,1]$.

Theorem 10.1 is proved.
\end{proof}

\begin{remark} 
Other inverse problems have been reduced to the overdetermined Cauchy problem 
studied in this section (see \cite{RS}, \cite{R5}, \cite{R}).
The idea of this reduction was used in \cite{RS}
for a numerical solution of some inverse problems.
\end{remark}

\section{Representation of I-function} 

 The $I(k)$ function (2.1) equals to the Weyl function (2.3). Our aim in 
this section is to derive the following formula  (\cite{R2}):
\begin{equation}
  I(k) =ik + \sum^J_{j=0} \frac{ir_j}{k-ik_j} + \widetilde a,
  \qquad \widetilde a:=\int^\infty_0 a(t)e^{ikt}\,dt,
  \tag{11.1}\end{equation}
where $k_0 := 0, \quad r_j = const>0$, $1 \leq j \leq J$,
$ r_0>0\ iff\ f(0) = 0$, $a(t) = \overline{a(t)}$ is a real-valued function,
\begin{equation}
  a(t) \in L^1(\R_+)\ \hbox{if}\ f(0) \neq 0
  \ \hbox{and}\ q(x) \in L_{1,1} (\R_+),
  \tag{11.2}\end{equation}
\begin{equation}
  a(t) \in L^1(\R_+)\ \hbox{if}\ f(0)=0\ \hbox{and}\ q\in L_{1,3}(\R_+).
  \tag{11.3}\end{equation}

We will discuss the inverse problem of finding $q(x)$ given 
$I(k)\ \forall k>0$.
Uniqueness of the solution to this problem is proved in
Theorem 2.1. Here we discuss a reconstruction algorithm and give examples.
Formula (11.1) appeared in \cite{R2}.

Using (2.17) and (2.1) one gets
\begin{equation}
  I(k) = \frac{ik-A(0) + \int^\infty_0 A_1(y)e^{iky}\,dy}
  {1+ \int^\infty_0 e^{iky} A(y)\, dy},
  \quad A(y) := A(0,y), \quad A_1(y) := A_x(0,y).
  \tag{11.4}\end{equation}
The function
\begin{equation}
  f(k) = 1 + \int^\infty_0 A(y)e^{iky}\,dy = f_0(k)
  \frac{k}{k+i} \prod^J_{j=1}
  \frac{k-ik_j}{k+ik_j},
  \tag{11.5}\end{equation}
where $f_0(k)$ is analytic in $\C_+, \quad f_0 (\infty) = 1$ in $\C_+$, that
is, 
\begin{equation}
  f_0(k) \to 1\ \hbox{as}\ |k| \to \infty, \quad\ \hbox{and}\ 
  f_0(k) \neq 0, \quad \forall k \in \overline{\C}_+ := \{k : Im k \geq 0 \}.
\tag{11.6}
\end{equation}

Let us prove

\begin{lemma} 
If $f(0) \neq 0$ and $q \in L_{1,1} (\R_+)$ then
\begin{equation}
  f_0(k) = 1 + \int^\infty_0 b_0 (t) e^{ikt}\,dt := 1+\widetilde b_0,
  \quad   b_0 \in W^{1,1} (\R_+).
  \tag{11.7}\end{equation}
\end{lemma}

Here $W^{1,1}(\R_+)$ is the Sobolev space of functions with the finite norm
\begin{equation}
  \Vert b_0 \Vert_{W^{1,1}} := \int^\infty_0 (|b_0(t)| +
  |b_0^\prime  (t)|)\, dt < \infty.
  \notag\end{equation}

\begin{proof}
It is sufficient to prove that, for any $1 \leq j \leq J$, the function
\begin{equation}
  \frac{k+ik_j}{k-ik_j} f(k) = 1+ \int^\infty_0 g_j (t) e^{ikt}\, dt, \quad
  g_j \in W^{1,1} (\R_+).
  \tag{11.8}\end{equation}
Since
$\frac{k+ik_j}{k-ik_j} = 1+ \frac{2ik_j}{k-ik_j}$, and since
$A(y) \in W^{1,1} (\R_+)$ provided that $q \in L_{1,1}(\R_+)$
(see (2.18),
(2.19)), it is sufficient to check that
\begin{equation}
  \frac{f(k)}{k-ik_j} = \int^\infty_0 g(t)e^{ikt}\,dt,
  \quad g \in W^{1,1} (\R_+).
  \tag{11.9}\end{equation}
One has $f(ik_j)=0$, thus
\begin{equation}
 \begin{align}
 \frac{f(k)}{k-ik_j}& = \frac{f(k) -f(ik)}{k-ik_j}
    = \int^\infty_0 dy A(y)             
 \frac{e^{i(k-ik_j)y}-1}{k-ik_j} e^{-k_jy}\,dy \notag \\
   & = \int^\infty_0 A(y)e^{-k_jy} i\int^y_0 e^{i(k-ik_j)s}\, ds
   = \int^\infty_0 e^{iks} h_j(s)\,ds   \tag{11.10} \end{align}
  \notag \end{equation}
where
\begin{equation}
  h_j(s) :=
  i \int^\infty_s A(y)e^{-k_j(y-s)}\,dy=i\int^\infty_0 A(t+s)e^{-k_jt}\,dt
  \tag{11.11}\end{equation}
From (11.11)one obtains (11.9) since $A(y) \in W^{1,1}(\R_+)$.

Lemma 11.1 is proved.
\end{proof}

\begin{lemma}
If $f(0)=0$ and $q\in L_{1,2} (\R_+)$, then (11.7) holds.
\end{lemma}

\begin{proof}
The proof goes as above with one difference : if $f(0)=0$ then $k_0 =0$ is 
present in formula (11.1) and in formulas (11.10) and (11.11) with $k_0 =0$
one has
\begin{equation}
  h_0 (s) = i \int^\infty_0 A(t+s)\, dt.
  \tag{11.12}\end{equation}

Thus, using (2.18), one gets
\begin{equation}
  \begin{align}
  &\int^\infty_0 | h_0(s)|\,ds \leq c\int^\infty_0\,ds \int^\infty_0\, dt
    \int^\infty_{\frac{t+s}{2}} |q(u)|\, du  \notag \\
  &\quad = 2c\int^\infty_0\, ds
    \int^\infty_{\frac{s}{2}}\, dv
    \int^\infty_v |q(u)|\, du
  \leq 2c \int^\infty_0\, ds
    \int^\infty_{\frac{s}{2}} | q(u)| u\,du      \tag{11.13}  \\
  &\quad = 4c\int^\infty_0 u^2 |q(u)|\, du < \infty
  \quad\ \hbox{if}\ \quad q \in L_{1,2}(\R_+), \notag\end{align} 
  \notag\end{equation}

where $c>0$ is a constant.
Similarly one checks that $h_0^\prime (s) \in L^1(\R_+)$ if 
$q \in L_{1,2}(\R_+)$.

Lemma 11.2 is proved.
\end{proof}

\begin{lemma}
Formula (11.1) holds.
\end{lemma}

\begin{proof}
Write
\begin{equation}
  \frac{1}{f(k)} = \frac{\frac{k+i}{k}
  \prod^J_{j=1}\frac{k+ik_j}{k-ik_j}} {f_0(k)}.
  \tag{11.14}\end{equation}
Clearly
\begin{equation}
  \frac{k+i}{k} \prod^J_{j=1} \frac{k+ik_j}{k-ik_j}
  = 1 + \sum^J_{j=0} \frac{c_j}{k-ik_j},\quad k_0 :=0, \quad k_j>0.
  \tag{11.15}\end{equation}

By the Wiener-Levy theorem \cite[\S 17]{GRS}, one has
\begin{equation}
  \frac{1}{f_0(k)} = 1 + \int^\infty_0 b(t)e^{ikt}\,dt,
  \quad b(t) \in W^{1,1} (\R_+).
  \tag{11.16}\end{equation}

Actually, the Wiener-Levy theorem yields $b(t) \in L^1(\R_+)$.

{\it However, since $b_0 \in W^{1,1}(\R_+)$,
one can prove that $b(t) \in W^{1,1} (\R_+)$.}

Indeed, $\widetilde{b}$ and $\widetilde b_0$ are related by the equation:
\begin{equation}
  (1 + \widetilde b_0)(1+ \widetilde{b}) = 1, \quad \forall k\in \R,
  \tag{11.17}\end{equation}
which implies
\begin{equation}
  \widetilde{b} = -\widetilde b_0 -\widetilde b_0 \widetilde{b},
  \tag{11.18}\end{equation}
or
\begin{equation}
  b(t) = -b_0(t) - \int^t_0 b_0(t-s) b(s)\, ds := -b_0 -b_0 \ast  b,
  \tag{11.19}\end{equation}
where $\ast$  is the convolution operation.

Since $b^\prime_0 \in L^1(\R_+)$ and $b\in L^1(\R_+)$ the convolution
$b_0^\prime  \ast  b \in L^1 (\R_+)$. So, differentiating (11.19) one sees that
$b^\prime  \in L^1 (\R_+)$, as claimed.

From (11.16), (11.15) and (11.4) one gets:
\begin{equation}
  I(k) = (ik-A(0) + \widetilde A_1) (1+ \widetilde{b})
  \left(1 + \sum^J_{j=0}\frac{c_j}{k-ik_j}\right)
  = ik+c + \sum^J_{j=0} \frac{a_j}{k-ik_j} + \widetilde{a},
  \tag{11.20}\end{equation}
where $c$ is a constant defined in (11.24) below, the constants
$a_j$ are defined in (11.25) and the function $\widetilde{a}$
is defined in (11.26).
We will prove that $c=0$ (see (11.28)).

To derive (11.20), we have used the formula:
\begin{equation}
  ik\widetilde{b} = ik
  \left[ \frac{e^{ikt}}{ik}b(t) \bigg|^\infty_0 - \frac{1}{ik}
  \int^\infty_0 e^{ikt} b^\prime(t) dt\right] =-b(0) -\widetilde b^\prime, 
  \tag{11.21}\end{equation}
and made the following transformations:
\begin{equation}
  \begin{align}
  & I(k) = ik-A(0)-b(0)-\widetilde b^\prime+\widetilde A_1-A(0)\widetilde{b}
    +\widetilde A_1 \widetilde{b}
   \sum^J_{j=0} \frac{c_j ik}{k-ik_j}
         \tag{11.22} \\
  &-\sum^J_{j=0} \frac{c_j[A(0) + b(0)]}{k-ik_j} + \sum^J_{j=0}
    \frac{\widetilde{g}(k) - \widetilde{g} (ik_j)}{k-ik_j} c_j
    + \sum^J_{j=0}
    \frac{\widetilde{g}(ik_j)c_j}{k-ik_j},
  \notag\end{align}\end{equation}
where
\begin{equation}
  \widetilde{g}(k) := -\widetilde b^\prime  + \widetilde A_1 - A(0)
  \widetilde{b} + \widetilde A_1\widetilde{b}.
  \tag{11.23}\end{equation}
Comparing (11.22) and (11.20) one concludes that
\begin{equation}
  c := -A(0) -b(0) +i\sum^J_{j=0} c_j,
  \tag{11.24}\end{equation}
\begin{equation}
  a_j := -c_j\left[k_j + A(0) + b(0) -\widetilde{g} (ik_j)\right],
  \tag{11.25}\end{equation}
\begin{equation}
  \widetilde{a}(k) := \widetilde{g}(k) + \sum^J_{j=0} 
  \frac{\widetilde{g} (k)-\widetilde{g} (ik_j)}{k-ik_j} c_j.
  \tag{11.26}\end{equation}

To complete the proof of Lemma 11.3 one has to prove that $c=0$, where $c$
is defined in (11.24). This is easily seen from the asymptotics of $I(k)$ as
$k \to \infty$. Namely, one has, as in (11.21):
\begin{equation}
  \widetilde{A}(k) = -\frac{A(0)}{ik} - \frac{1}{ik} \widetilde A^\prime 
  \tag{11.27}\end{equation}
From (11.27) and (11.4) it follows that
\begin{equation}
  \begin{align}
    I(k) &= (ik-A(0) + \widetilde A_1)
    \left[1-\frac{A(0)}{ik} + o\left(\frac{1}{k}\right)\right]^{-1}
    \notag\\
  & =(ik-A(0) + \widetilde A_1)
  \left(1+ \frac{A(0)}{ik} + o\left(\frac{1}{k}\right)\right)
    =ik+o(1), \quad k \to +\infty.
     \tag{11.28} \end{align}\end{equation}
From (11.28) and (11.20) it follows that $c=0$.

Lemma 11.3 is proved.
\end{proof}

\begin{lemma}
One has $a_j=ir_j$, $r_j>0$, $1 \leq j \leq J$, and $r_0=0 $ if
$f(0) \neq 0$, and $r_0>0$ if $f(0)=0$.
\end{lemma}

\begin{proof}
One has
\begin{equation}
  a_j=\operatorname*{Res}_{k=ik_j} I(k) = \frac{f^\prime (0,ik_j)}
  {\dotf(ik_j)}.
  \tag{11.29}\end{equation}
From (2.7) and (11.29) one gets:
\begin{equation}
  a_j= -\frac{c_j}{2ik_j} = i \frac{c_j}{2k_j} := ir_j ,
  \quad r_j :=\frac{c_j}{2k_j} >0, \quad j>0.
  \tag{11.30}\end{equation}
If $j=0$, then
\begin{equation}
  a_0= \operatorname*{Res}_{k=0} I(k)
  := \frac{f^\prime (0,0)}{\dotf(0)}.
  \tag{11.31}\end{equation}
Here by $\operatorname*{Res}_{k=0}I(k)$ we mean the right-hand side of
(11.31) since $I(k)$
is, in general, not analytic in a disc centered at $k=0$, it is analytic in 
$\C_+$ and, in general, cannot be continued analytically into $\C_-$.

Let us assume $q(x)\in L_{1,2}(\R_+)$.
In this case $f(k)$ is continuously differentiable in $\overline{\C}_+$.

From the Wronskian formula
\begin{equation}
  \frac{f^\prime (0,k)f(-k) - f^\prime (0-k)f(k)}{k} = 2i
  \tag{11.32}\end{equation}
taking $k \to 0$, one gets
\begin{equation}
  f^\prime (0,0) \dotf(0) = -i.
  \tag{11.33}\end{equation}
Therefore if $q\in L_{1,2}(\R_+)$ and $f(0)=0$, then
$\dotf(0)\not=0$ and $f^\prime(0,0)\not=0$.
One can prove \cite[pp.188-190]{M},
that if $q\in L_{1,1}(\R_+)$, then $\frac{k}{f(k)}$ is bounded as
$k\to 0$, $k\in\C_+$.

From (11.31) and (11.33) it follows that
\begin{equation}
  a_j= -\frac{i}{\left[ \dotf(0) \right]^2} = ir_0,
  \quad r_0 := -\frac{1}{[\dotf(0)]^2}.
  \tag{11.34}\end{equation}
From (2.17) one gets:
\begin{equation}
 \dotf(0) = i \int^\infty_0 A(y) \,y \,dy.
 \tag{11.35}\end{equation}

Since $A(y)$ is a real-valued function if $q(x)$ is real-valued
(this follows from the integral equation (5.62),
formula (11.35) shows that
\begin{equation}
  \left[\dotf(0) \right]^2 <0, \tag{11.36} \end{equation}
and (11.34) implies
\begin{equation}
  r_0 >0. \tag{11.37} \end{equation}

Lemma 11.4 is proved.
\end{proof}

One may be interested in the properties of function $a(t)$ in (11.1). These
can be obtained from (11.26), (11.16) and (11.7) as in the proof of Lemmas
11.1 and 11.2.

In particular (11.2) and (11.3) can be obtained.

Note that even
{\it if $q(x) \not\equiv 0$ is compactly supported, one cannot
claim that $a(t)$ is compactly supported}.

This can be proved as follows.

Assume for simplicity that $J=0$ and $f(0) \neq 0$. Then if $a(t)$ is 
compactly supported then $I(k)$ is an entire function of exponential type.
It is proved in \cite[p.278]{R} that if $q(x) \not\equiv 0$
is compactly supported,
$q \in L^1(\R_+)$, then $f(k)$ has infinitely many zeros in $\C$. The
function $f^\prime  (0,z) \neq 0$ if $f(z) = 0$.
Indeed, if $f(z) =0$ and $f^\prime (0,z) =0$
then $f(x,z) \equiv 0$ by the uniqueness of the solution of
the Cauchy problem for equation (1.1) with $k=z$.
Since $f(x,z) \not\equiv 0$
(see (1.3)), one has a contradiction,
which proves that $f^\prime (0,z) \neq 0$
if $f(z) =0$. Thus $I(k)$ cannot be an entire function if
$q(x) \not\equiv 0$, $q(x) \in L^1 (\R_+)$ and $q(x)$ is compactly supported.

Let us consider the following question:

{\it What are the potentials for which $a(t) =0$ in (11.1)?}

In other words, suppose
\begin{equation}
  I(k) = ik + \sum^J_{j=0} \frac{ir_j}{k-ik_j},
  \tag{11.38}\end{equation}
find $q(x)$ corresponding to $I$-function (11.38), and describe the decay 
properties of $q(x)$ as $x \to +\infty$.

We now show two ways of doing this.

By definition
\begin{equation}
  f^\prime (0,k) = I(k) f(k), \quad f^\prime (0,-k) = I(-k)f(-k),
  \quad k \in \R.
  \tag{11.39}\end{equation}
Using (11.39) and (2.23) one gets
\begin{equation}
  [I(k) -I(-k)] f(k)f(-k) = 2ik,
  \notag\end{equation}
or
\begin{equation}
  f(k)f(-k) = \frac{k}{Im I(k)}, \quad \forall k \in \R.
  \tag{11.40}\end{equation}

By (2.5), (2.6) and (11.30) one can write (see \cite{R2})
the spectral function
corresponding the  $I$-function (11.38) $(\sqrt{\lambda} = k)$:
\begin{equation}
 d\rho(\lambda)=\begin{cases}
 \frac{Im\,I(\lambda)}{\pi}\, d\lambda, & \lambda\geq 0,\\
 \sum^J_{j=1} 2k_jr_j \delta(\lambda+k^2_j)\,d\lambda, & \lambda<0, \end{cases}
  \tag{11.41}\end{equation}
where $\delta(\lambda)$ is the delta-function.

Knowing $d\rho (\lambda)$ one can recover $q(x)$ algorithmically by the
scheme (5.26).

Consider an example. Suppose $f(0) \neq 0, \quad J=1$,
\begin{equation}
  I(k) = ik + \frac{ir_1}{k-ik_1} = ik + \frac{ir_1(k+ik_1)}{k^2+k_1^2}=
  i \left(k+ \frac{r_1k}{k^2+k^2_1}\right) - \frac{r_1k_1}{k^2+k^2_1}.
  \tag{11.42}\end{equation}
Then (11.41) yields:
\begin{equation}
  d\rho (\lambda) = \begin{cases}
  \frac{d\lambda}{\pi} \left( \sqrt{\lambda}
    + \frac{r_1\sqrt{\lambda}}{\lambda+k^2_1} \right), & \lambda>0, \\
  2k_1r_1\delta(\lambda +k^2_1)\,d\lambda, & \lambda<0.
  \end{cases}\tag{11.43}\end{equation}
Thus (5.27) yields:
\begin{equation}
  L(x,y)=\frac{1}{\pi} \int^\infty_0 d\lambda \frac{r_1 \sqrt{\lambda}}
  {\lambda + k^2_1} \frac{\sin \sqrt{\lambda}x}{\sqrt{\lambda}}
  \frac{\sin \sqrt{\lambda} y}{\sqrt{\lambda}}
  + 2k_1r_1\frac{sh(k_1x)}{k_1}  \frac{sh(k_1y)}{k_1},
  \tag{11.44} \end{equation}
and, setting $\lambda = k^2$ and taking for simplicity $2k_1r_1=1$,
one finds:
\begin{equation}
  \begin{align}
  L_0(x,y)
  & :=   \frac{2r_1}{\pi} \int^\infty_0 \frac{dk k^2}{k^2 + k^2_1}
  \frac{\sin (kx) \sin (ky)}{k^2}
   \notag \\  
  &= \frac{2r_1}{\pi} \int^\infty_0
   \frac{dk \sin (kx) \sin (ky)}{k^2 + k^2_1}
\tag{11.45}\\
  & = \frac{r_1}{\pi} \int^\infty_0
    \frac{dk[\cos k (x-y) - \cos k(x+y)]}{k^2 +k^2_1}
   \notag \\
  & = \frac{r_1}{2k_1} \left(e^{-k_1|x-y|} -e^{-k_1(x+y)}\right),
  \quad k_1 >0,
  \notag \end{align}\end{equation}
where the known formula was used:
\begin{equation}
 \frac{1}{\pi}\int^\infty_0 \frac{\cos kx}{k^2+a^2}\,dk
 =\frac{1}{2a}\, e^{-a|x|},\qquad a>0, \qquad x\in \R.
 \tag{11.46} \end{equation}
Thus
\begin{equation}
  L(x,y)=\frac{r_1}{2k_1} \left[ e^{-k_1|x-y|} -e^{-k_1(x+y)} \right]
  + \frac{sh (k_1x)}{k_1}\ \frac{sh(k_1y)}{k_1}.
  \tag{11.47} \end{equation}

Equation (5.30) with kernel (11.47) is not an integral equation with
degenerate kernel:
\begin{equation}
  \begin{align}
  K(x,y) &+\int^x_0 K(x,t)
    \left[ \frac{ e^{-k_1|t-y|} -e^{-k_1(t+y)}}{2k_1/r_1}
    + \frac{sh(k_1t)}{k_1} \frac{sh(k_1y)}{k_1} \right]\, dt
  \tag{11.48}\\
  &= -\frac{e^{-k_1|x-y|} -e^{-k_1(x+y)} }{2k_1/r_1} -
  \frac{sh(k_1x)}{k_1} \frac{sh(k_1y)}{k_1}.
  \notag \end{align}
  \notag\end{equation}

This equation can be solved analytically \cite{R17}, but the solution 
requires space to present. Therefore we do not give the theory developed
in \cite{R17} but give another approach to a study of the properties 
of $q(x)$ given $I(k)$ of the form (11.42).
This approach is based on the theory of the Riemann problem \cite{G}.

Equations (11.40) and (11.42) imply
\begin{equation}
  f(k)f(-k)=\frac{k^2+k^2_1}{k^2+\nu^2_1},
  \qquad \nu^2_1:=k^2_1+r_1.
  \tag{11.49} \end{equation}
The function
\begin{equation}
  f_0(k):= f(k)\, \frac{k+ik_1}{k-ik_1} \not=0
  \quad\hbox{in}\quad \C_+.
  \tag{11.50} \end{equation}

Write (11.49) as
\begin{equation}
  f_0(k) \frac{k-ik_1}{k+ik_1} f_0(-k) \frac{k+ik_1}{k-ik_1}
  =\frac{k^2+k^2_1}{k^2+\nu^2_1}.
  \notag\end{equation}
Thus
\begin{equation}
  f_0(k)=\frac{k^2 +k^2_1}{k^2+\nu^2_1}\,h
  \qquad h(k):=\frac{1}{f_0(-k)}.
  \tag{11.51} \end{equation}
The function $f_0(-k)\not= 0$ in $\C_-$, $f_0(\infty)=1$ in $\C_-$,
so $h:=\frac{1}{f_0(-k)}$ is analytic in $\C_-$.

Consider (11.51) as a Riemann problem.
One has
\begin{equation}
  ind_{\R} \frac{k^2+k^2_1}{k^2+\nu^2_1}:= \frac{1}{2\pi i}
  \int^\infty_{-\infty} d\ln \frac{k^2+k^2_1}{k^2+\nu^2_1}=0.
  \tag{11.52}\end{equation}
Therefore (see \cite{G}) problem (11.51) is uniquely solvable.
Its solution is:
\begin{equation}
  f_0(k)=\frac{k+ik_1}{k+i\nu_1},\qquad
  h(k)=\frac{k-i\nu_1}{k-ik_1},
  \tag{11.53}\end{equation}
as one can check.

Thus, by (11.50),
\begin{equation}
  f(k)=\frac{k-ik_1}{k+i\nu_1}.
  \tag{11.54}\end{equation}

The corresponding $S$-matrix is:
\begin{equation}
  S(k)=\frac{f(-k)}{f(k)}=
  \frac{(k+ik_1)(k+i\nu_1)}{(k-ik_1)(k-i\nu_1)} 
  \tag{11.55}\end{equation}

Thus
\begin{equation}
  F_S(x):=\frac{1}{2\pi} \int^\infty_{-\infty} [1-S(k)]
   e^{ikx} dk=O\left(e^{-k_1x}\right) \quad\hbox{for}\quad x>0,
  \tag{11.56} \end{equation}
\begin{equation}
  F_d(x)=s_1\,e^{-k_1x},
  \notag\end{equation}
and
\begin{equation}
  F(x)=F_S(x)+F_d(x)=O\left(e^{-k_1x}\right).
  \tag{11.57} \end{equation}
Equation (5.50) implies  $A(x,x)=O\left(e^{-2k_1x}\right)$, so
\begin{equation}
  q(x)=O\left(e^{-2k_1x}\right),
  \qquad x\to +\infty.
  \tag{11.58} \end{equation}

Thus, if $f(0)\not=0$ and $a(t)=0$ then $q(x)$ decays exponentially
at the rate determined by the number $k_1$,
$k_1=\ds\operatorname*{min}_{1\leq j\leq J} k_j$.

If $f(0)=0$, $J=0$, and $a(t)=0$, then
\begin{equation}
  I(k)=ik+\frac{ir_0}{k},
  \tag{11.59} \end{equation}
\begin{equation}
  f(k)f(-k) =\frac{k^2}{k^2+r_0}, \qquad r_0>0.
  \tag{11.60} \end{equation}

Let $f_0(k)=\frac{(k+i)f(k)}{k}$. Then equation (11.60) implies:
\begin{equation}
  f_0(k)f_0(-k)= \frac{k^2+1}{k^2+\nu^2_0} ,
  \qquad \nu^2_0:=r_0,
  \tag{11.61} \end{equation}
and  $f_0(k)\not= 0\quad\hbox{in}\quad \C_+$.

Thus, since $ind_\R\frac{k^2+1}{k^2+\nu^2_0}=0$, $f_0(k)$ is uniquely
determined by the Riemann problem (11.61).

One has:
\begin{equation}
  f_0(k)=\frac{k+i}{k+i\nu_0},\qquad f_0(-k)=\frac{k-i}{k-i\nu_0},
  \notag \end{equation}
and
\begin{equation}
  f(k) =\frac{k}{k+i\nu_0},
  \quad S(k)=\frac{f(-k)}{f(k)}
       =\frac{k+i\nu_0}{k-i\nu_0},
 \tag{11.62}\end{equation}
\begin{equation}
  F_S(x)=\frac{1}{2\pi} \int^\infty_{-\infty}
  \left( 1-\frac{k+i\nu_0}{k-i\nu_0}\right)
  e^{ikx}dk=\frac{-2i\nu_0}{2\pi} \int^\infty_{-\infty}
  \frac{e^{ikx}dk}{k-i\nu_0}
  =2\nu_0 e^{-\nu_0x},
  \quad x>0,
  \notag \end{equation}
and $F_d(x)=0$.

So one gets:
\begin{equation}
  F(x)=F_S(x)=2\nu_0 e^{-\nu_0x},\qquad x>0.
  \tag{11.63} \end{equation}
Equation (5.50) yields:
\begin{equation}
  A(x,y)+2\nu_0 \int^\infty_x A(x,t) e^{-\nu_0(t+y)} dt
  = -2\nu_0 e^{-\nu_0(x+y)}, \qquad y\geq x\geq 0.
  \tag{11.64} \end{equation}
Solving (11.64) yields:
\begin{equation}
  A(x,y)=- 2\nu_0 e^{-\nu_0(x+y)} \frac{1}{1+ e^{-2\nu_0x}} .
  \tag{11.65} \end{equation}
The corresponding potential (5.51) is
\begin{equation}
  q(x)= O\left(e^{-2\nu_0x}\right),\qquad x\to\infty.
  \tag{11.66} \end{equation}
If $q(x)=O\left(e^{-kx}\right)$, $k>0$, then
$a(t)$ in (11.1) decays exponentially. Indeed, in this case
$b^\prime(t)$, $A_1(y)$, $b(t)$, $A_1\ast b$ decay expenentially,
so, by (11.23), $g(t)$ decays exponentially, and, by (11.26), the function
$\frac{\widetilde g(k)-\widetilde g(ik_j)}{k-ik_j}:=\widetilde h$
with $h(t)$ decaying exponentially. We leave the details to the reader.


\section{Algorithms for finding $q(x)$ from $I(k)$} 

One algorithm, discussed in section 11, 
is based on finding the spectral function $\rho (\lambda)$
from $I(k)$ by formula (11.41) and then finding $q(x)$ by the method
(5.26).

The second algorithm is based on finding the scattering data (2.10) and then
finding $q(x)$ by the method (5.49).

In both cases one has to find $k_j$, $1 \leq j \leq J$, and the number
$J$. In the second method one has to find $f(k)$ and $s_j$ also, and
$S(k) = \frac{f(-k)}{f(k)}$.

If $k_j$ and $f(k)$ are found then $s_j$ can be found from (2.12).
Indeed, by (11.1)
\begin{equation}
  ir_j := \operatorname*{Res}_{k=ik_j} I(k)
  = \frac{f^\prime(0, ik_j)}{\dotf(ik_j)}.
  \tag{12.1}\end{equation}
From (12.1) and (2.12) one finds
\begin{equation}
  s_j = -\frac{2ik_j}{ir_j [\dotf (ik_j)]^2}
  =-\frac{2k_j}{r_j[\dotf (ik_j)]^2}.
  \tag{12.2}\end{equation}

If $k_j$ are found, then one can find $f(k)$ from $I(k)$ as follows. Since
$f^\prime(0,k) = f(k) I(k)$, equation (2.23)
implies equation (11.40):
\begin{equation}
  f(k) f(-k) = \frac{k}{Im I(k)}.
  \tag{12.3}\end{equation}
Define
\begin{equation}
  w(k) := \prod^J_{j=1} \frac{k-ik_j}{k+ik_j}\quad \hbox{if}\quad
  I(0) < \infty,
  \tag{12.4}\end{equation}
and
\begin{equation}
  w(k) := \frac{k}{k+i} \prod^J_{j=1} \frac{k-ik_j}{k+ik_j}
  \quad \hbox{if}\quad I(0) =\infty.
  \tag{12.5}\end{equation}
One has $I(0) < \infty$ if $f(0) \neq 0$ and $I(0) = \infty$ if $f(0) =0$.
Note that if $q \in L_{1,2} (\R_+)$ and
$f(0) =0$ then $f^\prime(0,0) \neq 0$ and
$\dotf (0) \neq 0$.

Define
\begin{equation}
  h(k) := \frac{f(k)}{w(k)}.
  \tag{12.6}\end{equation}

Then $h(k)$ is analytic in $\C_+$, $h(k) \neq 0$ in $ \overline{\C}_+$, and
$h(\infty) = 1$ in $\overline{\C}_+$, while $h(-k)$ has similar properties
in $\C_-$. Denote $\frac{1}{h(-k)} := h_- (k)$. This function is analytic in
$\C_-, \quad h_- (k) \neq 0$ in $\overline{\C}_-$ and $h_-(\infty) = 1$ in
$\overline{C}_-$. Denote $h(k) := h_+(k)$.

Write (12.3) as the Riemann problem:
\begin{equation}
  h_+ (k) = g(k) h_-(k),
  \tag{12.7}\end{equation}
where
\begin{equation}
  g(k) = \frac{k}{Im I(k)}\quad \hbox{if}\quad I(0) < \infty,
  \tag{12.8}\end{equation}
and
\begin{equation}
  g(k) = \frac{k}{Im\,I(k)} \frac{k^2 + 1}{k^2} \quad\hbox{if}\quad
  I(0) = \infty.
  \tag{12.9}\end{equation}
We claim that the function $g(k)$ is positive
for all  $k>0$, bounded in a 
neighborhood of $k=0$
and has a finite limit at $k=0$ even if $I(0)=0$. 
Only the case $I(0)=0$ requires a comment. 
If $I(0)=0$, then $f'(0,0)=0$, 
$\dot f'(0,0)\neq 0$,  $f(0)\neq 0$, and one can see
from (12.3) that
the function $\frac{k}{Im\,I(k)} $ is bounded.
Thus, the claim is verified.  

The Riemann problem (12.7) can be solved analytically:
$\ln h_+ (k) - \ln h_-(k) = \ln g(k)$ and 
since $h_+(k)$ and $h_-(k)$ do not
vanish in $\overline{\C}_+$ and $\overline{\C}_-$ respectively, 
$\ln h_+(k)$
and $\ln h_-(k)$ are 
analytic in $\C_+$ and $\C_-$ respectively. Therefore
\begin{equation}
  h(k) = exp \left( \frac{1}{2 \pi i}
  \int^\infty_{-\infty} \frac{\ln g(t)}{t-k} \,dt\right),
  \tag{12.10}\end{equation}
\begin{equation}
  h(k) = h_+(k)\quad \hbox{if}\quad Im\, k>0,
  \quad h(k) = h_-(k) \quad\hbox{if}\quad Im \,k<0,
  \tag{12.11}\end{equation}
and
\begin{equation}
  f(k) = w(k)h(k), \quad Im \,k \geq 0.
  \tag{12.12}\end{equation}

Finally, let us explain how to find $k_j$ and $J$ given $I(k)$.

From (11.1) it follows that
\begin{equation}
  \frac{1}{2\pi} \int^\infty_{-\infty} (I(k) -ik)e^{-ikt}\, dk
  = -\sum^J_{j=1}
  r_je^{k_jt} -\frac{r_0}{2}\quad \hbox{for}\quad t<0.
  \tag{12.13}\end{equation}

Taking $t \to -\infty$ in (12.13) one can find step by step the numbers
$r_0$, $k_1$, $r_1$, $ k_2$, $r_2\dots$, $ r_J$, $k_J$. If $I(0) < \infty$,
then $r_0 =0$.

\section{Remarks.}

\subsection{Representation of the products of the solution to (1.1)}
In this subsection we follow \cite{Lt2}.
Consider equation (1.1) with $q=q_j$, $j=1,2$. The function
$u(x,y) := \varphi_1 (x,k) \varphi_2 (y,k)$ where $\varphi_j$,
$j=1,2$, satisfy the
first two conditions (1.4), solves the problem
\begin{equation}
  \left[\frac{\partial^2}{\partial x^2} - q_1(x)\right]u (x,y) =
  \left[\frac{\partial^2}{\partial y^2} - q_2 (y)\right] u(x,y),
  \tag{13.1}\end{equation}
\begin{equation}
  u(0,y) =0, \quad u_x(0,y) = \varphi_2 (y,k),
  \tag{13.2}\end{equation}
\begin{equation}
  u(x,0) =0, \quad u_y (x,0) = \varphi_1(x,k).
  \tag{13.3}\end{equation}

Let us write (13.1) as
\begin{equation}
  \left( \frac{\partial^2}{\partial x^2}
  - \frac{\partial^2}{\partial y^2}
  \right) u(x,y) = [q_1(x) - q_2(y)] u(x,y)
  \tag{13.4}\end{equation}
and use the known D'Alembert's formula to solve (13.3)-(13.4):
\begin{equation}
  u(x,y) = \frac{1}{2} \int_{D_{xy}} [q_1(s) - q_2(t)]
  u(s,t)\, ds\,dt
  + \frac{1}{2}\int^{x+y}_{x-y} \varphi_1(s)\, ds,
  \tag{13.5}\end{equation}
where $D_{xy}$ is the triangle $0<t<y$, $ x-y+t < s < x+y-t$.

Function (13.5)  satifies (13.3) and (13.1). Equation (13.5)
is uniquely solvable
by iterations:
\begin{equation}
  u(x,y) = \sum^\infty_{m=0} u_m (x,y),
  \quad u_0 (x,y) := \frac{1}{2}
  \int^{x+y}_{x-y} \varphi_1 (s,x)\,ds,
  \tag{13.6}\end{equation}
\begin{equation}
  u_{m+1} (x,y) = \frac{1}{2}
  \int_{D_{xy}}[q_1(s)-q_2(t)] u_m(s,t)\,ds\,dt.
  \tag{13.7}\end{equation}
Note that
\begin{equation}
  u_m(x,y) = \frac{1}{2} \int^{x+y}_{x-y} w_m(x,y,s) \varphi_1(s)\,ds.
  \tag{13.8}\end{equation}

If $m=0$ this is clear from (13.6). If it is true for some $m>0$,
then it is true for $m+1$:
\begin{equation}
  \begin{align}
    u_{m+1} (x,y) &= \frac{1}{2}
    \int^y_0 dt \int^{x+y-t}_{x-y+t}\, ds
    [q_1(s)-q_2 (t)] \frac{1}{2}
    \int^{s+t}_{s-t} w_m(s,t, \sigma) \varphi_1(\sigma)\,d\sigma
  \notag \\
  &= \frac{1}{2} \int^y_0 \,dt
    \int^{x+y}_{x-y} d\sigma\varphi_1 (\sigma)
       \widetilde{w}_m (x,y,t,\sigma)
  \tag{13.9}\\
   & = \frac{1}{2} \int^{x+y}_{x-y}\,d\sigma \varphi_1 (\sigma)
    w_{m+1} (x,y, \sigma),
  \notag \end{align}
  \notag\end{equation}

where $\widetilde{w}_m$ and $w_{m+1}$ are some functions.

Thus, by induction, one gets (13.8) for all $m$, and (13.6) implies
\begin{equation}
  u(x,y) = \frac{1}{2}
  \int^{x+y}_{x-y} w(x,y,s) \varphi_1 (s)\, ds,
  \tag{13.10}\end{equation}
where
\begin{equation}
  w(x,y,s) := \sum^\infty_{m=0} w_m (x,y,s).
  \tag{13.11}\end{equation}
To satisfy (13.2) one has to satisfy the equations:
\begin{equation}
  \begin{align}
  0&= \int^y_{-y} w (0,y,s) \varphi_1 (s,k) \,ds,
    \notag\\
  \varphi _2 (y,k)&  = \frac{1}{2} [w(0,y,y) \varphi_1(y) - w(0,y,-y)
  \varphi_1 (-y)]
  \tag{13.12}\\
  &\quad + \frac{1}{2} \int^y_{-y} w_x (0,y,s) \varphi_1 (s) \,ds.
  \notag\end{align}
  \notag\end{equation}
Formula (13.10) yields
\begin{equation}
  \varphi_1 (x,k) \varphi_2 (y,k) = \frac{1}{2} \int^{x+y}_{x-y} w(x,y,s)
  \varphi_1 (s,k) \,ds.
  \tag{13.13}\end{equation}
If $x=y$, then
\begin{equation}
  \varphi_1(x,k) \varphi_2 (x,k) = \frac{1}{2}
  \int^{2x}_0 w(x,x,s) \varphi_1(s,k)\,ds.
  \tag{13.14}\end{equation}
Therefore, if
\begin{equation}
  \int^a_0 h(x) \varphi_1 (x,k) \varphi_2 (x,k)\,dx = 0
  \quad \forall k>0,
  \notag\end{equation}
then
\begin{equation}
  \begin{align}
  0&= \int^a_0 h(x) \int^{2x}_0 w(x,x,s) \varphi_1 (s,k)\,ds\,dx
   \notag\\
  &= \int^{2a}_0\, ds
  \varphi_1(s,k) \int^a_{\frac{s}{2}}\, dx h(x) w(x,x,s)
  \quad \forall k>0.
  \notag\end{align}\end{equation}
Since the set $\{\varphi_1(s,k)\}_{\forall k>0}$ is
complete in $L^2(0,2a)$, it follows that
$0= \int^a_{\frac{s}{2}}\, dx h(x) w(x,x,s)$
for all $s\in[0,2a]$.
Differentiate with respect to $s$ and get
\begin{equation}
  w\left(\frac{s}{2},\frac{s}{2},s\right)
  \frac{1}{2}h \left( \frac{s}{2} \right)
  - \int^a_{\frac{s}{2}}\, ds h(x) w_s (x,x,s)=0.
  \tag{13.15}\end{equation}

From Volterra equation (13.15) it follows $h(x)=0$ if the kernel
$w_s(x,x,s) w^{-1}\left( \frac{s}{2},\frac{s}{2},s\right) := t(x,s)$
is summable. From the definition (13.11) of $w$ it
follows that if $\int^b_0 |q(x)|\, dx < \infty$  $\forall b>0$, then
$w_s(x,y,s)$ is summable.
The function $w(x,y,s)$ has $m$ summable derivatives
with respect to $x,y$ and $s$ if $q(x)$ has $m-1$ summable derivatives.
Thus one can derive from (13.15) that $h(x)=0$ if
$w\left( \frac{s}{2},\frac{s}{2},s \right)>0$ for all $s\in[0,2a]$.

If the boundary conditions at $x=0$ are different,
for example,
$\varphi_j^\prime (0,k) - h_0 \varphi_j (0,k) =0$, $j=1,2,$
then conditions
\begin{equation}
  u_x - h_0 u \big|_{x=0} =0, \quad u_y - h_0 u \big|_{y=0} =0, \quad
  h_0 = const>0
  \tag{13.16}\end{equation}
replace the first conditions (13.2) and (13.3).
One can normalize $\varphi_j(x,k)$ by setting
\begin{equation}
  \varphi_j (0,k) =1.
  \tag{13.17}\end{equation}
Then
\begin{equation}
  \varphi^\prime_j (0,k) = h_0,
  \tag{13.18}\end{equation}
\begin{equation}
  u(0,y) = \varphi_2(y,k), \quad u(x,0) = \varphi_1(x,k),
  \tag{13.19}\end{equation}
\begin{equation}
  u_x (0,k) = h_0 \varphi_2 (y,k), \quad u_y (x,0) = h_0 \varphi_1 (x,k),
  \tag{13.20}\end{equation}
and (13.5) is replaced by
\begin{equation}
  \begin{align}
  u(x,y) &= \frac{1}{2} \int_{D_{xy}} [q_1 (s)-q_2 (t)] u(s,t)\, ds\,dt
  \notag\\
  &+ \frac{1}{2}
  \int^{x+y}_{x-y} h_0 \varphi_1(s,k) ds + \frac{1}{2} [\varphi_1 (x+y,k) +
  \varphi_1(x-y,k)].
  \tag{13.21}\end{align}
  \notag\end{equation}
Note that
\begin{equation}
  \frac{1}{2} [\varphi_1(x+y)+\varphi_1(x-y)]
  = \frac{1}{2} \frac{\partial}{\partial y} \int^{x+y}_{x-y}\varphi_1 (s)\,ds
  = \frac{1}{2} \int^{x+y}_{x-y} \varphi_s (s)\, ds.
  \tag{13.22}\end{equation}
Equation (13.21) is uniquely solvable by iterations, as above, and its
solution is given by the first formula (13.6) with
\begin{equation}
  u_0 (x,y) = \frac{h_0}{2} \int^{x+y}_{x-y} \varphi_1 (s,k)\,ds
  + \frac{1}{2}\int^{x+y}_{x-y} \varphi_{1s}(s,k)\,ds.
  \tag{13.23}\end{equation}

The rest of the argument is as above: one proves existence and uniqueness of
the solution to equation (13.21) and the analog of formula (13.10):
\begin{equation}
  u(x,y) = \frac{1}{2} \int^{x+y}_{x-y} w(x,y,s) \Phi (s) \,ds,
  \quad\Phi := h_0 \varphi_1 (s,k) + \varphi_{1s} (s,k).
  \tag{13.24}\end{equation}

Thus
\begin{equation}
  u(x,x) = \varphi_1(x,k) \varphi_2 (x,k)
  = \int^{2x}_0  t(x,s) [h_0 \varphi_1 (s,k) + \varphi _{1s} (s,k)]\,ds,
  \tag{13.25}\end{equation}
where
\begin{equation}
  t(x,s) := \frac{1}{2} w(x,x,s),
  \tag{13.26}\end{equation}
and $t(x,s)$ is summable.

Thus, as before, completeness of the set of products
$\{\varphi_1 (x,k) \varphi_2 (x,k)\}_{\forall k>0}$
can be studied.

\subsection{Characterization of Weyl's solutions}

The standard definition of Weyl's solution to (1.1) is given by (2.2).

In \cite{M1} it is proved that
\begin{equation}
  W(x,k) = e^{ikx} (1+ o(1))\quad \hbox{as}\quad |k| \to \infty,
  \quad |x| \leq b, \quad k^2 \in \Delta,
  \tag{13.27}\end{equation}
where $\Delta :=\{\lambda :|Im\,\lambda|>\varepsilon$,
$dist(\lambda, S)>\varepsilon\}$,
\begin{equation}
  S := \R \cup [i\gamma_-, i\gamma_+],
  \quad \gamma_{\pm} :=
  \operatorname*{inf}_{\substack{u\in H^1(\R_\pm) \\u(0)=0} }
  \pm\int^{\pm\infty}_0
  [u^{\prime 2} + q|u|^2]\,dx.
  \tag{13.28}\end{equation}

The relation (13.27) gives a definition of the Weyl solution
by its behavior on
compact sets in the $x$-space as $|k| \to \infty$, as opposite
to (2.2), where
$k$ is fixed, and $x \to \infty$. For multidimensional Schr\"odinger equation
similar definition was proposed in \cite[p.356, problem 8]{R}.

We want to derive (13.27) for potentials in $L_{1,1}(\R_+)$ and
for $k>0, \, k\to +\infty$.

The idea is simple. For any $q=\overline{q} \in L^1_{loc} (\R_+)$, one can
construct $\varphi (x,k)$ and $\psi (x,k)$, the solutions to (1.1) and (1.4),
for any $|x| \leq b$, where $b>0$ is an arbitrary large fixed number, by
solving the Volterra equations
\begin{equation}
  \varphi (x,k) = \frac{\sin (kx)}{k}
  + \int^x_0 \frac{\sin[k(x-y)]}{k} q(y) \varphi (y,k) \,dy
  \tag{13.29}\end{equation}
\begin{equation}
  \psi (x,k) = \cos(kx) + \int^x_0 \frac{\sin [k(x-y)]}{k}
  q(y)\psi(y,k)\,dy.
  \tag{13.30}\end{equation}

One can also write an equation for the Weyl solution $W$:
\begin{equation}
  W (x,k) = \cos(kx) + m(k) \frac{\sin(kx)}{k}
  + \int^x_0 \frac{\sin[k(x-y)]}{k}
  q(y) W(y,k)\,dy.
  \tag{13.31}\end{equation}
This equation is uniquely solvable by iterations for $|x| \leq b$.

It is known that
\begin{equation}
  m(k) = ik + o(1), \quad |k| \to \infty,
  \quad Imk >\varepsilon|Rek|, \quad \varepsilon>0.
  \tag{13.32}\end{equation}
For $q\in L_{1,1}(\R_+)$ the above formula holds when $k>0, k\to +\infty$.
From (13.31) and (13.32) one gets, assuming $k>0$,
\begin{equation}
  W(x,k) = e^{ikx} \left(1 + O\left(\frac{1}{k}\right)\right)
  + \int^x_0 \frac{\sin[k(x-y)]}{k} q(y) W(y,k) \,dy.
  \tag{13.33}\end{equation}
Solving (13.33) by iterations yields (13.27) for $k>0, k\to +\infty$.
For $q\in L_{1,1}(\R_+)$ the Weyl solution is the Jost solution.
Therefore the above result for $k>0, k\to +\infty$ is just the standard
asymptotics for the Jost solution. It would be of interest
to generalize the above approach to the case of complex $k$
in the region (13.27).
\qed

One can look for an asymptotic representation of the 
solution to (1.1) for large
$|k|$, $Imk >\varepsilon|Re\,k|$, $\varepsilon>0$,
of the following form:
\begin{equation}
  u(x,k) = e^{ikx+\int^x_0 \sigma (t,k)\,dt},
  \tag{13.34}\end{equation}
where
\begin{equation}
  \sigma^\prime + 2ik \sigma + \sigma^2 - q(x) = 0,
  \quad \sigma  = \frac{q(x)}{2ik}
  + o\left(\frac{1}{k}\right), \quad |k| \to \infty.
  \tag{13.35}\end{equation}

From (13.34) one finds, assuming $q(x)$ continuous at $x=0$,
\begin{equation}
  \frac{u^\prime(0,k)}{u(0,k)} = ik + \frac{q(0)}{2ik} 
  + o\left(\frac{1}{k}\right),
  \qquad |k|\to\infty,\qquad k\in \C_+.
  \tag{13.36}\end{equation}
If $q(x)$ has $n$ derivatives, more terms of the asymptotics can be written
(see \cite[p.55]{M}).

\subsection{ Representation of the Weyl function
via the Green function} 

The Green function of the Dirichlet operator
$L_q=-\frac{d^2}{dx^2} +q(x)$ in $L^2(\R_+)$ can be written as:
\begin{equation}
  G(x,y,z)= \varphi(y,\sqrt{z}) W(x,\sqrt{z}),\quad x\geq y
  \tag{13.37}\end{equation}
where $\varphi(x,k)$, $k:=\sqrt{z}$, solves (1.1) and satisfies the first
two conditions (1.4), and $W(x,\sqrt{z})$ is the Weyl solution (2.2),
which satisfies the conditions:
\begin{equation}
  W(0,\sqrt{z})=1,\qquad W^\prime(0,\sqrt{z})=m(\sqrt{z}).
  \tag{13.38}\end{equation}

From (1.4), (13.37) and (13.38) it follows that:
\begin{equation}
  \frac{\partial^2 G(x,y,k)}{\partial x \partial y}
  \Big|_{x=y=0} =m(k).
  \tag{13.39}\end{equation}

If $q\in L_{1,1}(\R_+)$ then $W(x,k)=\frac{f(x,k)}{f(k)}$,
where $f(x,k)$ is the Jost solution (1.3), $k\in\C_+$.
Note that (13.17) and (13.38) imply:
\begin{equation}
  \frac{1}{\pi} Im\, G(x,y,\lambda+i0)
  =\frac{\varphi (x,\sqrt{\lambda})\varphi(y,\sqrt{\lambda})}{\pi}
  Im\,m(\sqrt{\lambda+i0})
  \tag{13.40}\end{equation}
and
\begin{equation}
  G(x,y,z)=\int^\infty_{-\infty} \frac{\theta(x,y,t)}{t-z}\,d\rho(t),
  \qquad \theta(x,y,t)=\varphi(x,\sqrt{t}) \varphi(y,\sqrt{t}).
  \tag{13.41}\end{equation}
Thus
\begin{equation}
  \frac{1}{\pi} Im\,G(x,y,t+i0)dt=\theta(x,y,t) d\rho(t).
  \tag{13.42}\end{equation}
From (13.42) and (13.40) one gets, assuming $\rho(-\infty)=0$,
\begin{equation}
  \rho(t)=\frac{1}{\pi} \int^t_{-\infty} Im\,m(\sqrt{\lambda+i0})\,d\lambda.
  \tag{13.43}\end{equation}

If $\lambda<0$ then $Im\,m\left( \sqrt{\lambda+i0} \right)=0$
except at the points $\lambda=-k^2_j$ at which $f(ik_j)=0$,
so that $m( \sqrt{-k^2_j+i0})=\infty$.
Thus, if $t$ and $a$ are continuity points of $\rho(t)$, then
\begin{equation}
  \rho(t)-\rho(a)=\frac{1}{\pi} \int^t_a Im\,m(\sqrt{\lambda +i0})
  \,d\lambda, \qquad a\geq 0.
  \tag{13.44}\end{equation}
Let us recall the Stieltjes inversion formula:

If $z=\sigma+ i\tau$, $\tau>0$, $\rho(t)$ is a function of bounded
variation on $\R$,
\begin{equation}
  \varphi(z):=\int^\infty_{-\infty} \frac{d\rho(t)}{t-z},
  \tag{13.45}\end{equation}
and if $a$ and $b$ are continuity points of $\rho(t)$, then
\begin{equation}
  \frac{1}{\pi} \int^b_a Im\,\varphi(\lambda+i0)\,d\lambda
  =\rho(b)-\rho(a).
  \tag{13.46}\end{equation}
Therefore (13.44) implies
\begin{equation}
  m(\sqrt{z})=\int^\infty_{-\infty} \frac{d\rho(t)}{t-z}.
  \tag{13.47}\end{equation}

The spectral function $d\rho(t)$ does not have 
a bounded variation globally, on the whole real axis,
and integral (13.47) diverges in the classical sense.
We want to reduce it to a convergent integral by
subtracting the classically divergent part of it.

If $q(x)=0$, then $\rho:=\rho_0(t)$ for $t<0$,
$m(\sqrt{\lambda})=i\sqrt{\lambda}$,
and formula (13.44) with $a=0$ yields
\begin{equation}
  \rho_0(\lambda)=\frac{2\lambda^{\frac{3}{2}}}{3\pi}.
  \tag{13.48}\end{equation}
If $q(x)=0$ then $G(x,y,\lambda)=\frac{\sin (\sqrt{\lambda}y)
}{\sqrt{\lambda}}
\,e^{i\sqrt{\lambda}x},\qquad y\leq x$,
so (13.39) yields $m(\sqrt{\lambda})=i\sqrt{\lambda}$.
Formula (13.47) yields formally
\begin{equation}
  i\sqrt{\lambda}=\frac{1}{\pi} \int^\infty_0 \frac{\sqrt{t}\,dt}{t-\lambda}. 
  \tag{13.49}\end{equation}
This integral diverges from the classical point of view.
Let us interpret (13.49) as follows.
Let $Im\,\lambda>0$.
Differentiate (13.49) formally and get
\begin{equation}
   \frac{i}{2\sqrt{\lambda}} =\frac{1}{\pi} \int^\infty_0
   \frac{\sqrt{t}\,dt}{(t-\lambda)^2},\qquad Im\,\lambda>0.
  \tag{13.50}\end{equation}
This is an identity, so (13.49) can be interpreted as an integral
from $0$ to $\lambda$ of (13.50).
The integral $\int^\infty_0 t^{-\frac{1}{2}} dt$ which one obtains
in the process of integration, is interpreted as zero, as an integral
of a hyperfunction or Hadamard finite part integral.

Subtract from (13.47) the divergent part (13.49) and get:
\begin{equation}
 m(\sqrt{z})-i\sqrt{z}=\int^\infty_{-\infty} \frac{d\sigma(t)}{t-z},
  \tag{13.47'}\end{equation}
where
\begin{equation}
  d\sigma(\lambda)=d\rho(\lambda)-d\rho_0(\lambda),
  \qquad d\rho_0(\lambda):=
  \begin{cases}
    \frac{\sqrt{\lambda} d\lambda}{\pi}, & \lambda\geq 0,\\
    0,& \lambda<0.\end{cases}
  \tag{13.47"}\end{equation}
Integral (13.47') converges in the classical sense
if $q\in L_{1,1}(\R_+)$. Indeed, by (2.5) and (13.47")
one has $d\sigma (t)=\frac {\sqrt{t}}{\pi}(\frac
1{|f(\sqrt{t}|^2)}-1)dt$.
By (5.65) one has $f(\sqrt{t})=1+O(\frac 1{\sqrt{t}})$ as
$t\to +\infty$. Thus $d\sigma(t)=O(\frac 1{\sqrt{t}})dt$
as $t\to +\infty$. Therefore integral (13.47') converges
in the classical sense, absolutely, if $Im z\neq 0$, otherwise
it converges in the sense of the Cauchy principal value.

Let us write (11.1) as
\begin{equation}
  m(k)-ik= \int^\infty_{-\infty} e^{ikt}
  \left[ -\sum^J_{j=0} r_j e^{k_jt} H(-t) +a(t)H(t) \right]\,dt,
  \qquad H(t)=\begin{cases} 1, & t\geq 0, \\ 0,&t<0.\end{cases}
  \tag{13.51}\end{equation}

From (13.51) and (13.47') one gets
\begin{equation}
  \int^\infty_{-\infty} \frac{d\sigma(s)}{s-\lambda}
  =\int^\infty_{-\infty} e^{ikt} \alpha(t)\,dt,
  \qquad \lambda=k^2+i0,
  \tag{13.52}\end{equation}
where
\begin{equation}
  \alpha(t):= -\sum^J_{j=0} r_j e^{k_jt} H(-t)+a(t)H(t).
  \tag{13.53}\end{equation}

Taking the inverse Fourier transform of (13.52) one 
can find $\alpha(t)$ in terms of $\sigma(s)$.
If $k>0$ then $k=\sqrt {k^2+i0}$ and if $k<0$ then $k=\sqrt {k^2-i0}$.
Thus:
$$
  \alpha(t)=\frac{1}{2\pi}
     \int^\infty_{-\infty} dk\, e^{-ikt}
  \int^\infty_{-\infty} \frac{d\sigma(s)}{s-k^2-i0}
$$
\begin{equation}
  =-\frac{1}{2\pi} \int^\infty_{-\infty} d\sigma(s)
 [ \int^\infty_{0} dk\,\frac{e^{-ikt}}{k^2+i0-s}+
  \int^0_{-\infty} dk\,\frac{e^{-ikt}}{k^2-i0-s} ].
  \tag{13.54}\end{equation}
Let us calculate the interior integral in the right-hand side
of the above formula. One has to consider two cases: $s>0$ and $s<0$.
Assume first that $s>0$. Then
$$
 \int^\infty_{0} dk\,\frac{e^{-ikt}}{k^2+i0-s}+
  \int^0_{-\infty} dk\,\frac{e^{-ikt}}{k^2-i0-s}
    =  \int^\infty_{-\infty} \frac{e^{-ikt} dk}{k^2-s}+
i\pi\Biggl[\int^\infty_{0} e^{-ikt}\delta (k^2-s) dk-
$$
\begin{equation}
   -\int^{-\infty}_{0}
e^{-ikt}\delta (k^2-s) dk \Biggl]
 =-\frac{\pi}{\sqrt{s}} \sin(\sqrt{s}t)H(s)+J,
  \tag{13.55}
\end{equation}
where 
$$
J:=\int^\infty_{-\infty} \frac{e^{-ikt} dk}{k^2-s}.
$$
If $s<0$, then 
\begin{equation}
 J= \frac{\pi}{ \sqrt{|s|}}e^{-\sqrt{|s|}|t|}.
\tag{13.56}  \end{equation}
If $s>0$, then
\begin{equation}
   J =-\frac{\pi}{\sqrt{s}} \sin (|t|\sqrt{s}).
  \tag{13.57}  \end{equation}
From (13.54)-(13.57) one gets
\begin{equation}
  \alpha(t)= \int^\infty_0 d\sigma(s)
    \frac{\sin(t\sqrt{s}) }{\sqrt{s}} H(t)
    -\frac{1}{2} \int^0_{-\infty}
    \frac{d\sigma(s)}{\sqrt{|s|}}
  e^{-|t|\sqrt{|s|}}.
  \tag{13.58}\end{equation}
Formula (13.58) agrees with (13.53): the second 
integral in (13.58) for $t>0$ is an $L^1(\R_+)$ function,
while for $t<0$ it reduces to the sum in (13.53)
because $d\sigma(s)=d\rho(s)$ for $s<0$,
$d\rho(s)$ for $s<0$ is given by formula (2.5)
and the relation between $c_j$ and $r_j$ is given by 
formula (2.7).

\end{document}